\documentclass[3p,twocolumn,usenatbib]{elsarticle}
\usepackage{lineno,hyperref}
\usepackage{xcolor}
\usepackage{graphicx}
\modulolinenumbers[5]

\usepackage{amssymb}

\journal{Journal of \LaTeX\ Templates}

\biboptions{authoryear}

\newcommand{\Nh}{N_{\rm H}}

\begin{document}

\begin{frontmatter}

\title{INTEGRAL view of AGN}

\author[oas]{Angela Malizia\corref{mycorrespondingauthor}}
\address[oas]{OAS-INAF, Via P. Gobetti 101, 40129 Bologna, Italy}
\address[iki]{Space Research Institute, Russian Academy of Sciences, Profsoyuznaya 84/32, 117997 Moscow, Russia}
\address[CNRS]{Institut National de Physique Nucléaire et de Physique des Particules (IN2P3) CNRS, Paris, France}
\address[esa]{ESA/ESAC, Camino Bajo del Castillo, 28692 Villanueva de la Canada, Madrid, Spain}

\cortext[mycorrespondingauthor]{Corresponding author}
\ead{angela.malizia@inaf.it}
\author[iki]{Sergey Sazonov}
\author[oas]{Loredana Bassani}
\author[oas]{Elena Pian}
\author[CNRS]{Volker Beckmann}
\author[oas]{Manuela Molina}
\author[iki]{Ilya Mereminskiy}
\author[esa]{Guillaume Belanger}

\begin{abstract}
AGN are among the most energetic phenomena in the Universe and in the last two decades {\it 
INTEGRAL}'s contribution in their study has had a significant impact.
Thanks to the {\it INTEGRAL} extragalactic sky surveys, all classes of soft X-ray detected (in 
the 2-10 keV band) AGN have been observed at higher energies as well. 
Up to now, around 450 AGN have been catalogued and a conspicuous part of them are either objects 
observed at high-energies for the first time or newly discovered AGN. 
The high-energy domain (20-200 keV) represents an important window for spectral studies of AGN 
and it is also the most appropriate for AGN population 
studies, since it is almost unbiased against obscuration and therefore free of the limitations 
which affect surveys at other frequencies.
Over the years, {\it INTEGRAL} data have allowed to characterise AGN spectra at high energies, to
investigate their absorption properties, to test the AGN unification scheme and to perform 
population studies. 
In this review the main results are reported and {\it INTEGRAL}'s 
contribution to AGN science is highlighted for each class of AGN.
Finally, new perspectives are provided, connecting {\it INTEGRAL}'s science with that at other 
wavelengths and in particular to the GeV/TeV regime which is still poorly explored.

\end{abstract}


\end{frontmatter}

\tableofcontents

\section{INTRODUCTION}
Active Galactic Nuclei (AGN) are one of the most powerful phenomena in the Universe. 
After many decades of observations and studies, our knowledge of these objects  has made enormous leaps forward.
These galaxies are defined active because they have in their centre an accreting  supermassive black hole (SMBH) of masses higher than 
$\geq$10$^{6}$ M$_{\odot}$ which radiates across the entire electromagnetic spectrum, from the radio up to gamma-rays.
Accretion, i.e. the extraction of gravitational energy from matter infalling onto a black hole, 
is in fact the most efficient bulk mass-energy conversion process known. 
The accreting matter orbits around the black hole and, having some angular momentum, through dissipation of energy, 
flattens to form a disc where magnetic viscosity transfers the angular momentum outward and the mass inward; the accretion disc together with the central black hole makes up the central engine of  an AGN.

It is likely, whether the accretion rate is high or low, that the gravitational energy liberated by this process is 
radiated locally with a large fraction in the form of thermal radiation from the surface of the disc, peaking in 
the optical/UV bands.
A significant amount of these optical/UV photons are reprocessed by a) dust located beyond the sublimation 
radius and re-emitted in the infrared band (IR); and b) by a corona of hot electrons close to the accretion disc
that up-scatters them via inverse Compton in the soft/hard X-ray bands where AGN emit a non negligible fraction 
of their luminosity \citep{1997ASPC..121..101M, Zdziarski_1998}.

However,  disc and corona are just the inner part of the nucleus of an AGN, and the proof that a complex environment 
surrounds this region is that a large fraction of the radiation may be absorbed by interstellar gas and dust 
close to the accretion disc and  likely  re-radiated in  other wavelengths. 
In the classical picture of an AGN, surrounding the accretion disc and the hot corona 
on 0.01--0.1 parsec scale, there is a region of high velocity gas of 1000 -- 5000 km/s, usually referred to 
as the Broad Line Region (BLR), which determines permitted and intercombination broad emission lines in the 
optical spectra and  can  cause ionised absorption producing characteristic features in
the UV and soft X-ray bands (the so-called warm absorber).

At a distance of 1-100 pc from the BH is located an obscuring optically thick torus of gas and dust.  
Whether we consider this torus as dusty-compact or dusty-cloudy, it is the main structure responsible 
for the
absorption of the primary continuum; it can be so thick that it completely hides the primary emission up to several keV.

Going to distances greater than 100 pc up to kpc scales, a biconical shaped and highly structured region of lower velocity gas ($<$ 900 km/s),  the so-called the Narrow Line Region (NLR), is located;
radiation which passes through this region produces permitted, intercombination and forbidden narrow lines in the optical-UV band.

Finally, the interaction between the supermassive black hole's rotating magnetic field and the accretion disc can also create powerful magnetic jets that eject material perpendicular 
to the disc at relativistic speeds and extend for hundreds to thousands of parsecs.

All these ingredients have contributed over the years to explain a wide variety of AGN classes at all wavebands and led to the postulation of the Unified Theory of AGN \citep{1993ApJ...414..506A,Urry_1995} which in
its simplest version hypothesises that the diversity of AGN can be largely explained as a viewing angle effect, although also accretion rate and efficiency play an important role. 

A main ingredient of this orientation-based model is the absorbing material, principally the optically and
geometrically thick torus, which obscures the nuclear regions of an active galaxy (the accretion disc and the hot corona 
as well as the BLR). 
We optically classify an AGN as type 2 or type 1 depending on whether  our line of sight intercepts or not this
obscuring material. Furthermore,  we classify an AGN as radio loud or radio quiet if its emission at radio frequencies 
is either strong or weak.
Within the radio-loud class, differences in typologies (BL Lac objects, Flat Spectrum Radio Quasars and Radio Galaxies) have 
also been explained in terms of orientation, i.e. referring to one or the other type if  the jet axis is perfectly 
aligned with the observer's line of sight  or progressively misaligned.

What clearly emerges from this AGN complex structure  is that absorption is a key ingredient to understand the physics 
of these objects. For this reason, the hard X-ray band  is the most appropriate for AGN population 
studies since it is almost unbiased\footnote{Note that only objects with N$_{H} >$ 10$^{24}$ cm$^{-2}$ could be 
missed in these surveys due to their much dimmer flux which prevents detections by current hard X-ray telescopes} 
against obscuration and therefore free of the limitations which affect surveys at other frequencies, i.e. from optical to soft X-rays. 

Furthermore, the hard X-ray band represents an important window for spectral studies of AGN.
The continuum of active galaxies  in the X-ray band  is well explained by the Comptonisation process which is described by a power law of photon index ($\Gamma$)
in the range 1.5 -- 2 showing  an exponential cut-off (E$_{c}$) at around 100 keV. 
Reprocessing of X-ray photons from the surface of the disc, or from more distant material, can produce in addition 
to  fluorescent emission lines, also a hump at 20-30 keV due to Compton reflection.
Therefore high-energy data are crucial to estimate the slope of the continuum emission over a wide energy band
but also to measure the high energy cut-off and the reflection fraction. These are important parameters 
because they enable us to understand the physical characteristics and the geometry of the region around 
the central nucleus. In other words, in the framework of the disc-corona system, 
while the cut-off energy is related essentially to the temperature $kT_{e}$ of the electrons in the corona, a combination of the temperature and optical depth, $\tau$, of the scattering electrons determines the spectral slope.  
Thus simultaneous measurements of $\Gamma$ and E$_{c}$ allow us to understand the physical parameters of 
the Comptonising region.
Therefore, the  more accurate are the measurements of these parameters, the  better we can determine the geometry and the physical properties of the inner region of AGN.\\
Finally, high-energy data provide  also important information  about the AGN contribution to the 
cosmic X-ray background (CXB). While the fraction of the CXB resolved into discrete sources is $\sim 80$\%
\citep{2017ApJS..228....2L}, $\sim 90$\% \citep{2017ApJS..228....2L}, $\sim 60$\% \citep{2013A&A...555A..42R} 
and $\sim 35$\% \citep{2016ApJ...831..185H} in the 0.5--2, 2--7, 5--10 and 8--24 keV energy bands, respectively, 
it becomes much lower (see section 4.4) at higher energies, near the peak of the CXB spectral intensity at around 30 keV. 
In order to reproduce the shape of the CXB, synthesis models (e.g., \cite{Comastri_2006}) need to use several parameters, 
such as the fraction of heavily obscured sources (the so-called Compton thick AGN characterised by 
N$_{H}$  $\geq$  10$^{24}$ cm$^{-2}$), the coverage and the geometry of the cold gas distributed around the black 
hole responsible for the reflection hump, the photon index and high-energy cut-off of the primary continuum emission 
as well as the luminosity function in the energy range of interest. 
Therefore, the determination of such parameters, in particular photon indices and cut-off energies, their mean values, 
and their distributions over a wide sample of sources, covering a broad
range of energies (above 100 keV), is essential to obtain a much firmer estimate of the AGN contribution to the CXB at high energy. 

In the last decades both {\it INTEGRAL}/IBIS \citep{2003A&A...411L.131U} and {\it Swift}/BAT \citep{Barthelmy_2005}, having good
sensitivity and  wide-field sky coverage, were able to make significant progress in the study of the high-energy 
domain (20-200 keV). 
In particular they have provided a great improvement in our knowledge of the high-energy extragalactic sky by detecting 
more than 1000 (mostly local) AGN at energies above 15 keV. 
It is worth noting that, due to the observational strategy,  {\it INTEGRAL} plays a key role in detecting
new absorbed objects and in particular AGN along the Galactic Plane, 
while {\it Swift}/BAT is more effective at higher Galactic latitudes. 
This  makes the two observatories fully complementary also in the case of extragalactic studies.\\
In this work we will review the contribution of {\it INTEGRAL} and in particular of the imager IBIS to AGN science, 
highlighting the most important results reached in its 17 years in orbit.

\section{From first detections to AGN  catalogues}
As mentioned before, the most important application of {\it INTEGRAL} has been for finding hard X-ray 
emitting sources along the Galactic Plane.
The so-called "Zone of Avoidance"  refers to the area comprised between $\pm$ 10-15$^\circ$, above and below the Galactic plane. 
Gas and dust obscure starlight within this region and screen nearly all background extragalactic objects from 
traditional optical-wavelength surveys; in the optical, as much as 20$\%$  of the extragalactic sky is obscured by the Galaxy.
As a consequence, historically the Galactic Plane has not been a focus for  extragalactic astronomy. 
Hard X-rays ($\ge$ 10 keV)  are however able to penetrate this zone providing a window that is virtually  free of 
obscuration relative to optical wavelengths and partly also to soft X-rays. 

The hard X-ray band has been poorly explored before the {\it INTEGRAL} and {\it Swift} satellites and the only previous truly all-sky survey conducted,  dates back to the eighties. 
This pioneering work, made with the {\it HEAO1}-A4 instrument \citep{Levine_1984},  yielded a catalogue of about  
70 sources down to flux  level of typically 1/75 of the Crab (or 2-3 $\times$ 10$^{-10}$ erg cm$^{-2}$ s$^{-1}$) 
in the 13-80 keV band. 
Only 7 extragalactic objects were reported in this survey: none of these objects is within 10$^\circ$ of the Galactic 
Plane and only two (Centaurus~A and the Perseus cluster) are located below 20$^\circ$ in Galactic latitude. 
Pointed observations by {\it BeppoSAX}/PDS \citep{Frontera_1997} have unveiled more sources  but  observations were 
sometime limited by the non imaging capability of the high energy instrument (PDS)
which is particularly crucial  in the Galactic Plane region. 

A decisive step forward in the exploitation of the entire hard X-ray sky has been possible  thanks to the imager IBIS 
on board {\it INTEGRAL},  which  since the beginnings of the mission, allowed the detection of AGN with a sensitivity 
up to a few mCrab in the most exposed regions (i.e. the Galactic centre) 
with an angular resolution of 12 arcmin and a point source location accuracy of 2-3 arcmin \citep{2003A&A...411L.131U}. 

\begin{figure}
    \includegraphics[width=\columnwidth]{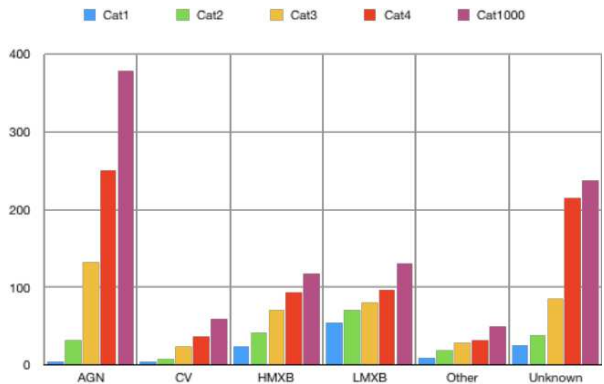}
  \caption{Evolution of source type and number through the five {\it INTEGRAL} IBIS/ISGRI catalogs produced to date (courtesy of A. J. Bird)}
  \label{histo}
\end{figure}

The  capabilities of {\it INTEGRAL}/IBIS in studying extragalactic sources, were  revealed soon after 
the launch of the satellite and in particular  during the Core Programme which lasted for the first 
five  years of the mission and consisted of   45 - 35\% of the total observing time. 
The Core Program (CP) was dedicated to key investigations and devoted to regular surveys of the 
Galactic Plane and selected deep sky fields at high galactic latitudes.
By observing around 9000 square degrees of the sky, the CP allowed the detection of a dozen  AGN as well as 
a quite large number of new unidentified objects firstly detected at high energies.
Most of these firstly detected AGN were bright Seyfert 2 systems, i.e. absorbed objects, 3 of which were 
Compton thick, all located in the Galactic Plane \citep{2004cosp...35.4369B,2006ESASP.604..667S,2004ESASP.552..139B}.
Many of these AGN have been previously studied at energies above 20 keV and for them {\it INTEGRAL} largely confirmed previous findings.
Among these, there is also the first blazar detected by {\it INTEGRAL}: PKS~1830-211 at a relatively 
high redshift ($z$ = 2.507) \citep{2004ESASP.552..139B}.   

Furthermore, IBIS/ISGRI data of the brightest AGN have been used alone or in conjunction with soft 2--10 keV 
data to perform dedicated studies \citep{2004A&A...421L..21S,2005A&A...444..431S,Beckmann_2005} 
exploring their spectral characterisation. 

It became soon evident that the population of AGN emitting above 20 keV was growing thanks to the discovery  
that many of the new detections were indeed  active galaxies.

\begin{figure*}[h]
\centering
\begin{minipage}{0.75\textwidth}
\includegraphics[trim= 0mm 0cm 0mm 0cm,
  width=1\textwidth,clip=t,angle=0.,scale=0.98]{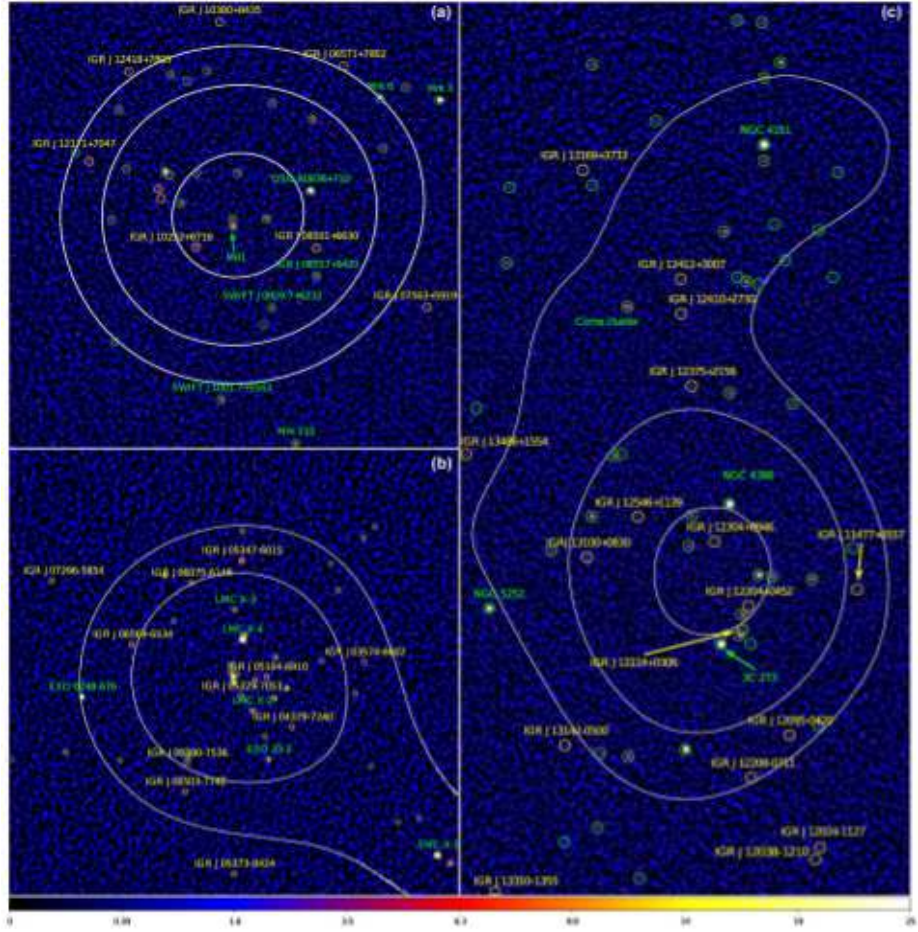}
\end{minipage}
  \caption{Hard X-ray maps of the M81, LMC and 3C 273/Coma fields as in \cite{Mereminskiy_2016}}
  \label{sky}
\end{figure*}

The first IBIS survey \citep{Bird_2004}, based on the first year of {\it INTEGRAL} observations, counted only 5 AGN 
which became 33 (almost 20\% of the entire catalogue) in the second survey \citep{Bird_2006}. 
A couple of dedicated AGN surveys were published by \citet{Beckmann_2006} and \citet{2006ApJ...636L..65B} which listed 
42 and 66 sources respectively.
Both surveys highlighted the capability of {\it INTEGRAL} to probe the extragalactic high energy sky  and most of 
all to find new and/or absorbed active galaxies.
Since 2004, a sequence of IBIS all-sky survey catalogues \citep{Bird_2004,Bird_2006,Bird_2007,Bird_2009, Bird_2016} 
based on data from the ISGRI detector have been published at regular intervals, making use of
an ever-increasing data set as new observations become publicly available. 
The last edition of the IBIS all sky survey \citep{Bird_2016} lists 939 sources: is clearly visible from Figure 
\ref{histo}, the majority of these sources are AGN and newly discovered (i.e. those with an IGR designation) sources 
which, in large part, are expected to be AGN after proper follow up.
On the other hand, also deep extragalactic surveys have been produced over the years thanks to long pointings at
specific sky areas  such as the Large Magellanic Cloud \citep{Grebenev_2012,Mereminskiy_2016},
the  3C~273/Coma and M81 regions \citep{2008A&A...485..707P,Mereminskiy_2016}.
Figure \ref{sky} taken from \citet{Mereminskiy_2016}, shows these  three fields (M81 (exposure of 9.7 Ms), Large Magellanic Cloud (6.8 Ms) and 3C~273/Coma (9.3 Ms)) in the 17-60 keV band as seen by {\it INTEGRAL}/IBIS  after  12 years (2003-2015) of observations.

As the total number of known or newly discovered AGN grew,  it was possible to assemble them for population studies as done over the years by \cite{2012MNRAS.426.1750M} and \cite{Malizia_2016}.  So far the number of AGN listed in this dataset amounts to more than 400 objects.

It is also worth mentioning here the IBIS high energy catalogues i.e. those collecting sources detected above 100 keV. 
There were 10 AGN listed in \cite{2006ApJ...649L...9B} and 28 AGN in \cite{2015MNRAS.448.3766K}, 
7 of which detected also in the 150--300 keV band.
Furthermore, a catalogue produced in 2008 by  \citet{Bouchet_2008}  based on SPI (the other primary wide field instrument 
on {\it INTEGRAL}) data, reported 34 AGN,   10 of  which detected up to 200 keV and 4 in the 200--600 keV band.

As mentioned above {\it INTEGRAL} gave a fundamental contribution in finding new high energy emitters, their number 
increasing more and more along surveys. The {\it INTEGRAL} community made a huge effort
in the identification and classification of these new IGR sources. 
This led to the discovery of new AGN as well as of new classes of objects emitting at high-energies.
However, the classification of a new, high-energy detected source is by no means trivial.  
First of all, one has to reduce the positional uncertainty associated with the high energy detection, 
which in some cases can be as high as 4--5 arcminutes. 
To do this, 2-10 keV data are important because they provide a unique tool to associate the high energy source with 
a single/multiple X-ray counterpart/s. 
When archival observations were not available, follow-up campaigns  have been performed with all  X-ray instruments such 
as {\it XMM}, {\it Chandra} and {\it Swift}/XRT which allowed also the spectral characterisation of the 
sources \citep{2005A&A...444L..37S,Malizia_2007,2008A&A...487..509S,Landi_2010,2017MNRAS.470.1107L}.

Association of the X-ray source with an object detected at other wavelength is a fundamental step in the analysis as it allows to pinpoint the correct counterpart, locate it with 
arcsec accuracy and therefore provide a way to study the source at other wavelengths. Optical, IR or radio catalogues 
are then searched  in order  to find the appropriate classification of the object. 
If this search does not yield any result, follow-up observations, above all in the optical, are then planned and carried out.

A parallel effort regarding the spectroscopic identification of newly-detected or poorly studied hard X-ray sources has been 
performed by several groups worldwide soon after the publication of the 1st IBIS survey.

To this date, this task has allowed the determination of the nature of around  150 AGN, with nearly 60\% of them 
classified as broad emission line nuclei. 
The bulk ($\sim$90\%) of these identifications stemmed from the program of Masetti and collaborators which encompassed the 
use of at least a dozen telescopes across the globe  (see \citet{2013A&A...556A.120M} and references therein).
Further AGN identifications from other groups have also been reported in the literature like in 
\citet{Bikmaev_2006, Bikmaev_2008, 2009A&A...502..787Z,Karasev_2018,2018A&A...618A.150F}.
It is worth noting that an overlap of detected sources is present across the {\it INTEGRAL} and {\it Swift}/BAT 
(e.g. \citet{Oh_2018}) catalogues. Therefore, several identifications of {\it Swift} hard X-ray 
emitters may be also accommodated in the {\it INTEGRAL} surveys (e.g., \citet{2017A&A...602A.124R} 
and references therein; \citet{Karasev_2018}).
Thus, the total number of identifications reported above should actually be considered as a strict lower limit.

This association/classification method has also been used to construct {\it INTEGRAL} AGN catalogues and so guarantees
that all objects in these catalogues are fully characterised in terms of optical identification/classification  
as well as  fully studied in terms of  X-ray spectral properties.

 \begin{figure}[t]
    \includegraphics[width=\columnwidth]{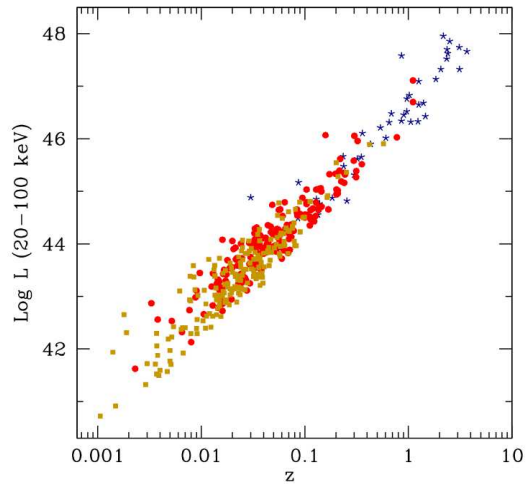}
  \caption{Observed hard X-ray (20-100 keV) luminosity versus redshift for the whole {\it INTEGRAL} AGN sample. Gold filled circles are narrow line AGN, red filled squares are broad line AGN and blue stars are balzars.}
  \label{lum}
\end{figure}

\section{AGN Types and Population studies }
The total number of  AGN so far detected by {\it INTEGRAL}, including recent additions, amounts to  440 objects.
In Figure \ref{lum}  their 20--100 keV observed luminosity is plotted against redshift, differentiating objects in 
three main optical classes: broad line AGN (red filled squares), narrow line AGN (gold filled circles) and blazars 
(blue stars).
The luminosities have been calculated for all sources assuming H$_{0}$=69.6 km s$^{-1}$ Mpc$^{-1}$ and q$_{0}$ = 0.

We find that the source redshifts span a range from 0.00084 to 3.7 with a median of  $z$=0.035,
while the Log of 20-100 keV luminosities in ergs s$^{-1}$ (assuming isotropic emission) ranges  from  40.23 to $\sim$48 with a mean at around 44.
M81 (a Seyfert 1.8/LINER)  is the closest and least luminous AGN seen by {\it INTEGRAL}, while IGR~J22517+2218 (a broad line QSO) 
is the farthest and most luminous object so far detected; the former hosts a  black hole  of mass  M = 7 $\times$ 10$^{7}$ M$_\odot$ while the latter houses a more massive one  (M = 10$^9$ M$_\odot$, \cite{2012MNRAS.421..390L}).
{\it INTEGRAL} also detected  NGC~4395, a Seyfert 2 galaxy which hosts a black hole  of about 10$^{4}$ M$_\odot$; 
this mass has  recently been estimated through reverberation mapping of the broad line region and resulted 
to be among the smallest central black hole masses ever reported for an AGN \citep{Woo_2019}.
In conclusion, the {\it INTEGRAL} sample spans a large range in source parameters and is therefore representative of 
the  population of AGN selected in the hard X-ray band.

\begin{figure}
    \includegraphics[width=\columnwidth]{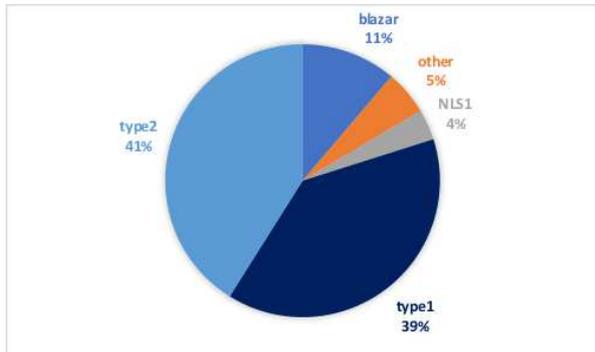}
  \caption{Pie chart of the main classes of AGN in the {\it INTEGRAL} sample.}
  \label{pie}
\end{figure}

After 17 years of {\it INTEGRAL} surveying the extragalactic sky, we can now say that all classes of AGN 
that are  seen in the 2-10 keV band, are also  detected at higher energies. 
In the pie chart of Figure \ref{pie} the main classes seen by {\it INTEGRAL}  are highlighted: it is evident from the 
figure that a large fraction is made up of Seyfert galaxies, equally divided in type 1 and 2; the second most numerous 
class is that of blazars, followed by a small number of objects of other classifications which are also 
interesting to study. 
The large database accumulated,  allowed over the years to probe new AGN classes,  to investigate the absorption 
properties of active galaxies, to test the AGN unification scheme and to perform population studies.

For example {\it INTEGRAL} has detected for the first time at high energies Narrow Line Seyfert 1 galaxies (NLS1) 
\citep{2008MNRAS.389.1360M}. These are interesting targets as they are characterised by unique properties 
when compared to their broad line analogues, both in the optical (see e.g. \cite{1985ApJ...297..166O}) and in the X-rays, 
where they show stronger variability (both in flux and spectral shape) and steeper power law spectra.
The most widely accepted explanation for these differences is that NLS1 have smaller black hole masses than normal Seyfert 1s;
however their luminosities are comparable \citep{1995MNRAS.277L...5P}, suggesting that they
must be emitting at higher fractions of their Eddington luminosity and therefore should also have higher fractional accretion
rates. A plausible scenario suggests that black holes in NLS1 have not yet been fed enough to become massive and are in a
rapidly growing phase \citep{2000MNRAS.314L..17M}; if NLS1 are indeed in an early phase of black hole evolution, 
then they are key targets for the study of AGN  formation and evolution. 
Within the  INTEGRAL AGN sample only 15 objects  are classified as NLS1 (or 3$\%$ of the sample). Most of these objects  have
been studied in detail by \citet{2011MNRAS.417.2426P} who found that hard X-ray selected NLS1 show variability 
over a broad range of X-ray frequencies, lack a strong soft excess, and often  display fully or partially covering absorption.
As expected, NLS1 detected at high energies by {\it INTEGRAL} are also associated to small black hole masses and occupy the 
lower tail of the Eddington ratios distribution with respect to classical NLS1.

Another interesting type of AGN  first detected in hard X-rays by {\it INTEGRAL} are the  XBONGs, i.e. 
X-ray Bright Optically Normal Galaxies \citep{2002ApJ...571..771C} which are  bright in X-rays (X-ray luminosity 
of 10$^{43}$-10$^{44}$ erg s$^{-1}$) but are  optically dull, i.e. they are hosted by normal galaxies whose optical 
spectra show no emission lines. 
Over the years, XBONG have shown to be a mixed bag of objects primarily including normal elliptical galaxies and AGN whose optical 
nuclear spectrum is probably diluted by the strong stellar continuum.
{\it INTEGRAL} XBONG (6 detected so far) are all heavily absorbed in X-rays (Log N$_{H}$  from 23 to more than 24 cm$^{-2}$)
and are quite bright above 20 keV (L$_{20-100}$ keV in the range 10$^{42}$ -- 10$^{44}$ erg s$^{-1}$).
These are luminosities typical of AGN which implies that {\it INTEGRAL} can detect heavily absorbed objects whose hosting 
galaxy outshines the active nucleus; these AGN would normally be missed in optical surveys.

Equally absorbed and bright are the majority of LINERs (Low-Ionization Nuclear Emission Regions) seen by {\it INTEGRAL} 
and in fact many of them are classified as LINERs of type 2 i.e. showing only narrow lines in their optical spectra.
There are 20 LINERs in the {\it INTEGRAL} AGN sample and only 2 are of type 1, i.e. unabsorbed.
Also LINERs have been detected at high energies for the first time by {\it INTEGRAL} (first identifications 
by \citet{2008A&A...482..113M}), clarifying the controversial origin (AGN versus starburst) of the ultimate power
source of these objects. 
Given their high 20-100 keV  luminosities, it is almost certain that all {\it INTEGRAL} LINERs are powered by an AGN, even if 
their mean luminosity ($\sim$1.7 $\times$ 10$^{43}$ erg s$^{−1}$) is slightly lower than that of classical Seyfert 
galaxies ($\sim$5.5 $\times$ 10$^{43}$ erg s$^{−1}$). 
The result that emerged by the {\it INTEGRAL} studies \citep{2012MNRAS.426.1750M}, is that LINERs are numerous as a class and like 
Seyfert galaxies,  come in two flavours: unabsorbed type 1 and absorbed type 2, with {\it INTEGRAL} having the capability 
of detecting  the second type in large numbers.

Regarding population studies, \citet{2009A&A...505..417B}  first used a sample of {\it INTEGRAL} detected AGN to study source parameters on a 
large scale. An interesting result that emerged from  this study is the significant correlation  found between the hard X-ray 
and optical luminosity and the mass of the central black hole in the sense that  more luminous objects appear also to be more massive.
This finding allowed to construct a  black hole fundamental plane similar to the one found using radio data
(L$_{V}$ $\propto$ L$_{X}^{0.6}$ M$_{BH}^{0.2}$).

The accurate study, at optical and X-rays frequencies of  all {\it INTEGRAL} AGN has also allowed the study of the correlation between
optical classification and X-ray absorption thus allowing to perform a strong test on the AGN unification model.  
Although the presence of a correlation is expected and indeed found, i.e. type 1 AGN are typically unabsorbed while type 2 AGN 
are often absorbed, the strength of the correlation is never 100\%. Using a large sample of {\it INTEGRAL} AGN, \citet{2012MNRAS.426.1750M} 
have  found that only a small  percentage of sources (12.5\%) does not fulfil the expectation of the Unified Theory and 
looking in depth at these outliers concluded that the standard-based AGN unification scheme is followed by the majority of bright AGN.
The only outliers are  absorbed type 1 AGN characterised by complex X-ray absorption likely due to ionised gas located in an accretion
disc wind or in the biconical structure associated to the central nucleus and so unrelated to the torus.  
The other outliers  are  type 2 AGN  which do not show X-ray absorption but this  could be either due to 
variability (meaning that their different optical/X-ray classifications can be explained in terms of state transitions and/or non
simultaneous X-ray and optical observations) or to  a high dust-to-gas ratio.

Finally, \citet{2015MNRAS.447.1289P} studied  the radio properties of a complete sample of {\it INTEGRAL} detected AGN, found 
a significant correlation between the radio flux at 0.8/1.4 GHz and the 20-100 keV flux,  with a slope between these two parameters
consistent with that expected for radiatively efficient accreting systems. 
This indicates that the high-energy emission coming from the inner accretion regions correlates with the radio emission averaged 
over hundreds of pc scales (i.e., thousands of years). 

\begin{figure}
    \includegraphics[width=\columnwidth]{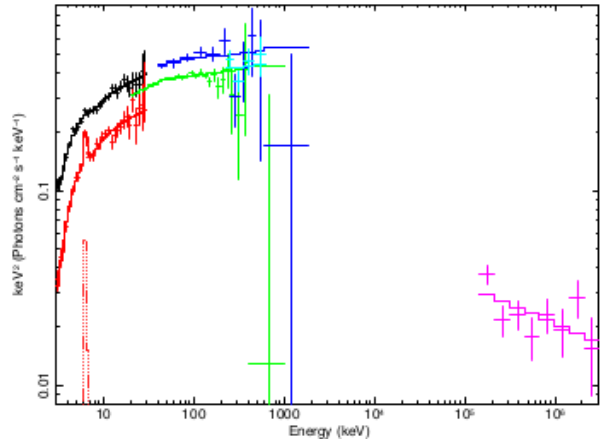}
  \caption{Combined {\it INTEGRAL} and (non-simultaneous) {\it Fermi}/LAT unfolded spectrum of Cen~A as in \cite{2011A&A...531A..70B}. The data are modelled by a double broken power-law with individual normalisation for the data of different epochs. The highest energy bins for IBIS/ISGRI and SPI are only upper limits.}
  \label{cena}
\end{figure}

\section{Main Results: Seyfert galaxies}
\subsection{Broad Band studies and high energy cutoff}
The study of the X/hard-X ray emission of AGN, i.e. from 2 keV to $\geq$100 keV,  is a fundamental tool in order
to have a direct probe of their innermost regions. As said before, the primary continuum over this broad 
band can be roughly represented by a cut-off power law  produced by the Comptonisation mechanism which 
is believed to arise in the corona, close to the central super-massive black hole. 
Clearly to have an overview of the physics and structure of the corona we need to study 
the broad-band spectra of a large sample of AGN in order to account for all spectral components, 
remove the degeneracy between parameters and therefore being able to obtain
a precise estimate of the photon index and high-energy cut-off for a large number of objects.
These two parameters are, as said, strictly linked to the temperature and the optical depth of the corona \citep{2001ApJ...556..716P}.\\
Before {\it INTEGRAL} and {\it Swift}/BAT, broad band spectra were available only for a limited number of bright AGN, 
basically due to the scarcity of measurements above 10-20 keV, with most of the information coming 
from broad-band spectra provided by the {\it BeppoSAX} satellite which had a broad energy coverage (0.1-100 keV) 
but no imaging capability above 10 keV.
Analysis of {\it BeppoSAX} data of types 1 and 2 AGN \citep{2002A&A...389..802P,2007A&A...461.1209D,2003ApJ...589L..17M} 
gave evidence for a wide range of values for the cut-off energy, ranging from 30 to 300 keV 
and higher, and an apparent different distribution of photon index and reflection parameter in different classes of AGN, e.g. \citet{2003ApJ...589L..17M}.\\
Spectral parameters of a few individual local bright AGN have been measured using {\it INTEGRAL} data with the addition of very high 
energy data as in the case of Cen~A \citep{2011A&A...531A..70B}, see Figure \ref{cena}. 
More sources have been studied using {\it INTEGRAL} data in conjunction with those of other
X-ray satellites, mainly {\it Swift}/XRT and {\it XMM-Newton}, e.g. NGC~4151 \citep{2010MNRAS.408.1851L}, NGC~4388
\citep{2004ApJ...614..641B}, NGC~2110 \citep{2010int..workE..81B}, NGC~4945 \citep{2016KPCB...32..172F} and many others.

Spectral studies have also been performed for the first time for AGN located in the Galactic 
Plane such as for GRS~1734-292 and other 6 objects as in \cite{2006MNRAS.371..821M}.
Furthermore, thanks to the good statistical quality of the {\it INTEGRAL} data, for these local AGN, 
flux variability and spectral changes have been studied for the first time for many of them 
(e.g. \citet{2008A&A...486..411S, 2016KPCB...32..172F,2013MNRAS.433.1687M}).   
Although snapshot observations in the soft X-rays (typically from {\it XMM-Newton} or from {\it Swift}/XRT) and
time-averaged (on timescales of years) measurements at high energies are not contemporaneous and acquired differently,  
the match between the two is generally good, with the cross-calibration constant between 
the two instruments being typically around 1 \citep{2008A&A...483..151P, 2012MNRAS.420.2087D, 2013MNRAS.433.1687M, 2011MNRAS.417.1140F}.
Broad-band spectral  analysis of the brightest Setyfert 1 galaxies as in   \citet{2008A&A...483..151P} 
and Seyfert 2 as in \citet{2012MNRAS.420.2087D}, have also been performed 
defining the characteristics of the two classes on  quite large samples of objects.

\begin{figure}
    \includegraphics[width=\columnwidth]{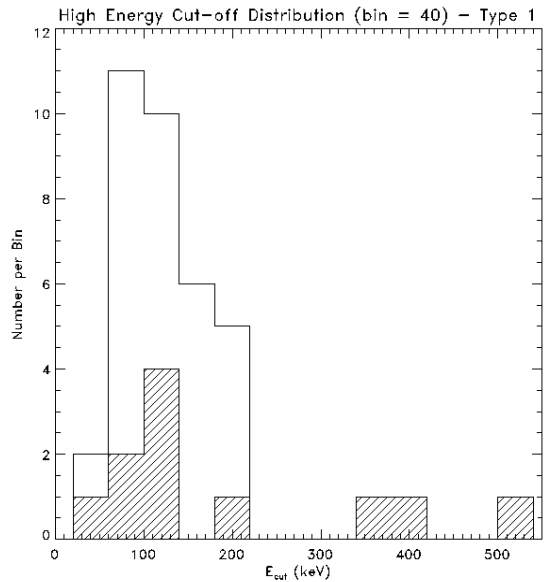}
  \caption{High-energy cut-off distribution of the entire sample from \cite{2009MNRAS.399..944M}. The diagonally
hatched histogram represents sources for which only lower limits of E$_{c}$ are available.}
  \label{cutoff}
\end{figure}

The broad-band spectral analysis of 41 Seyfert 1 galaxies belonging to the {\it INTEGRAL} complete 
sample \citep{2009MNRAS.399..944M} performed by fitting
together {\it XMM}, {\it Swift}/BAT, and {\it INTEGRAL}/IBIS data in the 0.3--100 keV energy band 
allowed \citet{2014ApJ...782L..25M} to confirm the distribution of photon indices ($\Gamma$= 1.73 
with standard deviation of 0.17) and, for the first time, to provide the high-energy cut-off distribution 
for a large sample of objects. \citet{2014ApJ...782L..25M} found that the mean high energy cut-off was
128 keV with a spread of 46 keV (see Figure \ref{cutoff}), clearly indicating that the 
primary continuum typically decays at much lower energies than previously thought. This value is more in line 
with the synthesis models of the cosmic diffuse background, which often assume an upper limit of $\sim$200 keV 
for the cut-off  (\citet{2007A&A...463...79G}, see also section 4.4).

In \cite{2014ApJ...782L..25M} the main parameters of the primary continuum have been estimated by employing a 
baseline phenomenological model (\texttt{PEXRAV} model in \texttt{XSPEC}) composed of an exponentially 
cut-off power-law reflected from neutral matter. At soft energies, when required by the data, intrinsic absorption 
in terms of simple or/and complex, cold or ionised absorbers have been added and when 
spectra showed clear signs of a soft excess, this has been generally fitted with 
a thermal component. A gaussian line has also been included, to take into account the  
presence of the iron k$\alpha$ line at around 6.4\,keV; and when present residuals around 7\,keV, 
these have also been fitted adding a second gaussian line to take into account the iron k$\beta$ feature.

With these spectral parameters \cite{2014ApJ...782L..25M}, following \cite{2001ApJ...556..716P}, 
have been able to evaluate the actual physical parameters of the Comptonising region by assuming the plasma temperature to be
kT$_{\rm e}$=E$_{\rm c}$/2, for optical depth $\tau$ less or $\sim$1, and kT$_{\rm e}$=E$_{\rm c}$/3 for $\tau$ $\gg$1.
The most probable range of plasma temperatures kT$_{\rm e}$ derived from this study was found to be in the range 20 
to 100 keV (or 2 - 12 $\times$10$^{8}$\,K).
Assuming the average value of $\Gamma$=1.73 and taking into account both low and high values of  E$_{\rm c}$, 
acceptable solutions for $\tau$ in the range 1 to 4 have been obtained. 
These results are in good agreement with those previously found by \citet{2001ApJ...556..716P} for a small sample of 
Seyfert 1 galaxies with {\it BeppoSAX} observations, and by \citet{2009A&A...505..417B} applying a cut-off power law model 
to  the stacked JEM-X plus IBIS spectral data  of Seyfert 1    (E$_{c}$ = 86$^{+21}_{-14}$ keV). \\
Furthermore, the high energy cut-off distribution of the {\it INTEGRAL} complete sample of type 1 AGN is in good 
agreement with what found by \citet{2016MNRAS.458.2454L} employing a more physical model  (\texttt{COMPPS} model) 
on a sample of 28 bright AGN (type 1 and 2) and analysing {\it INTEGRAL} data together with soft X-ray ones acquired 
by {\it XMM-Newton}, {\it Suzaku} and {\it RXTE}.
\citet{2016MNRAS.458.2454L} tested several model options assuming a thermal Comptonisation of the primary continuum
accompanied by a complex absorption and a Compton reflection. They accurately determined 
the mean temperature of the electron plasma to be 26 $\leq$ kTe $\leq$ 62 keV for the  majority of the sample objects 
with only two sources exhibiting   temperatures  kT$_{e}$ $>$ 200 keV. 
These low temperatures of the electron plasma obtained in all these studies, imply that the template Seyferts spectra 
used in the population synthesis models of AGN should be revised: the most important consequence of a shifted high-energy 
cut-off will be a considerably smaller fraction of CT AGN needed to explain the peak of the CXB spectrum.

Most of the works mentioned above, made use of non-simultaneous low versus high energy data, and despite 
the introduction of cross-calibration constants to properly take into account flux variability (and possible mismatches 
in the instruments calibration), some degree of uncertainty remains since spectral variability cannot be excluded 
{\it a priori}. 
Therefore, if one wants to remove this ambiguity, it is fundamental to have simultaneous observations in both the 
soft and hard X-ray bands and this is now achievable with {\it NuSTAR}.
In a recent work \cite{2019MNRAS.484.2735M} presented the 0.5 -- 78 keV spectral analysis of 18 broad line AGN belonging 
to the {\it INTEGRAL} complete sample, those for which simultaneous {\it Swift}/XRT and {\it NuSTAR} observations were available. 
Employing the same simple phenomenological model to fit the data as in \cite{2014ApJ...782L..25M}, these authors
found a mean high-energy cut-off of 111 keV ($\sigma$ = 45 keV) for the whole sample, in perfect agreement 
with what previously found employing {\it INTEGRAL} data.
These findings also confirm that simultaneity of the observations in the soft and hard X-ray band is not essential once flux and spectral
variability are properly accounted for.

Finally, compatible values of photon index and cut-off values have been found by \cite{2013MNRAS.433.1687M} using 
only high energy  data i.e. {\it INTEGRAL}/IBIS and {\it Swift}/BAT. Furthermore, this study allowed also a 
cross-calibration between the two instruments, finding general good agreement between BAT and IBIS spectra, 
despite a systematic mismatch of about 22 per cent in flux normalisation. 

These and other high-energy spectral studies prompted Fabian and collaborators \citep{2015MNRAS.451.4375F,2017MNRAS.467.2566F}
to propose the pair thermostat model to explain the observations. 
In the compactness/temperature diagram, AGN coronae which are hot and radiatively compact, are located 
close to the boundary of the region which is forbidden due to runaway pair production.
Pair production and annihilation can be considered essential ingredients in AGN corona physics 
and strongly affect the shape of the observed spectra. In fact, if photons are energetic enough, the subsequent increase in luminosity 
produces electron-positron pairs rather than an increase in temperatures, until a  point of  equilibrium is reached. 
At this point, pair production consumes all the available energy, therefore limiting the coronal temperature; electron-positron pair production 
becomes a runaway process thus acting as a sort of thermostat. 
Pair production from the non-thermal component (hybrid plasma compared to pure thermal plasma) can  reduce the temperature leading
to a much wider range of values, more consistent with present observations \citep{2017MNRAS.467.2566F}.

\subsection{Absorption and Compton thick fraction}
As said before, an obscuring 'torus' is believed to be responsible for the type 1 and 2 division of Seyfert galaxies 
and quasars. What is the actual structure of the torus and what is its physical relation to other structural components 
of AGN such as the accretion disc and broad-line region, is still unclear although progress in understanding this 
issue has been recently made \citep{Ricci_2017,2019ApJ...884..171H}.
These are some of  the central questions in AGN research and hard X-ray surveys provide a more direct answer than studies 
at lower energies because these surveys are not affected by absorption bias except for very heavy absorption.

When an X-ray telescope observes an AGN through the torus, the measured spectrum will exhibit a characteristic low-energy 
cutoff due to the photoabsorption of the soft radiation in the gas and dust of the torus (and perhaps also in the 
enclosed broad-line region), which allows one to estimate the absorption column density, $\Nh$, along the viewing direction. 
A statistical analysis of absorbing columns determined in this way (or by a similar but more model-dependent method 
for Compton-thick AGN) for a representative sample of objects makes it possible to infer the typical optical depth 
and covering fraction of the torii in the AGN population. With this understanding, a lot of efforts have been 
put into follow-up X-ray spectral observations of {\it INTEGRAL} (and {\it Swift}) detected AGN (see the previous sections),
which has eventually resulted in a unique and well characterised sample of local ($z\lesssim 0.2$) hard X-ray selected AGN. 

One of the earliest attempts to use {\it INTEGRAL} data for AGN absorption studies was undertaken by \cite{Beckmann_2006}. 
They made use of a sample of 38 Seyfert galaxies detected by IBIS in the first year of the mission, with absorption 
information available for 32 of them. Already this limited statistics provided a hint that the fraction of 
absorbed ($\Nh>10^{22}$~cm$^{-2}$) AGN decreases with increasing hard X-ray (20--40~keV) luminosity. 
Although the statistical significance of this result was very low, it was consistent with other emerging indications of 
such a trend both in the  local \citep{2004A&A...423..469S,2005ApJ...633L..77M} and distant
\citep{2003ApJ...596L..23S,2003ApJ...598..886U} Universe. The \citet{Beckmann_2006} sample included 4 Compton-thick
($\Nh>10^{24}$~cm$^{-2}$) objects, all previously known (NGC~1068, Mrk~3, NGC~4945 and Circinus galaxy), implying that the
observed fraction of such AGN is $\sim$10\% or somewhat higher, taking into account the absence of absorption information for
several objects. 

\cite{Malizia_2007} carried out a statistical analysis of 38 {\it new} AGN discovered by {\it INTEGRAL}/IBIS (some of these
objects have also been independently found by {\it Swift}/BAT) and followed up with {\it Swift}/XRT. Sixteen objects proved to
be absorbed AGN and three others were suggested to be Compton-thick based on the low ratios of the observed fluxes in the
2--10~keV and 20--100~keV energy bands. Therefore, the inferred fractions of absorbed and Compton-thick objects ($\sim 50$\% and
$\sim 10$\%) proved to be in good agreement with the results of \cite{Beckmann_2006} despite the largely different samples of
AGN (newly discovered vs. mostly previously known objects). 

A substantially larger sample of AGN, based on the first three and a half years of observations of the sky with IBIS
\citep{2007A&A...475..775K}, was analysed by \cite{2007A&A...462...57S}. Specifically, they used a set of 66 Seyfert 
galaxies located at $|b|>5^\circ$. The enhanced sample provided increased evidence that the fraction of absorbed 
AGN decreases with increasing luminosity (17--60~keV). The observed fraction of Compton-thick objects was again 
found to be $\sim 10$\% with an upper limit of $\sim 20$\% (taking into account missing information on the absorption 
columns for some of the objects).

\begin{figure}
\centering
    \includegraphics[width=\columnwidth,viewport=20 180 560 710]{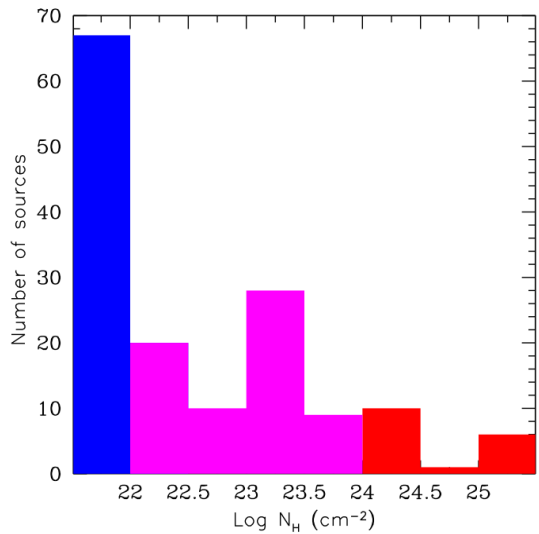}
  \caption{Observed distribution of absorption columns. Unabsorbed ($\Nh<10^{22}$~cm$^{−2}$), weakly absorbed ($10^{22}\le\Nh\le 10^{24}$~cm$^{−2}$) and Compton-thick ($\Nh\ge 10^{24}$~cm$^{−2}$) objects are shown in blue, magenta and red, respectively. From \citep{2015MNRAS.454.1202S}.
}
  \label{obs_nh_distr}
\end{figure}

Later on, several more studies \citep{2009MNRAS.399..944M,2009A&A...505..417B,2010int..workE...6S,2015MNRAS.454.1202S} 
have taken advantage of the ever increasing catalogue of {\it INTEGRAL} detected AGN and follow-up efforts to tighten 
the constraints on the absorption properties of the local AGN population.
In particular, using {\it INTEGRAL} selected AGN, it was assessed for the first time that even at high energy a bias 
in the estimation of the fraction of Compton thick sources still exists \citep{2009MNRAS.399..944M}. 
Their flux is reduced due to sensitivity limit, but if corrected their fraction turns out to be of $\sim$20\%,
which is more in line with estimates at other wavebands.
Afterwards this results has been confirmed using {\it Swift}/BAT data \citep{2011ApJ...728...58B}.

An important step forward has been made by \cite{2015MNRAS.454.1202S} using  a sample of 151 local Seyfert galaxies detected in
the 17--60~keV energy band by IBIS (at $|b|>5^\circ$), from the 7-year all-sky catalogue of \citet{2010A&A...523A..61K}. 
This sample is highly complete in terms of supplementary information (optical types, distances and X-ray absorption columns) 
and consists of 67 unabsorbed ($\Nh<10^{22}$~cm$^{-2}$) and 84 absorbed ($\Nh>10^{22}$~cm$^{-2}$) objects including 17 
proven or likely Compton-thick ($\Nh>10^{24}$~cm$^{-2}$) AGN (see Figure~\ref{obs_nh_distr} and \cite{2019AstL...45..490S}). 
The observed fractions, i.e. without the correction for the absorption bias, of absorbed and Compton-thick objects turned out to
be nearly the same ($\sim 60$\% and $\sim 10$\%) as in the previous analyses of smaller samples, despite the increased depth of
the IBIS survey making it possible to probe a broader (in terms of luminosity and distance) population of AGN. A similar
fraction ($\sim 8$\%) of Compton-thick AGN is also observed in the {\it Swift}/BAT survey \citep{2015ApJ...815L..13R}.

\begin{figure}
\centering
    \includegraphics[width=\columnwidth,viewport=0 150 590 730]{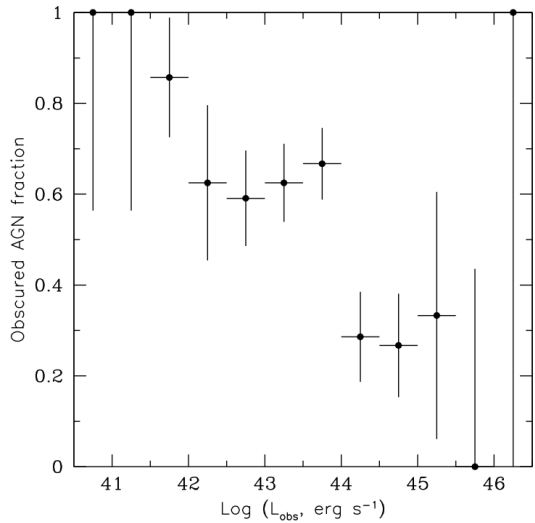}
  \caption{Observed fraction of absorbed AGN as a function of observed hard X-ray luminosity. From \citep{2015MNRAS.454.1202S}.}
  \label{obs_obsc_frac}
\end{figure}

\cite{2015MNRAS.454.1202S} decisively ascertained the declining trend of the observed fraction of absorbed AGN with increasing
luminosity (see Figure~\ref{obs_obsc_frac}). A similar dependence has been independently established for the local AGN population
using the {\it Swift}/BAT hard X-ray survey \citep{2011ApJ...728...58B} and for higher-redshift AGN with surveys conducted in
the standard X-ray band (corresponding to the hard X-ray band in the rest-frame of quasars at $z\gtrsim 1$)
\citep{2014ApJ...786..104U}.
\cite{2015MNRAS.454.1202S} suggested that this may be at least partially a selection effect, because not only 
hard X-ray, flux-limited surveys are negatively biased with respect to Compton-thick AGN, but hard X-ray selection is also
positively biased with respect to unabsorbed ones (due to the reflection of part of the central source's hard X-ray radiation
towards the observer). This implies that the intrinsic fraction of absorbed sources at a given luminosity must be higher than
the observed one. \cite{2015MNRAS.454.1202S} further speculated that there is possibly no intrinsic declining trend of this
fraction with luminosity if the hard X-ray emission from the accretion disc's corona is weakly collimated along its axis, as can
well be the case. If so, the covering fractions of the torii in the local Seyfert galaxies should typically be as high as $\sim
80$--90\%. Finally, \cite{2015MNRAS.454.1202S} demonstrated that Compton-thick objects intrinsically amount to nearly half of
all absorbed AGN. Perhaps surprisingly, these hard X-ray findings are in good agreement with conclusions reached two decades ago
based on optically selected samples of Seyfert galaxies  \citep{1995ApJ...454...95M,1999ApJ...522..157R}.

\subsection{Luminosity Function}
Together with {\it Swift}/BAT, {\it INTEGRAL}/IBIS multi-year observations have enabled an accurate determination of the hard
X-ray luminosity function of local Seyfert galaxies. Apart from statistically characterising the supermassive black hole
activity in present-day galactic nuclei, such a measurement is crucial for studying the cosmic history of black hole growth,
since it provides an all-important $z=0$ reference point for the models of AGN evolution based on deep, pencil-beam X-ray
surveys.

As already mentioned, shortly preceding the {\it INTEGRAL} measurements in the hard X-ray band, the AGN luminosity function was
measured in the softer band of 3–-20 keV in the  XSS survey \citep{2004A&A...423..469S}. This result was based on 95 local
Seyfert galaxies and represented a substantial improvement over previous estimates obtained at energies below 10~keV thanks to
the significantly reduced bias with respect to absorbed AGN and fairly large sample of objects. Nevertheless, XSS was still
strongly biased against sources with $\Nh>10^{23}$~cm$^{-2}$ and it was clear that yet harder (above 15~keV) large-area surveys
were needed for a robust measurement of the luminosity function of local AGN.

The first AGN luminosity function based on {\it INTEGRAL} results was computed by \cite{Beckmann_2006} in the 20--40~keV energy
band using the aforementioned sample of 38 non-blazar AGN detected by IBIS over the first year of the mission. Most of the
objects were nearby (the average redshift is 0.022). The derived luminosity function could be described by a broken power law,
an empirical model commonly used in fitting AGN luminosity functions in various wavebands. 
Afterwards, \citet{2007A&A...462...57S} made a more precise measurement of the hard X-ray luminosity function using a larger
sample of IBIS-detected AGN (66 Seyfert galaxies). 

\begin{figure}
\centering
    \includegraphics[width=\columnwidth,viewport=0 180 580 650]{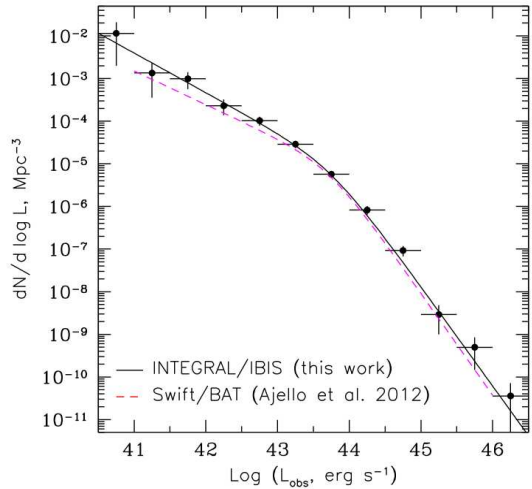}
  \caption{Observed hard X-ray luminosity function of local AGN (circles) fitted by a broken power law (solid line). For comparison, the luminosity function based on the {\it Swift}/BAT survey \citep{2012ApJ...749...21A} is shown by the dashed line. From \citep{2015MNRAS.454.1202S}.}
  \label{obs_lumfunc}
\end{figure}

The most recent version of the hard X-ray (17--60~keV) luminosity function of local AGN based on {\it INTEGRAL}/IBIS data is
presented in \citet{2015MNRAS.454.1202S} (see Figure~\ref{obs_lumfunc}). As mentioned before, these authors used a sample of 151
Seyfert galaxies, which, compared to previous ones, covers broader ranges of distances and luminosities: $z\sim 0.001$--$0.4$
and $L\sim 3\times 10^{40}$--$2\times 10^{46}$~erg~s$^{-1}$. It may nevertheless be considered a local one since most of the
objects are located at $z<0.1$. The derived luminosity function is in good agreement with that based on the {\it Swift}/BAT
survey \citep{2012ApJ...749...21A}. Apart from further tightening the parameters of the {\it observed} hard X-ray luminosity
function, \citet{2015MNRAS.454.1202S} have also reconstructed the {\it intrinsic} luminosity function of local AGN by correcting
for the observational biases related to X-ray absorption  and reflection discussed in the preceding section. As a consequence,
they determined the total intrinsic hard X-ray (17--60~keV) luminosity density of local AGN (with luminosities between
$10^{40.5}$ and $10^{46.5}$~erg~s$^{-1}$): $\sim 1.8\times 10^{39}$~erg~s$^{-1}$~Mpc$^{-1}$. 

It is important to note that although there is substantial room for further improvement of local AGN statistics at the low end
of the luminosity function (below $\sim 10^{41}$~erg~s$^{-1}$), in particular with the  eROSITA and ART-XC telescopes aboard the
recently launched {\it Spektr-RG} satellite, there is no such possibility at the high end ($L\gtrsim 10^{45}$~erg~s$^{-1}$)
since {\it INTEGRAL}/IBIS and {\it Swift}/BAT have already probed the whole Universe out to $z\sim 0.2$ at these luminosities
and found just a few such quasar-like objects. This low number density of luminous AGN at the present epoch reflects the
well-established fact that the main stage of growth of supermassive black holes took place in the remote past. 

\subsection{Links to the cosmic X-ray background}
For a significant fraction of the extragalactic sky, in particular in the M81, LMC and 3C~273/Coma fields, IBIS observations have
now reached a depth of $\sim 0.2$~mCrab. This has made it possible to measure the log $N$--log $S$ function of local Seyfert
galaxies down to $\sim 3\times 10^{-12}$~erg~s$^{-1}$~cm$^{-2}$ (17--60~keV) (see Figure~10,
\citet{Mereminskiy_2016}). 
Although just $\sim 2$\% of the CXB is resolved into point sources at these fluxes, the census of AGN conducted locally 
(at $z\lesssim 0.2$) by {\it INTEGRAL} and {\it Swift} is now nicely complemented by {\it NuSTAR} at higher redshifts
($z\lesssim 1$) (see Figure~10). Together with findings from deep extragalactic surveys performed in the standard
X-ray band, these results have substantially tightened the constraints on the composition of the CXB.

\begin{figure*}[h]
\centering
\begin{minipage}{0.9\textwidth}
\includegraphics[trim= 0mm 0cm 0mm 0cm,
  width=1\textwidth,clip=t,angle=0.,scale=0.98]{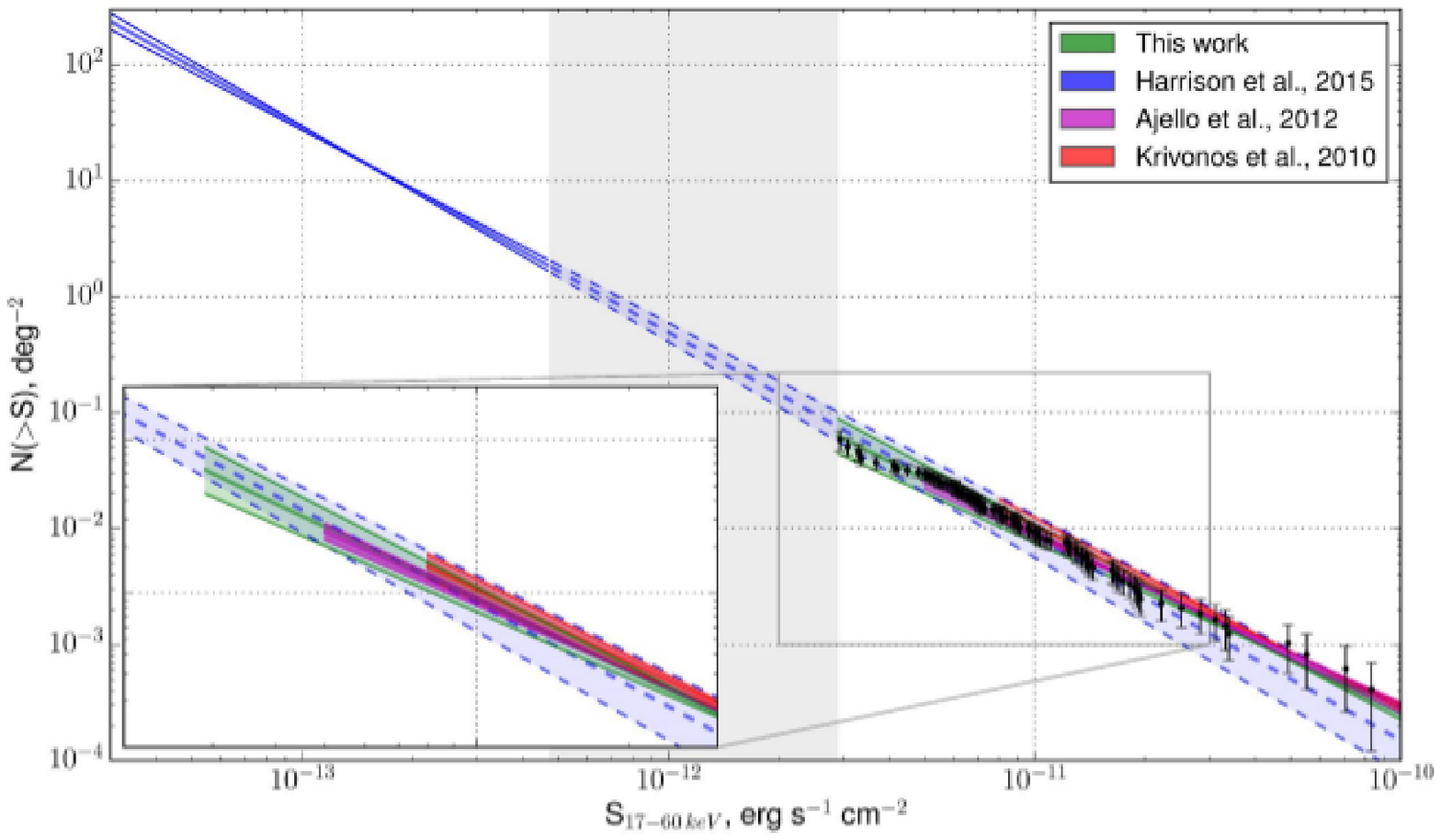}
  \caption{AGN number-flux relations in hard X-rays measured by {\it NuSTAR} \citep{2016ApJ...831..185H}, {\it Swift}/BAT \citep{2012ApJ...749...21A}, in the {\it INTEGRAL}/IBIS all-sky survey \citep{2010A&A...523A..61K} and in the {\it INTEGRAL}/IBIS deep fields \citep{Mereminskiy_2016}. The shaded area demonstrates a flux region not yet probed by hard X-ray missions. The inset is a zoom of the range of the used data, see \citet{Mereminskiy_2016}.}
  \end{minipage}
  \label{fig10}
\end{figure*}

\begin{figure}
\centering
    \includegraphics[width=\columnwidth,viewport=0 160 580 680]{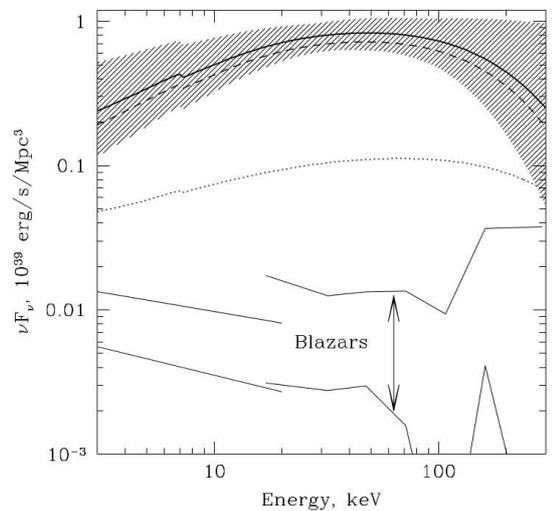}
  \caption{Cumulative spectrum of the local Seyfert galaxies, based on {\it INTEGRAL} and {\it RXTE} data (solid line with the corresponding hatched uncertainty region) The dashed and dotted lines show the contributions of low- ($<10^{43.5}$~erg~s$^{-1}$) and high- ($>10^{43.5}$~erg~s$^{-1}$) luminosity AGN, respectively. The estimated contribution of blazars is also shown. From \citep{2008A&A...482..517S}.}
  \label{cum_spec}
\end{figure}

In this context, an important result was presented by \citet{2008A&A...482..517S}, who performed a stacking analysis of the
X-ray spectra of AGN detected in the all-sky surveys performed by {\it INTEGRAL}/IBIS and {\it RXTE} (the aforementioned XSS survey),
taking into account the space densities of AGN with different luminosities and absorbing column densities, i.e. the luminosity
function and N$_{H}$ distributions discussed above. They obtained the broad-band (3--300~keV) spectral energy distribution of
the summed emission of the local AGN (see Figure~\ref{cum_spec}). It exhibits (albeit with low significance) a cutoff at energies
above 100--200 keV, in line with results obtained from broad band spectral studies of local AGN (see previous section (4.1)). 
It turned out that this locally determined spectrum is consistent with the CXB spectrum if the AGN population has experienced 
a pure luminosity evolution (so-called downsizing) between $z\sim 1$ and $z=0$, as is indeed indicated by the results of 
{\it Chandra} and {\it XMM-Newton} deep X-ray surveys \citep{2005AJ....129..578B}. 
This nicely demonstrates that the popular concept of the CXB being a superposition of AGN is generally correct. 

\subsection{Spatial distribution}
In the present-day Universe, matter is distributed very inhomogeneously on scales smaller than 100--200~Mpc. The contrast in
matter density between large galaxy concentrations (superclusters) and voids can reach an order of magnitude and more. 
The vast majority of galaxies in the local Universe is believed to contain supermassive black holes in their nuclei. 
Most of them are currently dormant or only weakly active but a significant fraction ($\sim 1$\%) are accreting matter 
at high rates and manifest themselves as AGN (Seyfert galaxies). It is reasonable to expect the space density of AGN 
to be approximately proportional to that of normal galaxies. As the {\it INTEGRAL} all-sky hard X-ray survey provides 
a virtually unobscured view of the AGN population out to a few hundred Mpc, we have a unique opportunity to verify that 
AGN trace the cosmic large-scale structure. 

\begin{figure}
\centering
    \includegraphics[width=\columnwidth,viewport=0 220 600 600]{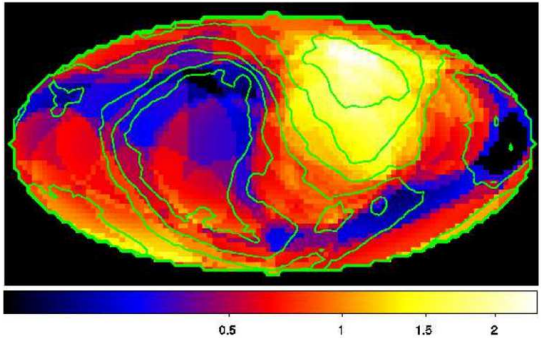}
  \caption{2D-map (in Galactic coordinates) of the AGN number density in the local Universe compared to that of normal galaxies. See \citep{2007A&A...475..775K} for details.}
  \label{anisotropy}
\end{figure}

Such a test has been done by \citet{2007A&A...475..775K}. They focused on the local volume of 70~Mpc radius, where maximal
contrasts in galaxy counts are expected. Based on a subsample of $\sim 40$ Seyfert galaxies detected by IBIS, 
they demonstrated that AGN do follow closely the large-scale structure of the Universe, strongly concentrating in such
well-known structures as the Virgo cluster, Great Attractor and Perseus-Pisces supercluster (see Figure~\ref{anisotropy}). 
This result has subsequently been confirmed using the {\it Swift}/BAT survey \citep{2012ApJ...749...21A}. 

\subsection{Bolometric properties}
Hard X-ray observations primarily probe the emission produced in the hot corona of the accretion disc. 
Although this high-energy component constitutes a significant fraction of the bolometric luminosity of an accreting
supermassive black hole, it is subdominant with respect to the softer (mainly UV) emission produced in the disc itself. 
A large fraction of the latter is in turn converted into even softer, infrared radiation in the surrounding dusty torus. 
In order to better understand the internal structure of AGN and physical processes taking place there, 
it is important to measure how the AGN bolometric luminosity is partitioned between these three main emission components 
(in radio galaxies and blazars, there may be an additional significant contribution from a relativistic jet, see below). 

To this end, \citet{2012ApJ...757..181S} have utilised proprietary and archival data of {\it Spitzer} infrared observations 
for a sample of 68 local Seyfert galaxies detected by IBIS. They found a clear  correlation between their hard X-ray 
and mid-infrared (MIR) luminosities: $L_{15\mu m}\propto L_{\rm HX}^{0.74\pm 0.06}$, where $L_{15\mu m}$ 
is the monochromatic luminosity at 15~$\mu$m and $L_{\rm HX}$ is the luminosity in the 17--60~keV energy band. 
Assuming that the observed MIR emission is radiation from an accretion disc reprocessed in a torus that subtends 
a solid angle decreasing with increasing luminosity (as inferred from the declining fraction of absorbed AGN, see above), 
the authors inferred that the intrinsic disc luminosity, $L_{\rm disc}$, is approximately proportional to the luminosity 
of the corona, $L_{\rm corona}$, namely $L_{\rm disc}$/$L_{\rm corona}\sim 2$. 
This ratio is a factor of $\sim 2$ smaller than inferred for typical (more distant and luminous) quasars producing 
the CXB \citep{2004MNRAS.347..144S}. 
The authors further demonstrated that the hard X-ray (17--60 keV) luminosity is a good proxy of the bolometric luminosity,
$L_{\rm bol}$, of Seyfert galaxies, with a typical ratio $L_{\rm bol}/L_{\rm HX}\sim 9$. 
This, together with black hole mass estimates available for the same sample of AGN, was used by \citet{2012AstL...38..475K} 
to infer the Eddington ratios of these objects, which turned out to lie between 1 and 100\% for the majority of them. 
Finally, \citet{2012ApJ...757..181S} estimated the cumulative bolometric luminosity density of local AGN, which turns out 
to be $\sim 2\times 10^{40}$~erg~s$^{-1}$~Mpc$^{-3}$.

\section{Main Results: Radio Galaxies}
Radio galaxies are sources  showing on radio maps an extended structure with lobes and jets.
Historically, they have been divided into two classes on the basis of their radio morphology: FRI, having bright jets 
in the centre and  low total luminosity and FRII, having faint jets but bright hot spots at the ends of the lobes and high 
total luminosities \citep{1974MNRAS.167P..31F}. 
The different morphology probably  reflects the method of energy transport in the radio source: 
FRIIs appear to be able to transport energy efficiently to the ends of the lobes, while FRI beams are inefficient 
in the sense that they radiate a significant amount of their energy away as they travel.  
The cause of the FRI/FRII difference is still unknown and both external properties (environment, host galaxy, 
merging history, etc.) or intrinsic factors  (different accretion processes) have been used to explain this 
dichotomy without reaching up to now firm conclusions. 

A further subdivision among radio galaxies has recently come from their optical spectroscopic properties 
\citep{2010A&A...509A...6B}: in general, objects with and without high-excitation emission lines in their optical 
spectra are referred to as High Excitation Radio Galaxies (HERG) and Low Excitation Radio Galaxies (LERG) respectively.
HEG  accrete in a radiatively efficient manner due to high Eddington ratios ($\ge$0.01 and up to 1) while LEG are 
known to exhibit radiatively  inefficient accretion related to low Eddington ratios ($\le$ 0.01). 

The jets of radio galaxies are oriented at inclination angles typically greater than 10 degrees with respect to the line 
of sight, making these objects intermediate between radio loud blazars and classical radio quiet 
Seyferts: it is for this reason that radio galaxies are unique laboratories where to study at the same time jets 
and accretion processes and search for a connection between the two.

A recent investigation done by \cite{2016MNRAS.461.3165B} show that  radio galaxies are not common among hard X-ray 
selected AGN: for example within the sample of around 400 AGN detected by {\it INTEGRAL}/IBIS up to 2016, \citep{Malizia_2016}
only 32 (i.e. 8$\%$ of the sample) are radio galaxies. 
These {\it INTEGRAL} selected radio galaxies cover all optical classes, are characterised by high 20-100 keV luminosities
(10$^{42}$--10$^{46}$ erg s$^{-1}$) and high Eddington ratios (typically larger than 0.01).  
Most of these objects display a FRII radio morphology and are classified as HERG. 
Several studies have been performed on {\it INTEGRAL} high energy selected sample of radio galaxies in order to investigate
their spectral characteristics.

\citet{2016MNRAS.461.3153P} studied the absorption properties of this sample with the addition of radio galaxies 
detected by {\it Swift}/BAT and found that the column density distribution is consistent with the unified model of AGN 
with those optically classified as type 2 being absorbed and those optically classified as type 1 not. However, there seems to be no evidence for the presence of Compton thick absorption in hard X-ray selected radio galaxies \citep{2018MNRAS.474.5684U}. 
Also a significant anti-correlation between the radio core dominance parameter (taken as an orientation  indicator) 
and the X-ray column density is found, again in line with expectations from the AGN unified theory: 
core dominated sources are unabsorbed in X-rays since they emit their radiation in a direction closer to the line 
of sight and therefore not intercepted by the torus.

The broad band spectra of some {\it INTEGRAL} detected radio galaxies have been studied over the years by various authors 
\citep{2011A&A...531A..70B, 2014ApJ...782L..25M,2014A&A...565A...2M, 2016MNRAS.458.2454L,2015MNRAS.451.2370M,2018MNRAS.481.4250U} 
and the overall result is that they behave very similarly 
to radio quiet AGN in terms of primary continuum,  presence of complex absorption and soft excess, with the possible 
exception of  the reflection features (10--30 keV bump and iron line) which tend to be weak in these objects.\\
These observational results confirm that the high energy  properties of these sources are consistent with an 
accretion-related  emission, likely originating from a hot corona coupled with a radiatively efficient accretion flow. 

\begin{figure}[h]
    \includegraphics[width=\columnwidth]{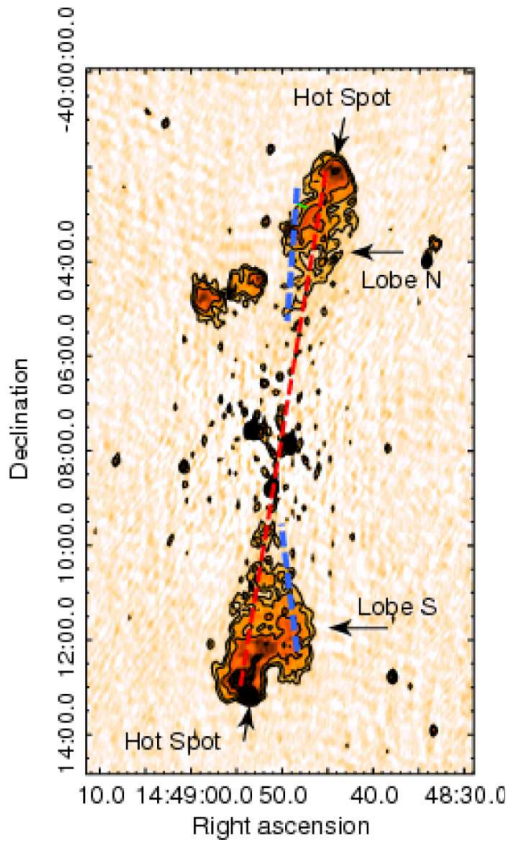}
  \caption{Radio 610MHz GMRT full resolution image of IGR~J14488-4008 as in \cite{2015MNRAS.451.2370M} }
  \label{giant}
\end{figure}

The radio size distribution of the entire {\it INTEGRAL} sample of radio galaxies shows an almost continuous coverage, 
from around 50 kpc (in PKS~0521-365) up to 1.5 Mpc (in IGR~J14488-4008), with many sources 
displaying  values above few hundred kpc; indeed 56$\%$ of the objects in the sample have sizes above 0.4 Mpc. 
If we consider the classical threshold to define a giant radio galaxy (GRG), i.e. a size larger than 0.7 Mpc, then in the 
{\it INTEGRAL} sample of 32 radio galaxies, 8  are giants, i.e. 25\% of the sample. 
It may be the case that high-energy surveys could be more efficient in searching for new GRG as compared to radio 
surveys, where, for example, in the well studied 3CRR sample, only  6$\%$ of the sources are identified as GRG 
\citep{2016MNRAS.461.3165B}.
That a large fraction of hard X-ray selected radio sources become radio giants  is also evident by the discovery 
of two completely new GRG among {\it INTEGRAL} AGN: IGR~J17488-2338 \citep{2014A&A...565A...2M} and 
IGR~J14488-4008 \citep{2015MNRAS.451.2370M}, which display peculiar and interesting radio as well as soft and hard X-ray properties.
In Figure~\ref{giant} the radio 610MHz GMRT full resolution image of IGR~J14488-4008 as published by \citet{2015MNRAS.451.2370M} 
is displayed: evident from the figure are the large  size of the source, its FRII morphology and  also its interesting environment.\\

All 8 {\it INTEGRAL} selected GRG have been the focus of an intense observational campaign especially at  
radio frequencies to probe the possibility of restarting activity in their nuclei.
In fact in these giants, the luminosity of the radio lobes and the estimated jet power are relatively low compared with the 
nuclear X-ray emission \citep{2018MNRAS.481.4250U}.
This indicates that either the nucleus is more powerful now than was in the past, consistent with a restarting of the central 
engine, or that the giant lobes are dimmer due to expansion losses.
The first scenario is backed up by the finding in the radio band that 6 objects (75$\%$ of the sample) host a core with a self 
absorbed spectrum (peaking in the range  from 2 to 10 GHz or above) typical of young radio sources
(ages of kyears) while their extended structure must be very  old and evolved (ages of Myears): these young nuclei are 
probably undergoing a restarting activity episode, suggesting a link between the detected hard X-ray  emission, due to 
the ongoing accretion, and the reactivation of their jets \citep{2019ApJ...875...88B}.
Of the two sources not showing a young  radio core spectrum, one has the radio lobes embedded in an extended low surface 
brightness cocoon, that is likely the  result of a previous period of activity 
\citep{2016AcA....66...85W}. The other source presents instead a discontinuity between the extended lobes and the arcsec 
core-jet structure again pointing towards the presence of different activity phases \citep{2007A&A...474..409G}. 
These findings support the scenario originally proposed by \citet{1996MNRAS.279..257S}, in which multiple episodes 
of activity would favour the growth of radio sources up to the extreme size of GRGs but also underline the importance 
of extracting hard X-ray selected samples of radio galaxies with the purpose of studying the duty cycle of AGN.

\section{Main Results: Blazars}
Among radio-loud AGN, blazars are the most luminous and variable.  
This is because strong relativistic aberration effects, primarily light magnification and time intervals foreshortening,
take place in the plasma flowing at speeds close to that of light in their powerful jets, that are oriented only a few
degrees away from the line of sight. 
They include both flat-spectrum radio quasars (FSRQ) and BL Lacertae objects (BL Lac). 
BL Lacs are further classified into low/intermediate/high-frequency-peaked BL Lac objects (LBL, IBL, HBL) depending on 
the location of the first characteristic peak frequency: below 10$^{14}$Hz, at 10$^{14-15}$Hz, or above 10$^{15}$ Hz,
respectively. 
Both blazar classes share the properties of a nonthermal continuum, but FSRQ have strong and broad optical emission lines, 
while in BL Lac the optical lines are weak or absent. 
FSRQ have higher bolometric luminosities than BL Lac \citep{1996ApJ...463..444S} and can exhibit signs of thermal activity possibly 
related to an accretion disc in their optical and UV spectra \citep{1988ApJ...326L..39S} in contrast to BL Lac, which have smooth continua.

Multiwavelength studies of blazars in the last 25 years have identified a characteristic broad-band spectral shape, with
two "humps" in a $\nu f_\nu$ representation \citep{2014A&ARv..22...73F}.  The first hump, peaking at mm to soft X-ray
frequencies (depending on the source, and varying even in the same source during different emission states), is produced by
synchrotron radiation, while the origin of the second hump, that has a maximum between  MeV and GeV energies, is complex,
with various scenarios contemplating a pure leptonic composition of the jet or a lepto-hadronic composition.  
In the leptonic case, the high energy hump can be due to inverse Compton scattering of relativistic electrons or positrons
off synchrotron photons (internal Compton or self-Compton) or off ambient photon fields, if these are relevant, as in the
case of FSRQ that, as said before, host luminous accretion discs and broad emission lines (external Compton).  
For the lepto-hadronic scenario, if sufficient energy is injected in the jet to trigger photo-pion production,
synchrotron-supported pair cascades will develop (\cite{2010arXiv1006.5048B,2017nacs.book...15M} and references therein). 
These initiate showers of mesons, leptons, neutrinos and high energy radiation.  

According to the location of the two radiation peaks, blazars form a "sequence" \citep{1998MNRAS.299..433F,1998MNRAS.301..451G} 
whereby sources with peaks at lower frequencies have larger luminosities 
and larger ratios between the high-energy and low-energy components (see however \citet{2012MNRAS.422L..48P}).  
The blazars with largest bolometric luminosities, largest dominance of gamma-ray luminosities, lowest frequencies 
of the synchrotron and highest energy Compton hump, generally coincide with the FSRQ, in whose  jets relativistic 
particles cool rapidly by losing energy in Compton upscattering disc and line optical-UV photons.  
The  less-efficiently cooling, less luminous sources coincide with the BL Lac objects.

Blazars represent almost 70\% of all sources detected at energies larger than 100 MeV by the {\it Fermi}/LAT instrument
\citep{2019arXiv190510771T}.
This, together with the fact that most blazars are bright X-ray sources, makes them also excellent targets for hard X-rays 
studies.   Indeed various {\it INTEGRAL}  surveys  list  a conspicuous number of them. \\
For example considering all the AGN detected by {\it INTEGRAL} (see \cite{2012MNRAS.426.1750M,Malizia_2016}
and more recent updates), we count 29 FSRQ and 19 BL Lac, around 11\% of  the entire {\it INTEGRAL} AGN population 
(see Figure~\ref{pie}). 
Interestingly all 8 blazars listed in  the complete sample of AGN discussed by Malizia et al. (2009) have now a counterpart at GeV energies, with 3 also having a TeV association: 
this indicates that these are  bright enough for detection up to the highest observable energies.  

\begin{figure}
    \includegraphics[width=\columnwidth]{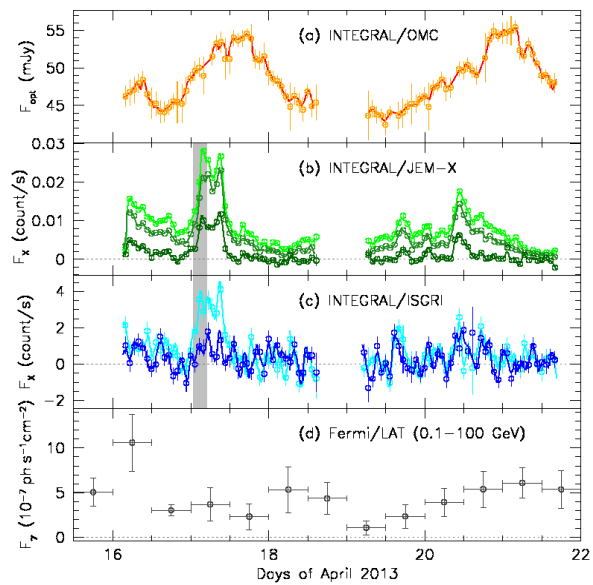}
  \caption{Light curve of MKN~421 in April 2013, see \cite{2014A&A...570A..77P}}
  \label{lc}
\end{figure}

Blazars detected by {\it INTEGRAL}  have been used  over the years to provide information on their high-energy variability patterns
\citep{2007A&A...475..827B}, to monitor individual targets over a period of intense activity (see for example Figure~\ref{lc} which shows the light curves of MKN~421 during a flare recorded in April 2013 by \citet{2014A&A...570A..77P}), or to study their broad band spectral characteristics such as in the case of 4C~04.42, where excess emission observed in the soft X-ray band was  interpreted as due to bulk Compton radiation of cold electrons \citep{2008MNRAS.388L..54D}.

According to the locations of the two spectral peaks in blazar energy distributions, the hard X-ray region represents 
either the tail of the synchrotron spectrum at energies higher than the cooling break, or the rising portion of the 
inverse Compton (leptonic case) or proton-synchrotron (lepto-hadronic case) spectrum.    
HBL, and in particular  those peaking in X/hard X-rays, are prominent in the 20-200 keV band, and having the lowest 
jet powers, represent the extreme end of the blazar sequence, opposite to FSRQ.  
One such example of extreme blazar was discovered by {\it INTEGRAL} in the source  IGR~J19443+2117 \citep{2009A&A...493..893L} 
which was later detected also by the Cherenkov telescope {\it HESS} as HESS~J1943+213 \citep{2011A&A...529A..49H,2018ApJ...862...41A}: 
the source displays a synchrotron peak at around 10 keV  and an inverse Compton peak  above few hundred of GeV.

During bright outbursts of "extreme" HBL, the  synchrotron spectrum, normally peaking at soft X-rays, flattens and reaches
a peak at energies equal or higher than 100 keV, as observed first in the BL Lac object MKN~501 in 1997 with {\it BeppoSAX} 
\citep{1998ApJ...492L..17P}.
Thus, depending on whether the blazar has a high-frequency or low-frequency synchrotron peak, IBIS observations will 
sample the rising part of the inverse Compton component or the tail of the synchrotron, thus allowing one to locate more 
precisely the peaks of these radiation components and  to extract precise information on the  energies of the 
emitting particles \citep{2016ApJ...832...17B}.

\begin{figure}
    \includegraphics[width=\columnwidth]{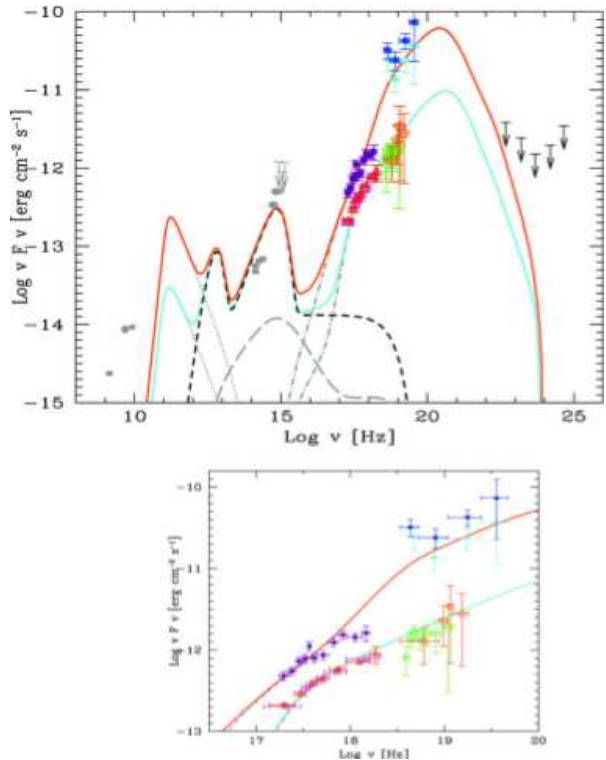}
  \caption{Top: spectral energy distribution of IGR J22517+2217. Grey circles and arrows represent archival radio/optical/UV data from NED. Red triangles and magenta squares  represent XIS 0 and XIS 3 data, green circles and orange pentagons represent PIN and BAT quiescent data respectively, while black arrows are Fermi upper limits in five bands. Filled violet squares  represent XRT data, filled cyan diamonds  and blue pentagons  represent IBIS and BAT flare data respectively. The solid cyan  and orange curves are the results of modelling of the quiescent and flaring states, respectively. With grey lines we show the different components of non-thermal emission: synchrotron (dotted), synchrotron self-Compton (long-dashed) and external Compton (dot–dashed). The black dashed line corresponds to the thermal emission of the disc, the IR torus and the X-ray disc corona. The model does not account for radio emission, produced from much larger regions of the jet. Bottom: enlargement in the X-ray energy range for the two SEDs. Symbols as in the top panel. Figure from \citet{2012MNRAS.421..390L}}
  \label{sed_giorgio}
\end{figure}

Equally important for hard X-ray studies are  FSRQ located at high redshifts displaying a Compton peak in the sub-MeV region 
which also favour their detection by instruments like {\it INTEGRAL}/IBIS. These sources have  the most powerful jets, 
the largest black hole masses and the most luminous accretion discs. {\it INTEGRAL} has  played a role in the discovery of such 
high redshift  blazars like IGR~J22517+2217 \citep{2007ApJ...669L...1B,2012MNRAS.421..390L}, 
Swift~J1656.3-3302 \citep{2008A&A...480..715M} and IGR~J12319-0749 \citep{2012A&A...543A...1B}.
So far 17 objects have been detected at redshift greater than 1, with 3 having $z$ above 3.
The most distant  AGN so far detected by {\it INTEGRAL}, the FSRQ IGR~J22517+2217,  has been the target of intense 
studies after the {\it INTEGRAL} detection \citep{2012MNRAS.421..390L} which lead to various results: the discovery of a strong flare episode, 
the study of the source SED over flaring and quiescent states (see Figure~\ref{sed_giorgio}) and the measurement of a flare jet power luminosity which turned out to be
around 30 times more powerful than the accretion disc luminosity.

{\it INTEGRAL}/IBIS's sensitivity of $\sim$10 mCrab in the 20-100 keV range implies that most blazars can be detected with a high
statistical significance  only when they are in a high emission state.  
Therefore, they are generally observed with {\it INTEGRAL} during outbursts, after a notification from a large field of view 
X-ray or gamma-ray camera, like e.g. {\it Swift}/BAT, {\it MAXI}, {\it Fermi}/LAT, or {\it AGILE}/GRID.  
This "Target-of-Opportunity" strategy is rather effective, but implies that {\it INTEGRAL} can
typically  cover the declining phase of the outburst, owing to the characteristic $\sim$1-2 days re-pointing time of the
satellite.  The simultaneous monitoring of {\it INTEGRAL}, primarily with IBIS, but in some cases also with the JEM-X cameras
(3-30 keV), and other multiwavelength facilities have set cogent constraints  on the physics that governs blazar 
correlated flux and spectral variability over the whole electromagnetic spectrum 
(e.g. \citet{2006A&A...449L..21P,2007MNRAS.382L..82G,2011ApJ...736L..38V,2010A&A...522A..66C,2011A&A...526A.125P,2017A&A...601A..30C}).

Owing to their luminosities, blazars are, after gamma-ray bursts, the farthest detectable sources at all wavelengths and
therefore potential beacons of the early Universe.   The cosmological relevance of blazars was not missed by the {\it INTEGRAL}
mission, that enabled not only the detection of previously unknown ones at  high redshift but also the monitoring of a 
few previously known distant ones as in
\citet{2005A&A...429..427P,2009A&A...496..423B,2010A&A...509A..69B,2011MNRAS.411.2137G}.

The {\it IceCUBE} detection of a $\sim$200 TeV neutrino from blazar TXS0506+056 ($z$ = 0.336) on 22 September 2017 confirmed 
many expectations that powerful extragalactic jetted sources may be the origin of high-energy (i.e. $E > 1$ TeV) neutrinos.
This detection also unambiguously points to the presence of a hadronic component in the jet composition, as only hadronic
showers can induce cascades that include neutrinos as by-products.  A multiwavelength observing campaign was launched as a
consequence of the detection, that included {\it INTEGRAL} among the many space-based facilities that were employed 
\citep{2018Sci...361.1378I}.  The resulting spectra and light curves and their correlations, combined with the neutrino event, 
were used to test blazar lepto-hadronic models, in an attempt to determine the mutual role of leptons and hadrons in producing radiation  
\citep{2018ApJ...864...84K,2019NatAs...3...88G,2019MNRAS.483L.127R}.
While neutrino events accompanying blazars are rarely detected with the present instrumentation, they represent a formidable 
diagnostic of the structure and mechanisms in relativistic jets.

\section{keV to GeV/TeV connection: a case study}
The connection between the hard X-ray region of the spectrum and the GeV/TeV domain is  critical for the study of many AGN, blazars {\it in primis} (see previous section), but also for radio galaxies and  Seyferts galaxies. 
In these last two types of objects, it is important to understand how  to connect the emission due to  the  disc/corona  which dominates at lower energies,  to the one related to the jet or some other mechanism (starburst?) which instead dominate  at higher frequencies.

\cite{2009ApJ...706L...7U} presented for the first time the result of the cross correlation between the fourth 
{\it INTEGRAL}/IBIS soft gamma-ray catalogue and the {\it Fermi}/LAT bright source list of objects emitting 
in the 100 MeV--100 GeV range.
Surprisingly, from this initial cross-correlation between low and high energy gamma-ray sources emerged that only 
an handful of objects were common to both surveys; they comprised blazars (10) both of FSRQ and BL Lac types, 
and no X-Ray Binaries, with the only exception of two microquasars.

This initial approach can now be applied further by finding  objects, among {\it INTEGRAL} detected AGN, 
that emit also at GeV and/or at TeV energies. 

This can be done by cross-correlating our list of 440 AGN  
with the most recent {\it Fermi}/LAT catalogue\footnote{8 year point source catalogue: 
\url{https://heasarc.gsfc.nasa.gov/W3Browse/fermi/fermilpsc.html}} and  the TeV on line 
catalogue\footnote{\url{http://tevcat.uchicago.edu}}.
From this cross correlation, interesting results have been found and reported here to highlight
{\it INTEGRAL}'s capability in this field.
The standard statistical technique developed by \citet{2005A&A...432L..49S} has been used. It
consists of simply calculating the number of GeV/TeV sources for which at least 
one INTEGRAL counterpart was found within a specified distance, out to a distance where all
GeV/TeV sources had at least one soft-gamma ray counterpart.
To have a control group, a list of 'anti GeV/TeV sources', mirrored in Galactic longitude and
latitude, has been created and the same correlation algorithm applied.  
In both cases a  correlation out to about  300 arcsec has been found, while
at further distances, chance correlations become increasingly more important.
At GeV energies 58 {\it INTEGRAL}  AGN (13$\%$ of the sample) have emission above  1 GeV, 
while only 15 objects (3$\%$ of the sample)  emit up to TeV energies, confirming  that 
the emission of the majority of hard X-ray  selected AGN  drops exponentially above 100-200 keV. 
The GeV sample is largely dominated by blazars: there are 18 BL Lac  and 24 FSRQ.  
Other sources include 9 radio galaxies, 3 Seyfert 2 (NGC~1068, NGC~4945 and the Circinus galaxy), 
one NLS1 galaxy (1H~0323+342), one peculiar object (IGR~J20569+4940)  and two sources of unclear  class
(IGR~J18249-3243 and IGR~J13109-5552).
Two radio galaxies detected in the GeV band are also TeV emitters, while
none of the 4 Seyfert galaxies have so far being detected at TeV energies. 
Interestingly of the 18 BL Lac objects detected by {\it Fermi}/LAT, 11 have been seen  also  
by Cherenkov telescopes: they are mostly high frequency peaked objects, 8 HBL and 2 IBL.  
Their Compton peak is generally just below 1 TeV but can be as high as 10 TeV  or more  
like in  H~1426+428 \citep{2019MNRAS.486.1741F}.

\begin{figure}
    \includegraphics[width=\columnwidth]{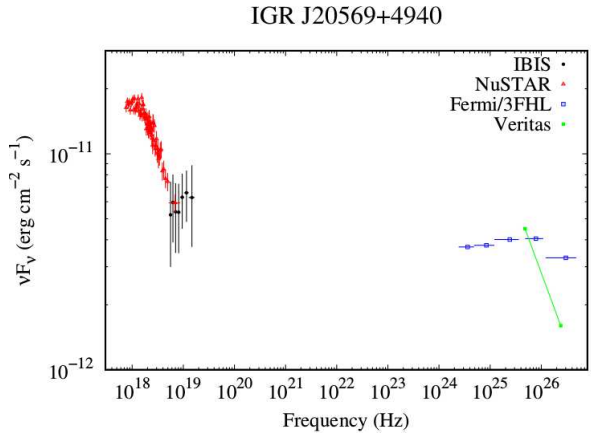}
  \caption{Spectral Energy Distribution of IGR~J20569+4940 obtained assembling unpublished  {\it NuSTAR} observation performed in November 2015  with the average {\it INTEGRAL} and 
  {\it Fermi}/LAT spectra with the addition of the reported  TeV data (courtesy of F. Ursini).}
  \label{nufnu}
\end{figure}

A peculiar source is IGR~J20569+4940. It is certainly a blazar but still  of unknown class and redshift: 
the source has recently been announced as a  VHE emitter \citep{2017ICRC...35..641B} after observations 
with {\it VERITAS}  performed in November 2016. 
It is  most likely a BL Lac \citep{2016MNRAS.462.3180C} with a high synchrotron peak \citep{2016ApJS..226...20F}. 
In Figure~\ref{nufnu},  we assembled high energy data on this source using an unpublished {\it NuSTAR} observation 
performed in November 2015  with the average {\it INTEGRAL} and {\it Fermi}/LAT spectra  plus the reported  TeV data. 
As evident  from the figure, two peaks can be located, one below  the {\it NuSTAR} low energy threshold, at around few keV, likely
associated to  the  synchrotron part of the SED, and the other at around  0.1-0.3 TeV, probably related to the Compton
part of the SED. This qualifies  IGR~J20569+4940 to be another  HBL or alternatively  another extreme  BL Lac.

As for the FSRQ, the set of {\it INTEGRAL} objects with GeV emission spans a wide range of redshifts (0.3-3.1) and black hole 
masses  (Log (M/M$_{\odot}$) = 8-9.5). As discussed in the previous section, these sources  display the most powerful  jets, 
they are very massive, host the largest black hole masses and the most 
luminous accretion discs; in other words  they are more extreme than blazars selected in other wavebands, like, for example, 
the only one explored by {\it Fermi}/LAT \citep{2013arXiv1302.2447B}. \\
Only 4 of the 24 GeV FSRQ  in our list have a detection at TeV energies, namely 3C~279, PKS~1510-089, 
PG~1222+216 and SWIFT~J0218+7348.
This low number  of detections is  expected since FSRQ are generally faint at very high energies for a number of reasons: 
the location of their  Compton peak at the lower end of the high energy gamma ray band implies a low flux at these energies; 
the compactness of the source emission region indicates strong attenuation by the broad line region photon field and, finally, 
their  rather high redshifts \citep{2019A&A...627A.159H}. \\
However,  the same reasons indicate that it is important to study the SED 
of these  peculiar FSRQ and  soft gamma ray observations can be of great  help to better  define their 
spectral energy distribution.

Among {\it INTEGRAL} AGN with GeV emission, there are a couple of  sources of unclear  class: 
IGR~J18249-3243 and IGR~J13109-5552. Optically they are both classified as Seyfert 1. 
They are both bright in the radio waveband and, although repeatedly observed, are 
poorly studied. Their structure seems to be extended but a dedicated observational campaign is necessary 
to confirm the presence of jet and related features.
The  broad band ({\it XMM} plus IBIS/BAT data) properties of both objects have been explored by \citet{2014ApJ...782L..25M}. 
Unlike other Seyfert galaxies  and similarly to what we observe in blazars neither source shows  a low energy cut-off 
($\ge$  357 keV in IGR~J13109-552 and not required  in IGR~J18249-3243). Furthermore, in IGR~J18249-3243 partial covering 
absorption  is present, like in an ordinary radio quiet Seyfert, making the picture even more confused.
Clearly both sources deserve further investigation, possibly through a multiwaveband campaign that
will be able to clarify their nature.

\subsection{TeV blazars}
Nearly all identified extragalactic TeV emitters are blazars (about 70 sources),
exceptions being a handful of nearby radio galaxies and starburst galaxies and a recently detected long gamma-ray 
burst at $z$ = 0.4. 
The  brightest blazars in the TeV domain are the HBL, i.e. those whose high energy component peaks in 
or just beyond that band. 
These sources are ideal targets of coordinated hard X-ray and  TeV observing campaigns 
(e.g. \cite{2016A&A...593A..91A,2018A&A...620A.181A,2019MNRAS.484..749C} and references therein).

Prototype blazars with these characteristic spectral energy distributions are MKN~421 and MKN~501, that are also the
brightest objects of this class.
They may reach "extreme" states whereby the synchortron spectrum extends to energies 
higher than 100 keV during flares  and the TeV emission increases by more than one order of magnitude with respect 
to the quiescent state. 
Recent simultaneous monitoring at hard X-rays and TeV energies with {\it NuSTAR} and the {\it Veritas} Cherenkov 
telescope respectively,
combined with observations at optical and radio frequencies, yielded detailed  insight into the physics of these sources,
with accurate mapping of critical parameters such as electron energies and Lorentz factor 
\citep{2015ApJ...812...65F,2015A&A...580A.100S,2016ApJ...819..156B}.
Repeated observations of HBLs in bright states, as triggered by {\it Swift}/XRT or atmospheric Cherenkov radiation telescopes,
and in particular of the nearby BL Lac object MKN~421  with {\it INTEGRAL}, sampled bright states and collected detailed light
curves and spectra \citep{2008A&A...486..721L,2014A&A...570A..77P}.
The type of X-ray variability suggests a complex behaviour and a correlation among different frequencies.  
However, so far, although IBIS detected rather hard spectra, no clear evidence was found  that these still rise (in $\nu
f_\nu$) beyond 15-20 keV.  Investigation with IBIS of the BL Lac object MKN~501, in search of a shift of the synchrotron
peak to hard X-rays during outbursts in 2013 and 2014 is underway.\\
A few dimmer BL Lac share the "extreme" properties of MKN~421 and MKN~501 and are potential ideal targets for coordinated
hard  X-ray and TeV monitoring \citep{2018MNRAS.477.4749C}.  In the near future, these can be followed-up with the sensitive
Cherenkov Telescope Array for more accurate assessment of their physical parameters than it is possible with present
state-of-the-art Cherenkov telescopes.   

Particularly interesting are the so called "orphan" gamma-ray flares, whereby some
sources exhibit strong TeV outbursts that are not accompanied by simultaneous X-ray flux increase at any appreciable 
level (e.g. \cite{2004ApJ...601..151K,2010ApJ...719L.162B,2017ApJ...850...58M}).

\section{INTEGRAL heritage and future perspectives}
{\it INTEGRAL}'s cumulative exposure has now exceeded 10 Ms in several
extragalactic fields with a total area of $\sim 2000$ sq. deg. and IBIS
has detected $\sim 100$ AGN (known or presumed) with fluxes down to
$3\times 10^{-12}$~erg~s$^{-1}$~cm$^{-2}$ (17--60 keV) in these
regions. This sample is a valuable part of the {\it INTEGRAL} -- {\it Swift} legacy,
since it widens the AGN parameter space to lower
luminosities and larger distances. In particular, a significant
fraction of AGN detected at these (low) fluxes are expected to be
located at $z\simeq 0.2$, which makes it possible to begin studying the
evolution of the AGN population over the last $\sim 2.5$ billion years
in a nearly unbiased way (thanks to the hard X-ray selection). To this
end, it is highly desirable to further lower the flux threshold in the
INTEGRAL extragalactic fields, which can be done by adding exposure
since the IBIS sensitivity is still largely determined by statistical
noise in the extragalactic sky.

Activities in this direction are already under way. In particular, in
2017--2018 there were extensive observations of the ultra-deep field
in the Virgo region (PI: Beckmann) and in 2019--2020 the deep field
around the M81 galaxy is to be observed for an additional 3.75 Ms (PI:
Mereminskiy), which will bring the total
exposure there to $\sim 15$ Ms and the corresponding flux limit to
$\sim 2\times 10^{-12}$~erg~s$^{-1}$~cm$^{-2}$ (17--60 keV). Some 50
new AGN are expected to be detected upon completion of this programme
(and there are suggestions to continue it afterwards). Moreover, there
is expected to be unique synergy between these {\it INTEGRAL} observations
and the upcoming all-sky X-ray survey by the
{\it Spektr-RG} observatory (with its eROSITA and ART-XC telescopes). The
latter is expected to reach record sensitivities $\sim 10^{-14}$,
$\sim 2\sim 10^{-13}$ and $\sim$ a few $10^{-13}$
erg~s$^{-1}$~cm$^{-2}$ in the 0.5--2, 2--10 and 5--11 keV energy
bands, respectively, after 4 years of observations, while shallower
maps will be available already after the first 6-month scan of the
sky. {\it INTEGRAL} hard X-ray observations and {\it Spektr-RG} data at lower
energies in the M81 field will be complementary in a number of ways,
in particular regarding the detection and identification of Compton
thick AGN.

It is worth noting that, although deep surveys on specific sky regions are important to probe lower
luminosity objects and further distance scales, it is only through large scale mapping that is 
possible to enlarge the {\it INTEGRAL} AGN database. 
By extrapolating from previous surveys, it is also possible to estimate how many AGN {\it INTEGRAL} will be able to detect in the future. 
From the sky coverage obtained in 2000 orbits (September 2018), we expect to double the number of the detected AGN and we foresee that we will be able to observe more than 1200 AGN by revolution 2500, likely by the end of the mission.
Even more interesting is the ongoing monitoring of the Galactic Plane region by {\it INTEGRAL}. 
As already mentioned, this is a region which is generally avoided by extragalactic studies and where 
{\it INTEGRAL} is playing an even greater role than {\it Swift}/BAT. 
In this region, deep sky coverage combined with large scale mapping, will allow  
a unique view of the extragalactic sky behind our Milky Way across the years, a legacy which will be extremely useful for future studies of this region at other wavelengths, especially for the {\it Cherenkov Telescope Array} (CTA) future  surveys.

On the other hand, the advent of the {\it CTA}  in the near future 
and the sensitivity upgrades of neutrino detectors will result in a boost of multiwavelength 
and multi-messenger studies of high energy extragalactic  sources, and help 
unveil the physics of their complex central engines.

\section*{List of abbreviations}
List of definitions of abbreviations used in the paper.\\
NLS1: Narrow Line Seyfert 1\\
XBONG: X-ray Bright Optically Normal Galaxies\\
LINERs: Low-Ionization  Nuclear  Emission  Regions\\
FSRQ: Flat Spectrum Radio Quasar\\
BL Lac: BL Lacertae \\
LBL, IBL, HBL: low/intermediate/high-frequency-peaked BL Lac objects\\
RG: Radio Galaxies\\
FRI:  Fanaroff type I\\
FRII:  Fanaroff  type II\\
HERG:  High Excitation  Radio  Galaxies \\
LERG:  Low  Excitation  Radio  Galaxies \\
GRG:   Giant Radio Galaxies

\section*{Acknowledgements}
We acknowledge all the scientists which contributed over the years to the analysis and the interpretation of {\it INTEGRAL} AGN data. In particular we would like to thank Nicola Masetti for his fundamental work in leading the optical follow-up campaigns of the unidentified {\it INTEGRAL} sources. 
We thank also J.~B. Stephen for his contribution in the correlation studies presented in this work.
AM and LB acknowledge financial
support from ASI under contract INTEGRAL ASI/INAF n.2019-35-HH; SS and IM acknowledge  support from the Russian Science Foundation’s grant 19-12-00396 in working on this review.

\section*{References}

\bibliography{mybiblio}

\begin{thebibliography}{167}
\expandafter\ifx\csname natexlab\endcsname\relax\def\natexlab#1{#1}\fi
\providecommand{\url}[1]{\texttt{#1}}
\providecommand{\href}[2]{#2}
\providecommand{\path}[1]{#1}
\providecommand{\DOIprefix}{doi:}
\providecommand{\ArXivprefix}{arXiv:}
\providecommand{\URLprefix}{URL: }
\providecommand{\Pubmedprefix}{pmid:}
\providecommand{\doi}[1]{\href{http://dx.doi.org/#1}{\path{#1}}}
\providecommand{\Pubmed}[1]{\href{pmid:#1}{\path{#1}}}
\providecommand{\bibinfo}[2]{#2}
\ifx\xfnm\relax \def\xfnm[#1]{\unskip,\space#1}\fi
\bibitem[{{Ahnen} et~al.(2016){Ahnen}, {Ansoldi}, {Antonelli}, {Antoranz},
  {Babic}, {Banerjee}, {Bangale}, {Barres de Almeida}, {Barrio}, {Becerra
  Gonz{\'a}lez}, {Bednarek}, {Bernardini}, {Biasuzzi}, {Biland}, {Blanch},
  {Bonnefoy}, {Bonnoli}, {Borracci}, {Bretz}, {Buson}, {Carosi}, {Chatterjee},
  {Clavero}, {Colin}, {Colombo}, {Contreras}, {Cortina}, {Covino}, {Da Vela},
  {Dazzi}, {De Angelis}, {De Lotto}, {de O{\~n}a Wilhelmi}, {Di Pierro},
  {Dom{\'\i}nguez}, {Dominis Prester}, {Dorner}, {Doro}, {Einecke}, {Eisenacher
  Glawion}, {Elsaesser}, {Fern{\'a}ndez-Barral}, {Fidalgo}, {Fonseca}, {Font},
  {Frantzen}, {Fruck}, {Galindo}, {Garc{\'\i}a L{\'o}pez}, {Garczarczyk},
  {Garrido Terrats}, {Gaug}, {Giammaria}, {Godinovi{\'c}}, {Gonz{\'a}lez
  Mu{\~n}oz}, {Gora}, {Guberman}, {Hadasch}, {Hahn}, {Hanabata}, {Hayashida},
  {Herrera}, {Hose}, {Hrupec}, {Hughes}, {Idec}, {Kodani}, {Konno}, {Kubo},
  {Kushida}, {La Barbera}, {Lelas}, {Lindfors}, {Lombardi}, {Longo},
  {L{\'o}pez}, {L{\'o}pez-Coto}, {Majumdar}, {Makariev}, {Mallot}, {Maneva},
  {Manganaro}, {Mannheim}, {Maraschi}, {Marcote}, {Mariotti}, {Mart{\'\i}nez},
  {Mazin}, {Menzel}, {Mirand a}, {Mirzoyan}, {Moralejo}, {Moretti}, {Nakajima},
  {Neustroev}, {Niedzwiecki}, {Nievas Rosillo}, {Nilsson}, {Nishijima}, {Noda},
  {Nogu{\'e}s}, {Orito}, {Overkemping}, {Paiano}, {Palacio}, {Palatiello},
  {Paneque}, {Paoletti}, {Paredes}, {Paredes-Fortuny}, {Pedaletti}, {Perri},
  {Persic}, {Poutanen}, {Prada Moroni}, {Prandini}, {Puljak}, {Rhode},
  {Rib{\'o}}, {Rico}, {Rodriguez Garcia}, {Saito}, {Satalecka}, {Schultz},
  {Schweizer}, {Shore}, {Sillanp{\"a}{\"a}}, {Sitarek}, {Snidaric},
  {Sobczynska}, {Stamerra}, {Steinbring}, {Strzys}, {Takalo}, {Takami},
  {Tavecchio}, {Temnikov}, {Terzi{\'c}}, {Tescaro}, {Teshima}, {Thaele},
  {Torres}, {Toyama}, {Treves}, {Verguilov}, {Vovk}, {Ward}, {Will}, {Wu} and
  {Zanin}}]{2016A&A...593A..91A}
\bibinfo{author}{{Ahnen}, M.L.}, \bibinfo{author}{{Ansoldi}, S.},
  \bibinfo{author}{{Antonelli}, L.A.}, \bibinfo{author}{{Antoranz}, P.},
  \bibinfo{author}{{Babic}, A.}, \bibinfo{author}{{Banerjee}, B.},
  \bibinfo{author}{{Bangale}, P.}, \bibinfo{author}{{Barres de Almeida}, U.},
  \bibinfo{author}{{Barrio}, J.A.}, \bibinfo{author}{{Becerra Gonz{\'a}lez},
  J.}, \bibinfo{author}{{Bednarek}, W.}, \bibinfo{author}{{Bernardini}, E.},
  \bibinfo{author}{{Biasuzzi}, B.}, \bibinfo{author}{{Biland}, A.},
  \bibinfo{author}{{Blanch}, O.}, \bibinfo{author}{{Bonnefoy}, S.},
  \bibinfo{author}{{Bonnoli}, G.}, \bibinfo{author}{{Borracci}, F.},
  \bibinfo{author}{{Bretz}, T.}, \bibinfo{author}{{Buson}, S.},
  \bibinfo{author}{{Carosi}, A.}, \bibinfo{author}{{Chatterjee}, A.},
  \bibinfo{author}{{Clavero}, R.}, \bibinfo{author}{{Colin}, P.},
  \bibinfo{author}{{Colombo}, E.}, \bibinfo{author}{{Contreras}, J.L.},
  \bibinfo{author}{{Cortina}, J.}, \bibinfo{author}{{Covino}, S.},
  \bibinfo{author}{{Da Vela}, P.}, \bibinfo{author}{{Dazzi}, F.},
  \bibinfo{author}{{De Angelis}, A.}, \bibinfo{author}{{De Lotto}, B.},
  \bibinfo{author}{{de O{\~n}a Wilhelmi}, E.}, \bibinfo{author}{{Di Pierro},
  F.}, \bibinfo{author}{{Dom{\'\i}nguez}, A.}, \bibinfo{author}{{Dominis
  Prester}, D.}, \bibinfo{author}{{Dorner}, D.}, \bibinfo{author}{{Doro}, M.},
  \bibinfo{author}{{Einecke}, S.}, \bibinfo{author}{{Eisenacher Glawion}, D.},
  \bibinfo{author}{{Elsaesser}, D.}, \bibinfo{author}{{Fern{\'a}ndez-Barral},
  A.}, \bibinfo{author}{{Fidalgo}, D.}, \bibinfo{author}{{Fonseca}, M.V.},
  \bibinfo{author}{{Font}, L.}, \bibinfo{author}{{Frantzen}, K.},
  \bibinfo{author}{{Fruck}, C.}, \bibinfo{author}{{Galindo}, D.},
  \bibinfo{author}{{Garc{\'\i}a L{\'o}pez}, R.J.},
  \bibinfo{author}{{Garczarczyk}, M.}, \bibinfo{author}{{Garrido Terrats}, D.},
  \bibinfo{author}{{Gaug}, M.}, \bibinfo{author}{{Giammaria}, P.},
  \bibinfo{author}{{Godinovi{\'c}}, N.}, \bibinfo{author}{{Gonz{\'a}lez
  Mu{\~n}oz}, A.}, \bibinfo{author}{{Gora}, D.}, \bibinfo{author}{{Guberman},
  D.}, \bibinfo{author}{{Hadasch}, D.}, \bibinfo{author}{{Hahn}, A.},
  \bibinfo{author}{{Hanabata}, Y.}, \bibinfo{author}{{Hayashida}, M.},
  \bibinfo{author}{{Herrera}, J.}, \bibinfo{author}{{Hose}, J.},
  \bibinfo{author}{{Hrupec}, D.}, \bibinfo{author}{{Hughes}, G.},
  \bibinfo{author}{{Idec}, W.}, \bibinfo{author}{{Kodani}, K.},
  \bibinfo{author}{{Konno}, Y.}, \bibinfo{author}{{Kubo}, H.},
  \bibinfo{author}{{Kushida}, J.}, \bibinfo{author}{{La Barbera}, A.},
  \bibinfo{author}{{Lelas}, D.}, \bibinfo{author}{{Lindfors}, E.},
  \bibinfo{author}{{Lombardi}, S.}, \bibinfo{author}{{Longo}, F.},
  \bibinfo{author}{{L{\'o}pez}, M.}, \bibinfo{author}{{L{\'o}pez-Coto}, R.},
  \bibinfo{author}{{Majumdar}, P.}, \bibinfo{author}{{Makariev}, M.},
  \bibinfo{author}{{Mallot}, K.}, \bibinfo{author}{{Maneva}, G.},
  \bibinfo{author}{{Manganaro}, M.}, \bibinfo{author}{{Mannheim}, K.},
  \bibinfo{author}{{Maraschi}, L.}, \bibinfo{author}{{Marcote}, B.},
  \bibinfo{author}{{Mariotti}, M.}, \bibinfo{author}{{Mart{\'\i}nez}, M.},
  \bibinfo{author}{{Mazin}, D.}, \bibinfo{author}{{Menzel}, U.},
  \bibinfo{author}{{Mirand a}, J.M.}, \bibinfo{author}{{Mirzoyan}, R.},
  \bibinfo{author}{{Moralejo}, A.}, \bibinfo{author}{{Moretti}, E.},
  \bibinfo{author}{{Nakajima}, D.}, \bibinfo{author}{{Neustroev}, V.},
  \bibinfo{author}{{Niedzwiecki}, A.}, \bibinfo{author}{{Nievas Rosillo}, M.},
  \bibinfo{author}{{Nilsson}, K.}, \bibinfo{author}{{Nishijima}, K.},
  \bibinfo{author}{{Noda}, K.}, \bibinfo{author}{{Nogu{\'e}s}, L.},
  \bibinfo{author}{{Orito}, R.}, \bibinfo{author}{{Overkemping}, A.},
  \bibinfo{author}{{Paiano}, S.}, \bibinfo{author}{{Palacio}, J.},
  \bibinfo{author}{{Palatiello}, M.}, \bibinfo{author}{{Paneque}, D.},
  \bibinfo{author}{{Paoletti}, R.}, \bibinfo{author}{{Paredes}, J.M.},
  \bibinfo{author}{{Paredes-Fortuny}, X.}, \bibinfo{author}{{Pedaletti}, G.},
  \bibinfo{author}{{Perri}, L.}, \bibinfo{author}{{Persic}, M.},
  \bibinfo{author}{{Poutanen}, J.}, \bibinfo{author}{{Prada Moroni}, P.G.},
  \bibinfo{author}{{Prandini}, E.}, \bibinfo{author}{{Puljak}, I.},
  \bibinfo{author}{{Rhode}, W.}, \bibinfo{author}{{Rib{\'o}}, M.},
  \bibinfo{author}{{Rico}, J.}, \bibinfo{author}{{Rodriguez Garcia}, J.},
  \bibinfo{author}{{Saito}, T.}, \bibinfo{author}{{Satalecka}, K.},
  \bibinfo{author}{{Schultz}, C.}, \bibinfo{author}{{Schweizer}, T.},
  \bibinfo{author}{{Shore}, S.N.}, \bibinfo{author}{{Sillanp{\"a}{\"a}}, A.},
  \bibinfo{author}{{Sitarek}, J.}, \bibinfo{author}{{Snidaric}, I.},
  \bibinfo{author}{{Sobczynska}, D.}, \bibinfo{author}{{Stamerra}, A.},
  \bibinfo{author}{{Steinbring}, T.}, \bibinfo{author}{{Strzys}, M.},
  \bibinfo{author}{{Takalo}, L.}, \bibinfo{author}{{Takami}, H.},
  \bibinfo{author}{{Tavecchio}, F.}, \bibinfo{author}{{Temnikov}, P.},
  \bibinfo{author}{{Terzi{\'c}}, T.}, \bibinfo{author}{{Tescaro}, D.},
  \bibinfo{author}{{Teshima}, M.}, \bibinfo{author}{{Thaele}, J.},
  \bibinfo{author}{{Torres}, D.F.}, \bibinfo{author}{{Toyama}, T.},
  \bibinfo{author}{{Treves}, A.}, \bibinfo{author}{{Verguilov}, V.},
  \bibinfo{author}{{Vovk}, I.}, \bibinfo{author}{{Ward}, J.E.},
  \bibinfo{author}{{Will}, M.}, \bibinfo{author}{{Wu}, M.H.},
  \bibinfo{author}{{Zanin}, R.}, \bibinfo{year}{2016}.
\newblock \bibinfo{title}{{Long-term multi-wavelength variability and
  correlation study of Markarian 421 from 2007 to 2009}}.
\newblock \bibinfo{journal}{\aap} \bibinfo{volume}{593}, \bibinfo{pages}{A91}.
\newblock \href{http://arxiv.org/abs/1605.09017}{\tt arXiv:1605.09017}.
\bibitem[{{Ahnen} et~al.(2018){Ahnen}, {Ansoldi}, {Antonelli}, {Arcaro},
  {Babi{\'c}}, {Banerjee}, {Bangale}, {Barres de Almeida}, {Barrio}, {Becerra
  Gonz{\'a}lez} and et~al.}]{2018A&A...620A.181A}
\bibinfo{author}{{Ahnen}, M.L.}, \bibinfo{author}{{Ansoldi}, S.},
  \bibinfo{author}{{Antonelli}, L.A.}, \bibinfo{author}{{Arcaro}, C.},
  \bibinfo{author}{{Babi{\'c}}, A.}, \bibinfo{author}{{Banerjee}, B.},
  \bibinfo{author}{{Bangale}, P.}, \bibinfo{author}{{Barres de Almeida}, U.},
  \bibinfo{author}{{Barrio}, J.A.}, \bibinfo{author}{{Becerra Gonz{\'a}lez},
  J.}, \bibinfo{author}{et~al.}, \bibinfo{year}{2018}.
\newblock \bibinfo{title}{{Extreme HBL behavior of Markarian 501 during 2012}}.
\newblock \bibinfo{journal}{\aap} \bibinfo{volume}{620}, \bibinfo{pages}{A181}.
\newblock \href{http://arxiv.org/abs/1808.04300}{\tt arXiv:1808.04300}.
\bibitem[{{Ajello} et~al.(2012){Ajello}, {Alexander}, {Greiner}, {Madejski},
  {Gehrels} and {Burlon}}]{2012ApJ...749...21A}
\bibinfo{author}{{Ajello}, M.}, \bibinfo{author}{{Alexander}, D.M.},
  \bibinfo{author}{{Greiner}, J.}, \bibinfo{author}{{Madejski}, G.M.},
  \bibinfo{author}{{Gehrels}, N.}, \bibinfo{author}{{Burlon}, D.},
  \bibinfo{year}{2012}.
\newblock \bibinfo{title}{{The 60 Month All-sky Burst Alert Telescope Survey of
  Active Galactic Nucleus and the Anisotropy of nearby AGNs}}.
\newblock \bibinfo{journal}{\apj} \bibinfo{volume}{749}, \bibinfo{pages}{21}.
\newblock \href{http://arxiv.org/abs/1202.3137}{\tt arXiv:1202.3137}.
\bibitem[{{Antonucci} et~al.(1993){Antonucci}, {Kinney} and
  {Hurt}}]{1993ApJ...414..506A}
\bibinfo{author}{{Antonucci}, R.}, \bibinfo{author}{{Kinney}, A.L.},
  \bibinfo{author}{{Hurt}, T.}, \bibinfo{year}{1993}.
\newblock \bibinfo{title}{{Hubble Space Telescope ultraviolet spectroscopy of
  the highly polarized but quiescent quasar OI 287}}.
\newblock \bibinfo{journal}{\apj} \bibinfo{volume}{414},
  \bibinfo{pages}{506--509}.
\bibitem[{{Archer} et~al.(2018){Archer}, {Benbow}, {Bird}, {Brose},
  {Buchovecky}, {Bugaev}, {Cui}, {Daniel}, {Falcone}, {Feng}, {Finley},
  {Flinders}, {Fortson}, {Furniss}, {Gillanders}, {H{\"u}tten}, {Hanna},
  {Hervet}, {Holder}, {Hughes}, {Humensky}, {Johnson}, {Kaaret}, {Kar},
  {Kelley-Hoskins}, {Kieda}, {Krause}, {Krennrich}, {Kumar}, {Lang}, {Lin},
  {McArthur}, {Moriarty}, {Mukherjee}, {Nieto}, {O'Brien}, {Ong}, {Otte},
  {Park}, {Petrashyk}, {Pohl}, {Popkow}, {Pueschel}, {Quinn}, {Ragan},
  {Reynolds}, {Richards}, {Roache}, {Rulten}, {Sadeh}, {Sembroski},
  {Shahinyan}, {Tyler}, {Wakely}, {Weiner}, {Weinstein}, {Wells}, {Wilcox},
  {Wilhelm}, {Williams}, {VERITAS Collaboration}, {Brisken} and
  {Pontrelli}}]{2018ApJ...862...41A}
\bibinfo{author}{{Archer}, A.}, \bibinfo{author}{{Benbow}, W.},
  \bibinfo{author}{{Bird}, R.}, \bibinfo{author}{{Brose}, R.},
  \bibinfo{author}{{Buchovecky}, M.}, \bibinfo{author}{{Bugaev}, V.},
  \bibinfo{author}{{Cui}, W.}, \bibinfo{author}{{Daniel}, M.K.},
  \bibinfo{author}{{Falcone}, A.}, \bibinfo{author}{{Feng}, Q.},
  \bibinfo{author}{{Finley}, J.P.}, \bibinfo{author}{{Flinders}, A.},
  \bibinfo{author}{{Fortson}, L.}, \bibinfo{author}{{Furniss}, A.},
  \bibinfo{author}{{Gillanders}, G.H.}, \bibinfo{author}{{H{\"u}tten}, M.},
  \bibinfo{author}{{Hanna}, D.}, \bibinfo{author}{{Hervet}, O.},
  \bibinfo{author}{{Holder}, J.}, \bibinfo{author}{{Hughes}, G.},
  \bibinfo{author}{{Humensky}, T.B.}, \bibinfo{author}{{Johnson}, C.A.},
  \bibinfo{author}{{Kaaret}, P.}, \bibinfo{author}{{Kar}, P.},
  \bibinfo{author}{{Kelley-Hoskins}, N.}, \bibinfo{author}{{Kieda}, D.},
  \bibinfo{author}{{Krause}, M.}, \bibinfo{author}{{Krennrich}, F.},
  \bibinfo{author}{{Kumar}, S.}, \bibinfo{author}{{Lang}, M.J.},
  \bibinfo{author}{{Lin}, T.T.Y.}, \bibinfo{author}{{McArthur}, S.},
  \bibinfo{author}{{Moriarty}, P.}, \bibinfo{author}{{Mukherjee}, R.},
  \bibinfo{author}{{Nieto}, D.}, \bibinfo{author}{{O'Brien}, S.},
  \bibinfo{author}{{Ong}, R.A.}, \bibinfo{author}{{Otte}, A.N.},
  \bibinfo{author}{{Park}, N.}, \bibinfo{author}{{Petrashyk}, A.},
  \bibinfo{author}{{Pohl}, M.}, \bibinfo{author}{{Popkow}, A.},
  \bibinfo{author}{{Pueschel}, E.}, \bibinfo{author}{{Quinn}, J.},
  \bibinfo{author}{{Ragan}, K.}, \bibinfo{author}{{Reynolds}, P.T.},
  \bibinfo{author}{{Richards}, G.T.}, \bibinfo{author}{{Roache}, E.},
  \bibinfo{author}{{Rulten}, C.}, \bibinfo{author}{{Sadeh}, I.},
  \bibinfo{author}{{Sembroski}, G.H.}, \bibinfo{author}{{Shahinyan}, K.},
  \bibinfo{author}{{Tyler}, J.}, \bibinfo{author}{{Wakely}, S.P.},
  \bibinfo{author}{{Weiner}, O.M.}, \bibinfo{author}{{Weinstein}, A.},
  \bibinfo{author}{{Wells}, R.M.}, \bibinfo{author}{{Wilcox}, P.},
  \bibinfo{author}{{Wilhelm}, A.}, \bibinfo{author}{{Williams}, D.A.},
  \bibinfo{author}{{VERITAS Collaboration}}, \bibinfo{author}{{Brisken}, W.F.},
  \bibinfo{author}{{Pontrelli}, P.}, \bibinfo{year}{2018}.
\newblock \bibinfo{title}{{HESS J1943+213: An Extreme Blazar Shining through
  the Galactic Plane}}.
\newblock \bibinfo{journal}{\apj} \bibinfo{volume}{862}, \bibinfo{pages}{41}.
\newblock \DOIprefix\doi{10.3847/1538-4357/aacbd0},
  \href{http://arxiv.org/abs/1806.04144}{\tt arXiv:1806.04144}.
\bibitem[{{Balokovi{\'c}} et~al.(2016){Balokovi{\'c}}, {Paneque}, {Madejski},
  {Furniss}, {Chiang}, {Ajello}, {Alexander}, {Barret}, {Blandford}, {Boggs}
  and et~al.}]{2016ApJ...819..156B}
\bibinfo{author}{{Balokovi{\'c}}, M.}, \bibinfo{author}{{Paneque}, D.},
  \bibinfo{author}{{Madejski}, G.}, \bibinfo{author}{{Furniss}, A.},
  \bibinfo{author}{{Chiang}, J.}, \bibinfo{author}{{Ajello}, M.},
  \bibinfo{author}{{Alexander}, D.M.}, \bibinfo{author}{{Barret}, D.},
  \bibinfo{author}{{Blandford}, R.D.}, \bibinfo{author}{{Boggs}, S.E.},
  \bibinfo{author}{et~al.}, \bibinfo{year}{2016}.
\newblock \bibinfo{title}{{Multiwavelength Study of Quiescent States of Mrk 421
  with Unprecedented Hard X-Ray Coverage Provided by NuSTAR in 2013}}.
\newblock \bibinfo{journal}{\apj} \bibinfo{volume}{819}, \bibinfo{pages}{156}.
\newblock \href{http://arxiv.org/abs/1512.02235}{\tt arXiv:1512.02235}.
\bibitem[{{Barger} et~al.(2005){Barger}, {Cowie}, {Mushotzky}, {Yang}, {Wang},
  {Steffen} and {Capak}}]{2005AJ....129..578B}
\bibinfo{author}{{Barger}, A.J.}, \bibinfo{author}{{Cowie}, L.L.},
  \bibinfo{author}{{Mushotzky}, R.F.}, \bibinfo{author}{{Yang}, Y.},
  \bibinfo{author}{{Wang}, W.H.}, \bibinfo{author}{{Steffen}, A.T.},
  \bibinfo{author}{{Capak}, P.}, \bibinfo{year}{2005}.
\newblock \bibinfo{title}{{The Cosmic Evolution of Hard X-Ray-selected Active
  Galactic Nuclei}}.
\newblock \bibinfo{journal}{\aj} \bibinfo{volume}{129},
  \bibinfo{pages}{578--609}.
\newblock \href{http://arxiv.org/abs/astro-ph/0410527}{\tt
  arXiv:astro-ph/0410527}.
\bibitem[{Barthelmy et~al.(2005)Barthelmy, Barbier, Cummings, Fenimore,
  Gehrels, Hullinger, Krimm, Markwardt, Palmer, Parsons and
  et~al.}]{Barthelmy_2005}
\bibinfo{author}{Barthelmy, S.D.}, \bibinfo{author}{Barbier, L.M.},
  \bibinfo{author}{Cummings, J.R.}, \bibinfo{author}{Fenimore, E.E.},
  \bibinfo{author}{Gehrels, N.}, \bibinfo{author}{Hullinger, D.},
  \bibinfo{author}{Krimm, H.A.}, \bibinfo{author}{Markwardt, C.B.},
  \bibinfo{author}{Palmer, D.M.}, \bibinfo{author}{Parsons, A.},
  \bibinfo{author}{et~al.}, \bibinfo{year}{2005}.
\newblock \bibinfo{title}{The burst alert telescope (bat) on the swift midex
  mission}.
\newblock \bibinfo{journal}{Space Science Reviews} \bibinfo{volume}{120},
  \bibinfo{pages}{143--164}.
\bibitem[{{Bassani} et~al.(2004a){Bassani}, {Alizia} and
  {Stephen}}]{2004cosp...35.4369B}
\bibinfo{author}{{Bassani}, L.}, \bibinfo{author}{{Alizia}, A.},
  \bibinfo{author}{{Stephen}, J.B.}, \bibinfo{year}{2004}a.
\newblock \bibinfo{title}{{Looking for AGN in the INTEGRAL Core Program}}, in:
  \bibinfo{editor}{{Paill{\'e}}, J.P.} (Ed.), \bibinfo{booktitle}{35th COSPAR
  Scientific Assembly}, p. \bibinfo{pages}{4369}.
\bibitem[{{Bassani} et~al.(2007){Bassani}, {Landi}, {Malizia}, {Fiocchi},
  {Bazzano}, {Bird}, {Dean}, {Gehrels}, {Giommi} and
  {Ubertini}}]{2007ApJ...669L...1B}
\bibinfo{author}{{Bassani}, L.}, \bibinfo{author}{{Landi}, R.},
  \bibinfo{author}{{Malizia}, A.}, \bibinfo{author}{{Fiocchi}, M.T.},
  \bibinfo{author}{{Bazzano}, A.}, \bibinfo{author}{{Bird}, A.J.},
  \bibinfo{author}{{Dean}, A.J.}, \bibinfo{author}{{Gehrels}, N.},
  \bibinfo{author}{{Giommi}, P.}, \bibinfo{author}{{Ubertini}, P.},
  \bibinfo{year}{2007}.
\newblock \bibinfo{title}{{IGR J22517+2218=MG3 J225155+2217: A New Gamma-Ray
  Lighthouse in the Distant Universe}}.
\newblock \bibinfo{journal}{\apjl} \bibinfo{volume}{669},
  \bibinfo{pages}{L1--L4}.
\newblock \href{http://arxiv.org/abs/0709.3023}{\tt arXiv:0709.3023}.
\bibitem[{{Bassani} et~al.(2012){Bassani}, {Landi}, {Marshall}, {Malizia},
  {Bazzano}, {Bird}, {Gehrels}, {Ubertini} and {Masetti}}]{2012A&A...543A...1B}
\bibinfo{author}{{Bassani}, L.}, \bibinfo{author}{{Landi}, R.},
  \bibinfo{author}{{Marshall}, F.E.}, \bibinfo{author}{{Malizia}, A.},
  \bibinfo{author}{{Bazzano}, A.}, \bibinfo{author}{{Bird}, A.J.},
  \bibinfo{author}{{Gehrels}, N.}, \bibinfo{author}{{Ubertini}, P.},
  \bibinfo{author}{{Masetti}, N.}, \bibinfo{year}{2012}.
\newblock \bibinfo{title}{{IGR J12319-0749: evidence for another extreme blazar
  found with INTEGRAL}}.
\newblock \bibinfo{journal}{\aap} \bibinfo{volume}{543}, \bibinfo{pages}{A1}.
\newblock \DOIprefix\doi{10.1051/0004-6361/201219243},
  \href{http://arxiv.org/abs/1205.3037}{\tt arXiv:1205.3037}.
\bibitem[{{Bassani} et~al.(2004b){Bassani}, {Malizia}, {Stephen}, {Gianotti},
  {Schiavone}, {Bazzano}, {Bird}, {Bouchet}, {Courvoisier}, {Dean}, {De
  Cesare}, {Del Santo}, {De Rosa}, {Hudec}, {Mirabel}, {Laurent}, {Piro},
  {Shaw} and {Zdziarski}}]{2004ESASP.552..139B}
\bibinfo{author}{{Bassani}, L.}, \bibinfo{author}{{Malizia}, A.},
  \bibinfo{author}{{Stephen}, J.B.}, \bibinfo{author}{{Gianotti}, F.},
  \bibinfo{author}{{Schiavone}, F.}, \bibinfo{author}{{Bazzano}, a.},
  \bibinfo{author}{{Bird}, A.J.}, \bibinfo{author}{{Bouchet}, L.},
  \bibinfo{author}{{Courvoisier}, T.}, \bibinfo{author}{{Dean}, A.J.},
  \bibinfo{author}{{De Cesare}, G.}, \bibinfo{author}{{Del Santo}, M.},
  \bibinfo{author}{{De Rosa}, A.}, \bibinfo{author}{{Hudec}, R.},
  \bibinfo{author}{{Mirabel}, F.}, \bibinfo{author}{{Laurent}, P.},
  \bibinfo{author}{{Piro}, L.}, \bibinfo{author}{{Shaw}, S.},
  \bibinfo{author}{{Zdziarski}, A.A.}, \bibinfo{year}{2004}b.
\newblock \bibinfo{title}{{The Sky Behind Our Galaxy as Seen by IBIS on
  INTEGRAL}}, in: \bibinfo{editor}{{Schoenfelder}, V.},
  \bibinfo{editor}{{Lichti}, G.}, \bibinfo{editor}{{Winkler}, C.} (Eds.),
  \bibinfo{booktitle}{5th INTEGRAL Workshop on the INTEGRAL Universe}, p.
  \bibinfo{pages}{139}.
\newblock \href{http://arxiv.org/abs/astro-ph/0404442}{\tt
  arXiv:astro-ph/0404442}.
\bibitem[{{Bassani} et~al.(2013){Bassani}, {Molina}, {Landi}, {Malizia},
  {Bird}, {Bazzano} and {Ubertini}}]{2013arXiv1302.2447B}
\bibinfo{author}{{Bassani}, L.}, \bibinfo{author}{{Molina}, M.},
  \bibinfo{author}{{Landi}, R.}, \bibinfo{author}{{Malizia}, A.},
  \bibinfo{author}{{Bird}, A.J.}, \bibinfo{author}{{Bazzano}, A.},
  \bibinfo{author}{{Ubertini}, P.}, \bibinfo{year}{2013}.
\newblock \bibinfo{title}{{Chasing extreme blazars with INTEGRAL}}.
\newblock \bibinfo{journal}{arXiv e-prints} ,
  \bibinfo{pages}{arXiv:1302.2447}\href{http://arxiv.org/abs/1302.2447}{\tt
  arXiv:1302.2447}.
\bibitem[{{Bassani} et~al.(2006){Bassani}, {Molina}, {Malizia}, {Stephen},
  {Bird}, {Bazzano}, {B{\'e}langer}, {Dean}, {De Rosa}, {Laurent}, {Lebrun},
  {Ubertini} and {Walter}}]{2006ApJ...636L..65B}
\bibinfo{author}{{Bassani}, L.}, \bibinfo{author}{{Molina}, M.},
  \bibinfo{author}{{Malizia}, A.}, \bibinfo{author}{{Stephen}, J.B.},
  \bibinfo{author}{{Bird}, A.J.}, \bibinfo{author}{{Bazzano}, A.},
  \bibinfo{author}{{B{\'e}langer}, G.}, \bibinfo{author}{{Dean}, A.J.},
  \bibinfo{author}{{De Rosa}, A.}, \bibinfo{author}{{Laurent}, P.},
  \bibinfo{author}{{Lebrun}, F.}, \bibinfo{author}{{Ubertini}, P.},
  \bibinfo{author}{{Walter}, R.}, \bibinfo{year}{2006}.
\newblock \bibinfo{title}{{INTEGRAL IBIS Extragalactic Survey: Active Galactic
  Nuclei Selected at 20-100 keV}}.
\newblock \bibinfo{journal}{\apjl} \bibinfo{volume}{636},
  \bibinfo{pages}{L65--L68}.
\newblock \DOIprefix\doi{10.1086/500132},
  \href{http://arxiv.org/abs/astro-ph/0512015}{\tt arXiv:astro-ph/0512015}.
\bibitem[{{Bassani} et~al.(2016){Bassani}, {Venturi}, {Molina}, {Malizia},
  {Dallacasa}, {Panessa}, {Bazzano} and {Ubertini}}]{2016MNRAS.461.3165B}
\bibinfo{author}{{Bassani}, L.}, \bibinfo{author}{{Venturi}, T.},
  \bibinfo{author}{{Molina}, M.}, \bibinfo{author}{{Malizia}, A.},
  \bibinfo{author}{{Dallacasa}, D.}, \bibinfo{author}{{Panessa}, F.},
  \bibinfo{author}{{Bazzano}, A.}, \bibinfo{author}{{Ubertini}, P.},
  \bibinfo{year}{2016}.
\newblock \bibinfo{title}{{Soft {\ensuremath{\gamma}}-ray selected radio
  galaxies: favouring giant size discovery}}.
\newblock \bibinfo{journal}{\mnras} \bibinfo{volume}{461},
  \bibinfo{pages}{3165--3171}.
\newblock \href{http://arxiv.org/abs/1606.05456}{\tt arXiv:1606.05456}.
\bibitem[{{Bazzano} et~al.(2006){Bazzano}, {Stephen}, {Fiocchi}, {Bird},
  {Bassani}, {Dean}, {Malizia}, {Ubertini}, {Lebrun} and
  {Walter}}]{2006ApJ...649L...9B}
\bibinfo{author}{{Bazzano}, A.}, \bibinfo{author}{{Stephen}, J.B.},
  \bibinfo{author}{{Fiocchi}, M.}, \bibinfo{author}{{Bird}, A.J.},
  \bibinfo{author}{{Bassani}, L.}, \bibinfo{author}{{Dean}, A.J.},
  \bibinfo{author}{{Malizia}, A.}, \bibinfo{author}{{Ubertini}, P.},
  \bibinfo{author}{{Lebrun}, F.}, \bibinfo{author}{{Walter}, R.},
  \bibinfo{year}{2006}.
\newblock \bibinfo{title}{{INTEGRAL IBIS Census of the Sky Beyond 100 keV}}.
\newblock \bibinfo{journal}{\apj} \bibinfo{volume}{649},
  \bibinfo{pages}{L9--L12}.
\newblock \href{http://arxiv.org/abs/astro-ph/0608164}{\tt
  arXiv:astro-ph/0608164}.
\bibitem[{{Beckmann} et~al.(2007){Beckmann}, {Barthelmy}, {Courvoisier},
  {Gehrels}, {Soldi}, {Tueller} and {Wendt}}]{2007A&A...475..827B}
\bibinfo{author}{{Beckmann}, V.}, \bibinfo{author}{{Barthelmy}, S.D.},
  \bibinfo{author}{{Courvoisier}, T.J.L.}, \bibinfo{author}{{Gehrels}, N.},
  \bibinfo{author}{{Soldi}, S.}, \bibinfo{author}{{Tueller}, J.},
  \bibinfo{author}{{Wendt}, G.}, \bibinfo{year}{2007}.
\newblock \bibinfo{title}{{Hard X-ray variability of active galactic nuclei}}.
\newblock \bibinfo{journal}{\aap} \bibinfo{volume}{475},
  \bibinfo{pages}{827--835}.
\newblock \href{http://arxiv.org/abs/0709.2230}{\tt arXiv:0709.2230}.
\bibitem[{{Beckmann} and {Do Cao}(2010)}]{2010int..workE..81B}
\bibinfo{author}{{Beckmann}, V.}, \bibinfo{author}{{Do Cao}, O.},
  \bibinfo{year}{2010}.
\newblock \bibinfo{title}{{The elusive radio loud Seyfert 2 galaxy NGC 2110}},
  in: \bibinfo{booktitle}{Eighth Integral Workshop. The Restless Gamma-ray
  Universe (INTEGRAL 2010)}, p.~\bibinfo{pages}{81}.
\newblock \href{http://arxiv.org/abs/1102.4974}{\tt arXiv:1102.4974}.
\bibitem[{{Beckmann} et~al.(2004){Beckmann}, {Gehrels}, {Favre}, {Walter},
  {Courvoisier}, {Petrucci} and {Malzac}}]{2004ApJ...614..641B}
\bibinfo{author}{{Beckmann}, V.}, \bibinfo{author}{{Gehrels}, N.},
  \bibinfo{author}{{Favre}, P.}, \bibinfo{author}{{Walter}, R.},
  \bibinfo{author}{{Courvoisier}, T.J.L.}, \bibinfo{author}{{Petrucci}, P.O.},
  \bibinfo{author}{{Malzac}, J.}, \bibinfo{year}{2004}.
\newblock \bibinfo{title}{{INTEGRAL and XMM-Newton Spectral Studies of NGC
  4388}}.
\newblock \bibinfo{journal}{\apj} \bibinfo{volume}{614},
  \bibinfo{pages}{641--647}.
\newblock \href{http://arxiv.org/abs/astro-ph/0406553}{\tt
  arXiv:astro-ph/0406553}.
\bibitem[{{Beckmann} et~al.(2011){Beckmann}, {Jean}, {Lubi{\'n}ski}, {Soldi}
  and {Terrier}}]{2011A&A...531A..70B}
\bibinfo{author}{{Beckmann}, V.}, \bibinfo{author}{{Jean}, P.},
  \bibinfo{author}{{Lubi{\'n}ski}, P.}, \bibinfo{author}{{Soldi}, S.},
  \bibinfo{author}{{Terrier}, R.}, \bibinfo{year}{2011}.
\newblock \bibinfo{title}{{The hard X-ray emission of Centaurus A}}.
\newblock \bibinfo{journal}{\aap} \bibinfo{volume}{531}, \bibinfo{pages}{A70}.
\newblock \href{http://arxiv.org/abs/1104.4253}{\tt arXiv:1104.4253}.
\bibitem[{Beckmann et~al.(2005)Beckmann, Shrader, Gehrels, Soldi, Lubi{\'n}ski,
  Zdziarski, Petrucci and Malzac}]{Beckmann_2005}
\bibinfo{author}{Beckmann, V.}, \bibinfo{author}{Shrader, C.R.},
  \bibinfo{author}{Gehrels, N.}, \bibinfo{author}{Soldi, S.},
  \bibinfo{author}{Lubi{\'n}ski, P.}, \bibinfo{author}{Zdziarski, A.A.},
  \bibinfo{author}{Petrucci, P.}, \bibinfo{author}{Malzac, J.},
  \bibinfo{year}{2005}.
\newblock \bibinfo{title}{The high‐energy spectrum of ngc 4151}.
\newblock \bibinfo{journal}{The Astrophysical Journal} \bibinfo{volume}{634},
  \bibinfo{pages}{939--946}.
\bibitem[{{Beckmann} et~al.(2009){Beckmann}, {Soldi}, {Ricci},
  {Alfonso-Garz{\'o}n}, {Courvoisier}, {Domingo}, {Gehrels}, {Lubi{\'n}ski},
  {Mas-Hesse} and {Zdziarski}}]{2009A&A...505..417B}
\bibinfo{author}{{Beckmann}, V.}, \bibinfo{author}{{Soldi}, S.},
  \bibinfo{author}{{Ricci}, C.}, \bibinfo{author}{{Alfonso-Garz{\'o}n}, J.},
  \bibinfo{author}{{Courvoisier}, T.J.L.}, \bibinfo{author}{{Domingo}, A.},
  \bibinfo{author}{{Gehrels}, N.}, \bibinfo{author}{{Lubi{\'n}ski}, P.},
  \bibinfo{author}{{Mas-Hesse}, J.M.}, \bibinfo{author}{{Zdziarski}, A.A.},
  \bibinfo{year}{2009}.
\newblock \bibinfo{title}{{The second INTEGRAL AGN catalogue}}.
\newblock \bibinfo{journal}{\aap} \bibinfo{volume}{505},
  \bibinfo{pages}{417--439}.
\newblock \href{http://arxiv.org/abs/0907.0654}{\tt arXiv:0907.0654}.
\bibitem[{Beckmann et~al.(2006)Beckmann, Soldi, Shrader, Gehrels and
  Produit}]{Beckmann_2006}
\bibinfo{author}{Beckmann, V.}, \bibinfo{author}{Soldi, S.},
  \bibinfo{author}{Shrader, C.R.}, \bibinfo{author}{Gehrels, N.},
  \bibinfo{author}{Produit, N.}, \bibinfo{year}{2006}.
\newblock \bibinfo{title}{The hard x‐ray 20--40 kev agn luminosity function}.
\newblock \bibinfo{journal}{The Astrophysical Journal} \bibinfo{volume}{652},
  \bibinfo{pages}{126--135}.
\bibitem[{{Benbow} and {VERITAS Collaboration}(2017)}]{2017ICRC...35..641B}
\bibinfo{author}{{Benbow}, W.}, \bibinfo{author}{{VERITAS Collaboration}},
  \bibinfo{year}{2017}.
\newblock \bibinfo{title}{{Highlights from the VERITAS AGN Observation
  Program}}.
\newblock \bibinfo{journal}{International Cosmic Ray Conference}
  \bibinfo{volume}{301}, \bibinfo{pages}{641}.
\newblock \href{http://arxiv.org/abs/1708.02374}{\tt arXiv:1708.02374}.
\bibitem[{{Bianchin} et~al.(2009){Bianchin}, {Foschini}, {Ghisellini},
  {Tagliaferri}, {Tavecchio}, {Treves}, {Di Cocco}, {Gliozzi}, {Pian},
  {Sambruna} and {Wolter}}]{2009A&A...496..423B}
\bibinfo{author}{{Bianchin}, V.}, \bibinfo{author}{{Foschini}, L.},
  \bibinfo{author}{{Ghisellini}, G.}, \bibinfo{author}{{Tagliaferri}, G.},
  \bibinfo{author}{{Tavecchio}, F.}, \bibinfo{author}{{Treves}, A.},
  \bibinfo{author}{{Di Cocco}, G.}, \bibinfo{author}{{Gliozzi}, M.},
  \bibinfo{author}{{Pian}, E.}, \bibinfo{author}{{Sambruna}, R.M.},
  \bibinfo{author}{{Wolter}, A.}, \bibinfo{year}{2009}.
\newblock \bibinfo{title}{{The changing look of PKS 2149-306}}.
\newblock \bibinfo{journal}{\aap} \bibinfo{volume}{496},
  \bibinfo{pages}{423--428}.
\newblock \href{http://arxiv.org/abs/0902.1789}{\tt arXiv:0902.1789}.
\bibitem[{Bikmaev et~al.(2008)Bikmaev, Burenin, Revnivtsev, Sazonov, Sunyaev,
  Pavlinsky and Sakhibullin}]{Bikmaev_2008}
\bibinfo{author}{Bikmaev, I.F.}, \bibinfo{author}{Burenin, R.A.},
  \bibinfo{author}{Revnivtsev, M.G.}, \bibinfo{author}{Sazonov, S.Y.},
  \bibinfo{author}{Sunyaev, R.A.}, \bibinfo{author}{Pavlinsky, M.N.},
  \bibinfo{author}{Sakhibullin, N.A.}, \bibinfo{year}{2008}.
\newblock \bibinfo{title}{Optical identifications of five integral hard x-ray
  sources in the galactic plane}.
\newblock \bibinfo{journal}{Astronomy Letters} \bibinfo{volume}{34},
  \bibinfo{pages}{653--663}.
\bibitem[{Bikmaev et~al.(2006)Bikmaev, Sunyaev, Revnivtsev and
  Burenin}]{Bikmaev_2006}
\bibinfo{author}{Bikmaev, I.F.}, \bibinfo{author}{Sunyaev, R.A.},
  \bibinfo{author}{Revnivtsev, M.G.}, \bibinfo{author}{Burenin, R.A.},
  \bibinfo{year}{2006}.
\newblock \bibinfo{title}{New nearby active galactic nuclei among integral and
  rxte x-ray sources}.
\newblock \bibinfo{journal}{Astronomy Letters} \bibinfo{volume}{32},
  \bibinfo{pages}{221--227}.
\bibitem[{Bird et~al.(2006)Bird, Barlow, Bassani, Bazzano, Belanger, Bodaghee,
  Capitanio, Dean, Fiocchi, Hill and et~al.}]{Bird_2006}
\bibinfo{author}{Bird, A.J.}, \bibinfo{author}{Barlow, E.J.},
  \bibinfo{author}{Bassani, L.}, \bibinfo{author}{Bazzano, A.},
  \bibinfo{author}{Belanger, G.}, \bibinfo{author}{Bodaghee, A.},
  \bibinfo{author}{Capitanio, F.}, \bibinfo{author}{Dean, A.J.},
  \bibinfo{author}{Fiocchi, M.}, \bibinfo{author}{Hill, A.B.},
  \bibinfo{author}{et~al.}, \bibinfo{year}{2006}.
\newblock \bibinfo{title}{The second ibis/isgri soft gamma‐ray survey
  catalog}.
\newblock \bibinfo{journal}{The Astrophysical Journal} \bibinfo{volume}{636},
  \bibinfo{pages}{765--776}.
\bibitem[{Bird et~al.(2004)Bird, Barlow, Bassani, Bazzano, Bodaghee, Capitanio,
  Cocchi, Del~Santo, Dean, Hill and et~al.}]{Bird_2004}
\bibinfo{author}{Bird, A.J.}, \bibinfo{author}{Barlow, E.J.},
  \bibinfo{author}{Bassani, L.}, \bibinfo{author}{Bazzano, A.},
  \bibinfo{author}{Bodaghee, A.}, \bibinfo{author}{Capitanio, F.},
  \bibinfo{author}{Cocchi, M.}, \bibinfo{author}{Del~Santo, M.},
  \bibinfo{author}{Dean, A.J.}, \bibinfo{author}{Hill, A.B.},
  \bibinfo{author}{et~al.}, \bibinfo{year}{2004}.
\newblock \bibinfo{title}{The first ibis/isgri soft gamma-ray galactic plane
  survey catalog}.
\newblock \bibinfo{journal}{The Astrophysical Journal} \bibinfo{volume}{607},
  \bibinfo{pages}{L33--L37}.
\bibitem[{Bird et~al.(2009)Bird, Bazzano, Bassani, Capitanio, Fiocchi, Hill,
  Malizia, McBride, Scaringi, Sguera and et~al.}]{Bird_2009}
\bibinfo{author}{Bird, A.J.}, \bibinfo{author}{Bazzano, A.},
  \bibinfo{author}{Bassani, L.}, \bibinfo{author}{Capitanio, F.},
  \bibinfo{author}{Fiocchi, M.}, \bibinfo{author}{Hill, A.B.},
  \bibinfo{author}{Malizia, A.}, \bibinfo{author}{McBride, V.A.},
  \bibinfo{author}{Scaringi, S.}, \bibinfo{author}{Sguera, V.},
  \bibinfo{author}{et~al.}, \bibinfo{year}{2009}.
\newblock \bibinfo{title}{The fourth ibis/isgri soft gamma-ray survey catalog}.
\newblock \bibinfo{journal}{The Astrophysical Journal Supplement Series}
  \bibinfo{volume}{186}, \bibinfo{pages}{1--9}.
\bibitem[{Bird et~al.(2016)Bird, Bazzano, Malizia, Fiocchi, Sguera, Bassani,
  Hill, Ubertini and Winkler}]{Bird_2016}
\bibinfo{author}{Bird, A.J.}, \bibinfo{author}{Bazzano, A.},
  \bibinfo{author}{Malizia, A.}, \bibinfo{author}{Fiocchi, M.},
  \bibinfo{author}{Sguera, V.}, \bibinfo{author}{Bassani, L.},
  \bibinfo{author}{Hill, A.B.}, \bibinfo{author}{Ubertini, P.},
  \bibinfo{author}{Winkler, C.}, \bibinfo{year}{2016}.
\newblock \bibinfo{title}{The ibis soft gamma-ray sky after
  1000integralorbits}.
\newblock \bibinfo{journal}{The Astrophysical Journal Supplement Series}
  \bibinfo{volume}{223}, \bibinfo{pages}{15}.
\bibitem[{Bird et~al.(2007)Bird, Malizia, Bazzano, Barlow, Bassani, Hill,
  Belanger, Capitanio, Clark, Dean and et~al.}]{Bird_2007}
\bibinfo{author}{Bird, A.J.}, \bibinfo{author}{Malizia, A.},
  \bibinfo{author}{Bazzano, A.}, \bibinfo{author}{Barlow, E.J.},
  \bibinfo{author}{Bassani, L.}, \bibinfo{author}{Hill, A.B.},
  \bibinfo{author}{Belanger, G.}, \bibinfo{author}{Capitanio, F.},
  \bibinfo{author}{Clark, D.J.}, \bibinfo{author}{Dean, A.J.},
  \bibinfo{author}{et~al.}, \bibinfo{year}{2007}.
\newblock \bibinfo{title}{The third ibis/isgri soft gamma‐ray survey
  catalog}.
\newblock \bibinfo{journal}{The Astrophysical Journal Supplement Series}
  \bibinfo{volume}{170}, \bibinfo{pages}{175--186}.
\bibitem[{{Boettcher}(2010)}]{2010arXiv1006.5048B}
\bibinfo{author}{{Boettcher}, M.}, \bibinfo{year}{2010}.
\newblock \bibinfo{title}{{Models for the Spectral Energy Distributions and
  Variability of Blazars}}.
\newblock \bibinfo{journal}{arXiv e-prints} ,
  \bibinfo{pages}{arXiv:1006.5048}\href{http://arxiv.org/abs/1006.5048}{\tt
  arXiv:1006.5048}.
\bibitem[{{Bottacini} et~al.(2010a){Bottacini}, {Ajello}, {Greiner}, {Pian},
  {Rau}, {Palazzi}, {Covino}, {Ghisellini}, {Kr{\"u}hler}, {K{\"u}pc{\"u}
  Yolda{\textcommabelow s}}, {Cappelluti} and {Afonso}}]{2010A&A...509A..69B}
\bibinfo{author}{{Bottacini}, E.}, \bibinfo{author}{{Ajello}, M.},
  \bibinfo{author}{{Greiner}, J.}, \bibinfo{author}{{Pian}, E.},
  \bibinfo{author}{{Rau}, A.}, \bibinfo{author}{{Palazzi}, E.},
  \bibinfo{author}{{Covino}, S.}, \bibinfo{author}{{Ghisellini}, G.},
  \bibinfo{author}{{Kr{\"u}hler}, T.}, \bibinfo{author}{{K{\"u}pc{\"u}
  Yolda{\textcommabelow s}}, A.}, \bibinfo{author}{{Cappelluti}, N.},
  \bibinfo{author}{{Afonso}, P.}, \bibinfo{year}{2010}a.
\newblock \bibinfo{title}{{PKS 0537-286, carrying the information of the
  environment of SMBHs in the early Universe}}.
\newblock \bibinfo{journal}{\aap} \bibinfo{volume}{509}, \bibinfo{pages}{A69}.
\newblock \href{http://arxiv.org/abs/0910.2463}{\tt arXiv:0910.2463}.
\bibitem[{{Bottacini} et~al.(2016){Bottacini}, {B{\"o}ttcher}, {Pian} and
  {Collmar}}]{2016ApJ...832...17B}
\bibinfo{author}{{Bottacini}, E.}, \bibinfo{author}{{B{\"o}ttcher}, M.},
  \bibinfo{author}{{Pian}, E.}, \bibinfo{author}{{Collmar}, W.},
  \bibinfo{year}{2016}.
\newblock \bibinfo{title}{{3C 279 in Outburst in 2015 June: A Broadband SED
  Study Based on the INTEGRAL Detection}}.
\newblock \bibinfo{journal}{\apj} \bibinfo{volume}{832}, \bibinfo{pages}{17}.
\newblock \href{http://arxiv.org/abs/1610.01617}{\tt arXiv:1610.01617}.
\bibitem[{{Bottacini} et~al.(2010b){Bottacini}, {B{\"o}ttcher}, {Schady},
  {Rau}, {Zhang}, {Ajello}, {Fendt} and {Greiner}}]{2010ApJ...719L.162B}
\bibinfo{author}{{Bottacini}, E.}, \bibinfo{author}{{B{\"o}ttcher}, M.},
  \bibinfo{author}{{Schady}, P.}, \bibinfo{author}{{Rau}, A.},
  \bibinfo{author}{{Zhang}, X.L.}, \bibinfo{author}{{Ajello}, M.},
  \bibinfo{author}{{Fendt}, C.}, \bibinfo{author}{{Greiner}, J.},
  \bibinfo{year}{2010}b.
\newblock \bibinfo{title}{{Probing the Transition Between the Synchrotron and
  Inverse-compton Spectral Components of 1ES 1959+650}}.
\newblock \bibinfo{journal}{\apjl} \bibinfo{volume}{719},
  \bibinfo{pages}{L162--L166}.
\newblock \href{http://arxiv.org/abs/1010.3259}{\tt arXiv:1010.3259}.
\bibitem[{Bouchet et~al.(2008)Bouchet, Jourdain, Roques, Strong, Diehl, Lebrun
  and Terrier}]{Bouchet_2008}
\bibinfo{author}{Bouchet, L.}, \bibinfo{author}{Jourdain, E.},
  \bibinfo{author}{Roques, J.}, \bibinfo{author}{Strong, A.},
  \bibinfo{author}{Diehl, R.}, \bibinfo{author}{Lebrun, F.},
  \bibinfo{author}{Terrier, R.}, \bibinfo{year}{2008}.
\newblock \bibinfo{title}{Integralspi all‐sky view in soft gamma rays: A
  study of point‐source and galactic diffuse emission}.
\newblock \bibinfo{journal}{The Astrophysical Journal} \bibinfo{volume}{679},
  \bibinfo{pages}{1315--1326}.
\bibitem[{{Bruni} et~al.(2019){Bruni}, {Panessa}, {Bassani}, {Chiaraluce},
  {Kraus}, {Dallacasa}, {Bazzano}, {Hern{\'a}ndez-Garc{\'\i}a}, {Malizia},
  {Ubertini}, {Ursini} and {Venturi}}]{2019ApJ...875...88B}
\bibinfo{author}{{Bruni}, G.}, \bibinfo{author}{{Panessa}, F.},
  \bibinfo{author}{{Bassani}, L.}, \bibinfo{author}{{Chiaraluce}, E.},
  \bibinfo{author}{{Kraus}, A.}, \bibinfo{author}{{Dallacasa}, D.},
  \bibinfo{author}{{Bazzano}, A.},
  \bibinfo{author}{{Hern{\'a}ndez-Garc{\'\i}a}, L.},
  \bibinfo{author}{{Malizia}, A.}, \bibinfo{author}{{Ubertini}, P.},
  \bibinfo{author}{{Ursini}, F.}, \bibinfo{author}{{Venturi}, T.},
  \bibinfo{year}{2019}.
\newblock \bibinfo{title}{{A Discovery of Young Radio Sources in the Cores of
  Giant Radio Galaxies Selected at Hard X-Rays}}.
\newblock \bibinfo{journal}{\apj} \bibinfo{volume}{875}, \bibinfo{pages}{88}.
\newblock \href{http://arxiv.org/abs/1903.05922}{\tt arXiv:1903.05922}.
\bibitem[{{Burlon} et~al.(2011){Burlon}, {Ajello}, {Greiner}, {Comastri},
  {Merloni} and {Gehrels}}]{2011ApJ...728...58B}
\bibinfo{author}{{Burlon}, D.}, \bibinfo{author}{{Ajello}, M.},
  \bibinfo{author}{{Greiner}, J.}, \bibinfo{author}{{Comastri}, A.},
  \bibinfo{author}{{Merloni}, A.}, \bibinfo{author}{{Gehrels}, N.},
  \bibinfo{year}{2011}.
\newblock \bibinfo{title}{{Three-year Swift-BAT Survey of Active Galactic
  Nuclei: Reconciling Theory and Observations?}}
\newblock \bibinfo{journal}{\apj} \bibinfo{volume}{728}, \bibinfo{pages}{58}.
\newblock \href{http://arxiv.org/abs/1012.0302}{\tt arXiv:1012.0302}.
\bibitem[{{Buttiglione} et~al.(2010){Buttiglione}, {Capetti}, {Celotti},
  {Axon}, {Chiaberge}, {Macchetto} and {Sparks}}]{2010A&A...509A...6B}
\bibinfo{author}{{Buttiglione}, S.}, \bibinfo{author}{{Capetti}, A.},
  \bibinfo{author}{{Celotti}, A.}, \bibinfo{author}{{Axon}, D.J.},
  \bibinfo{author}{{Chiaberge}, M.}, \bibinfo{author}{{Macchetto}, F.D.},
  \bibinfo{author}{{Sparks}, W.B.}, \bibinfo{year}{2010}.
\newblock \bibinfo{title}{{An optical spectroscopic survey of the 3CR sample of
  radio galaxies with z \&lt; 0.3 . II. Spectroscopic classes and accretion
  modes in radio-loud AGN}}.
\newblock \bibinfo{journal}{\aap} \bibinfo{volume}{509}, \bibinfo{pages}{A6}.
\newblock \href{http://arxiv.org/abs/0911.0536}{\tt arXiv:0911.0536}.
\bibitem[{{Castignani} et~al.(2017){Castignani}, {Pian}, {Belloni}, {D'Ammand
  o}, {Foschini}, {Ghisellini}, {Pursimo}, {Bazzano}, {Beckmann}, {Bianchin},
  {Fiocchi}, {Impiombato}, {Raiteri}, {Soldi}, {Tagliaferri}, {Treves} and
  {T{\"u}rler}}]{2017A&A...601A..30C}
\bibinfo{author}{{Castignani}, G.}, \bibinfo{author}{{Pian}, E.},
  \bibinfo{author}{{Belloni}, T.M.}, \bibinfo{author}{{D'Ammand o}, F.},
  \bibinfo{author}{{Foschini}, L.}, \bibinfo{author}{{Ghisellini}, G.},
  \bibinfo{author}{{Pursimo}, T.}, \bibinfo{author}{{Bazzano}, A.},
  \bibinfo{author}{{Beckmann}, V.}, \bibinfo{author}{{Bianchin}, V.},
  \bibinfo{author}{{Fiocchi}, M.T.}, \bibinfo{author}{{Impiombato}, D.},
  \bibinfo{author}{{Raiteri}, C.M.}, \bibinfo{author}{{Soldi}, S.},
  \bibinfo{author}{{Tagliaferri}, G.}, \bibinfo{author}{{Treves}, A.},
  \bibinfo{author}{{T{\"u}rler}, M.}, \bibinfo{year}{2017}.
\newblock \bibinfo{title}{{Multiwavelength variability study and search for
  periodicity of PKS 1510-089}}.
\newblock \bibinfo{journal}{\aap} \bibinfo{volume}{601}, \bibinfo{pages}{A30}.
\newblock \href{http://arxiv.org/abs/1612.05281}{\tt arXiv:1612.05281}.
\bibitem[{{Chevalier} et~al.(2019){Chevalier}, {Sanchez}, {Serpico}, {Lenain}
  and {Maurin}}]{2019MNRAS.484..749C}
\bibinfo{author}{{Chevalier}, J.}, \bibinfo{author}{{Sanchez}, D.A.},
  \bibinfo{author}{{Serpico}, P.D.}, \bibinfo{author}{{Lenain}, J.P.},
  \bibinfo{author}{{Maurin}, G.}, \bibinfo{year}{2019}.
\newblock \bibinfo{title}{{Variability studies and modelling of the blazar PKS
  2155-304 in the light of a decade of multi-wavelength observations}}.
\newblock \bibinfo{journal}{\mnras} \bibinfo{volume}{484},
  \bibinfo{pages}{749--759}.
\newblock \href{http://arxiv.org/abs/1901.01743}{\tt arXiv:1901.01743}.
\bibitem[{{Chiaro} et~al.(2016){Chiaro}, {Salvetti}, {La Mura}, {Giroletti},
  {Thompson} and {Bastieri}}]{2016MNRAS.462.3180C}
\bibinfo{author}{{Chiaro}, G.}, \bibinfo{author}{{Salvetti}, D.},
  \bibinfo{author}{{La Mura}, G.}, \bibinfo{author}{{Giroletti}, M.},
  \bibinfo{author}{{Thompson}, D.J.}, \bibinfo{author}{{Bastieri}, D.},
  \bibinfo{year}{2016}.
\newblock \bibinfo{title}{{Blazar flaring patterns (B-FlaP) classifying blazar
  candidate of uncertain type in the third Fermi-LAT catalogue by artificial
  neural networks}}.
\newblock \bibinfo{journal}{\mnras} \bibinfo{volume}{462},
  \bibinfo{pages}{3180--3195}.
\newblock \href{http://arxiv.org/abs/1607.07822}{\tt arXiv:1607.07822}.
\bibitem[{{Collmar} et~al.(2010){Collmar}, {B{\"o}ttcher}, {Krichbaum},
  {Agudo}, {Bottacini}, {Bremer}, {Burwitz}, {Cuccchiara}, {Grupe} and
  {Gurwell}}]{2010A&A...522A..66C}
\bibinfo{author}{{Collmar}, W.}, \bibinfo{author}{{B{\"o}ttcher}, M.},
  \bibinfo{author}{{Krichbaum}, T.P.}, \bibinfo{author}{{Agudo}, I.},
  \bibinfo{author}{{Bottacini}, E.}, \bibinfo{author}{{Bremer}, M.},
  \bibinfo{author}{{Burwitz}, V.}, \bibinfo{author}{{Cuccchiara}, A.},
  \bibinfo{author}{{Grupe}, D.}, \bibinfo{author}{{Gurwell}, M.},
  \bibinfo{year}{2010}.
\newblock \bibinfo{title}{{The multifrequency campaign on 3C 279 in January
  2006}}.
\newblock \bibinfo{journal}{\aap} \bibinfo{volume}{522}, \bibinfo{pages}{A66}.
\newblock \href{http://arxiv.org/abs/1008.1010}{\tt arXiv:1008.1010}.
\bibitem[{Comastri et~al.(2006)Comastri, Gilli and Hasinger}]{Comastri_2006}
\bibinfo{author}{Comastri, A.}, \bibinfo{author}{Gilli, R.},
  \bibinfo{author}{Hasinger, G.}, \bibinfo{year}{2006}.
\newblock \bibinfo{title}{Rolling down from the 30 kev peak: Modelling the hard
  x-ray and γ-ray backgrounds}.
\newblock \bibinfo{journal}{Experimental Astronomy} \bibinfo{volume}{20},
  \bibinfo{pages}{41--47}.
\bibitem[{{Comastri} et~al.(2002){Comastri}, {Mignoli}, {Ciliegi},
  {Severgnini}, {Maiolino}, {Brusa}, {Fiore}, {Baldi}, {Molendi}, {Morganti},
  {Vignali}, {La Franca}, {Matt} and {Perola}}]{2002ApJ...571..771C}
\bibinfo{author}{{Comastri}, A.}, \bibinfo{author}{{Mignoli}, M.},
  \bibinfo{author}{{Ciliegi}, P.}, \bibinfo{author}{{Severgnini}, P.},
  \bibinfo{author}{{Maiolino}, R.}, \bibinfo{author}{{Brusa}, M.},
  \bibinfo{author}{{Fiore}, F.}, \bibinfo{author}{{Baldi}, A.},
  \bibinfo{author}{{Molendi}, S.}, \bibinfo{author}{{Morganti}, R.},
  \bibinfo{author}{{Vignali}, C.}, \bibinfo{author}{{La Franca}, F.},
  \bibinfo{author}{{Matt}, G.}, \bibinfo{author}{{Perola}, G.C.},
  \bibinfo{year}{2002}.
\newblock \bibinfo{title}{{The HELLAS2XMM Survey. II. Multiwavelength
  Observations of P3: An X-Ray-bright, Optically Inactive Galaxy}}.
\newblock \bibinfo{journal}{\apj} \bibinfo{volume}{571},
  \bibinfo{pages}{771--778}.
\newblock \href{http://arxiv.org/abs/astro-ph/0202080}{\tt
  arXiv:astro-ph/0202080}.
\bibitem[{{Costamante} et~al.(2018){Costamante}, {Cutini}, {Tosti}, {Antolini}
  and {Tramacere}}]{2018MNRAS.477.4749C}
\bibinfo{author}{{Costamante}, L.}, \bibinfo{author}{{Cutini}, S.},
  \bibinfo{author}{{Tosti}, G.}, \bibinfo{author}{{Antolini}, E.},
  \bibinfo{author}{{Tramacere}, A.}, \bibinfo{year}{2018}.
\newblock \bibinfo{title}{{On the origin of gamma-rays in Fermi blazars:
  beyondthe broad-line region}}.
\newblock \bibinfo{journal}{\mnras} \bibinfo{volume}{477},
  \bibinfo{pages}{4749--4767}.
\newblock \href{http://arxiv.org/abs/1804.02408}{\tt arXiv:1804.02408}.
\bibitem[{{Dadina}(2007)}]{2007A&A...461.1209D}
\bibinfo{author}{{Dadina}, M.}, \bibinfo{year}{2007}.
\newblock \bibinfo{title}{{BeppoSAX observations in the 2-100 keV band of the
  nearby Seyfert galaxies: an atlas of spectra}}.
\newblock \bibinfo{journal}{\aap} \bibinfo{volume}{461},
  \bibinfo{pages}{1209--1252}.
\bibitem[{{de Rosa} et~al.(2008){de Rosa}, {Bassani}, {Ubertini}, {Malizia} and
  {Dean}}]{2008MNRAS.388L..54D}
\bibinfo{author}{{de Rosa}, A.}, \bibinfo{author}{{Bassani}, L.},
  \bibinfo{author}{{Ubertini}, P.}, \bibinfo{author}{{Malizia}, A.},
  \bibinfo{author}{{Dean}, A.J.}, \bibinfo{year}{2008}.
\newblock \bibinfo{title}{{Bulk Compton motion in the luminous quasar
  4C04.42?}}
\newblock \bibinfo{journal}{\mnras} \bibinfo{volume}{388},
  \bibinfo{pages}{L54--L58}.
\newblock \DOIprefix\doi{10.1111/j.1745-3933.2008.00498.x}.
\bibitem[{{de Rosa} et~al.(2012){de Rosa}, {Panessa}, {Bassani}, {Bazzano},
  {Bird}, {Landi}, {Malizia}, {Molina} and {Ubertini}}]{2012MNRAS.420.2087D}
\bibinfo{author}{{de Rosa}, A.}, \bibinfo{author}{{Panessa}, F.},
  \bibinfo{author}{{Bassani}, L.}, \bibinfo{author}{{Bazzano}, A.},
  \bibinfo{author}{{Bird}, A.}, \bibinfo{author}{{Landi}, R.},
  \bibinfo{author}{{Malizia}, A.}, \bibinfo{author}{{Molina}, M.},
  \bibinfo{author}{{Ubertini}, P.}, \bibinfo{year}{2012}.
\newblock \bibinfo{title}{{Broad-band study of hard X-ray-selected absorbed
  active galactic nuclei}}.
\newblock \bibinfo{journal}{\mnras} \bibinfo{volume}{420},
  \bibinfo{pages}{2087--2101}.
\newblock \href{http://arxiv.org/abs/1111.1946}{\tt arXiv:1111.1946}.
\bibitem[{{Fabian} et~al.(2017){Fabian}, {Lohfink}, {Belmont}, {Malzac} and
  {Coppi}}]{2017MNRAS.467.2566F}
\bibinfo{author}{{Fabian}, A.C.}, \bibinfo{author}{{Lohfink}, A.},
  \bibinfo{author}{{Belmont}, R.}, \bibinfo{author}{{Malzac}, J.},
  \bibinfo{author}{{Coppi}, P.}, \bibinfo{year}{2017}.
\newblock \bibinfo{title}{{Properties of AGN coronae in the NuSTAR era - II.
  Hybrid plasma}}.
\newblock \bibinfo{journal}{\mnras} \bibinfo{volume}{467},
  \bibinfo{pages}{2566--2570}.
\newblock \DOIprefix\doi{10.1093/mnras/stx221},
  \href{http://arxiv.org/abs/1701.06774}{\tt arXiv:1701.06774}.
\bibitem[{{Fabian} et~al.(2015){Fabian}, {Lohfink}, {Kara}, {Parker},
  {Vasudevan} and {Reynolds}}]{2015MNRAS.451.4375F}
\bibinfo{author}{{Fabian}, A.C.}, \bibinfo{author}{{Lohfink}, A.},
  \bibinfo{author}{{Kara}, E.}, \bibinfo{author}{{Parker}, M.L.},
  \bibinfo{author}{{Vasudevan}, R.}, \bibinfo{author}{{Reynolds}, C.S.},
  \bibinfo{year}{2015}.
\newblock \bibinfo{title}{{Properties of AGN coronae in the NuSTAR era}}.
\newblock \bibinfo{journal}{\mnras} \bibinfo{volume}{451},
  \bibinfo{pages}{4375--4383}.
\newblock \DOIprefix\doi{10.1093/mnras/stv1218},
  \href{http://arxiv.org/abs/1505.07603}{\tt arXiv:1505.07603}.
\bibitem[{{Falomo} et~al.(2014){Falomo}, {Pian} and
  {Treves}}]{2014A&ARv..22...73F}
\bibinfo{author}{{Falomo}, R.}, \bibinfo{author}{{Pian}, E.},
  \bibinfo{author}{{Treves}, A.}, \bibinfo{year}{2014}.
\newblock \bibinfo{title}{{An optical view of BL Lacertae objects}}.
\newblock \bibinfo{journal}{\aapr} \bibinfo{volume}{22}, \bibinfo{pages}{73}.
\newblock \href{http://arxiv.org/abs/1407.7615}{\tt arXiv:1407.7615}.
\bibitem[{{Fan} et~al.(2016){Fan}, {Yang}, {Liu}, {Luo}, {Lin}, {Yuan}, {Xiao},
  {Zhou}, {Hua} and {Pei}}]{2016ApJS..226...20F}
\bibinfo{author}{{Fan}, J.H.}, \bibinfo{author}{{Yang}, J.H.},
  \bibinfo{author}{{Liu}, Y.}, \bibinfo{author}{{Luo}, G.Y.},
  \bibinfo{author}{{Lin}, C.}, \bibinfo{author}{{Yuan}, Y.H.},
  \bibinfo{author}{{Xiao}, H.B.}, \bibinfo{author}{{Zhou}, A.Y.},
  \bibinfo{author}{{Hua}, T.X.}, \bibinfo{author}{{Pei}, Z.Y.},
  \bibinfo{year}{2016}.
\newblock \bibinfo{title}{{The Spectral Energy Distributions of Fermi
  Blazars}}.
\newblock \bibinfo{journal}{\apjs} \bibinfo{volume}{226}, \bibinfo{pages}{20}.
\newblock \href{http://arxiv.org/abs/1608.03958}{\tt arXiv:1608.03958}.
\bibitem[{{Fanaroff} and {Riley}(1974)}]{1974MNRAS.167P..31F}
\bibinfo{author}{{Fanaroff}, B.L.}, \bibinfo{author}{{Riley}, J.M.},
  \bibinfo{year}{1974}.
\newblock \bibinfo{title}{{The morphology of extragalactic radio sources of
  high and low luminosity}}.
\newblock \bibinfo{journal}{\mnras} \bibinfo{volume}{167},
  \bibinfo{pages}{31P--36P}.
\bibitem[{{Fedorova} et~al.(2011){Fedorova}, {Beckmann}, {Neronov} and
  {Soldi}}]{2011MNRAS.417.1140F}
\bibinfo{author}{{Fedorova}, E.V.}, \bibinfo{author}{{Beckmann}, V.},
  \bibinfo{author}{{Neronov}, A.}, \bibinfo{author}{{Soldi}, S.},
  \bibinfo{year}{2011}.
\newblock \bibinfo{title}{{Studying the long-time variability of the Seyfert 2
  galaxy NGC 4388 with INTEGRAL and Swift}}.
\newblock \bibinfo{journal}{\mnras} \bibinfo{volume}{417},
  \bibinfo{pages}{1140--1147}.
\bibitem[{{Fedorova} and {Zhdanov}(2016)}]{2016KPCB...32..172F}
\bibinfo{author}{{Fedorova}, E.V.}, \bibinfo{author}{{Zhdanov}, V.I.},
  \bibinfo{year}{2016}.
\newblock \bibinfo{title}{{Variations of the X-ray INTEGRAL spectrum of the
  active galactic nucleus of NGC 4945}}.
\newblock \bibinfo{journal}{Kinematics and Physics of Celestial Bodies}
  \bibinfo{volume}{32}, \bibinfo{pages}{172--180}.
\bibitem[{{Foffano} et~al.(2019){Foffano}, {Prandini}, {Franceschini} and
  {Paiano}}]{2019MNRAS.486.1741F}
\bibinfo{author}{{Foffano}, L.}, \bibinfo{author}{{Prandini}, E.},
  \bibinfo{author}{{Franceschini}, A.}, \bibinfo{author}{{Paiano}, S.},
  \bibinfo{year}{2019}.
\newblock \bibinfo{title}{{A new hard X-ray-selected sample of extreme
  high-energy peaked BL Lac objects and their TeV gamma-ray properties}}.
\newblock \bibinfo{journal}{\mnras} \bibinfo{volume}{486},
  \bibinfo{pages}{1741--1762}.
\newblock \DOIprefix\doi{10.1093/mnras/stz812},
  \href{http://arxiv.org/abs/1903.07972}{\tt arXiv:1903.07972}.
\bibitem[{{Fortin} et~al.(2018){Fortin}, {Chaty}, {Coleiro}, {Tomsick} and
  {Nitschelm}}]{2018A&A...618A.150F}
\bibinfo{author}{{Fortin}, F.}, \bibinfo{author}{{Chaty}, S.},
  \bibinfo{author}{{Coleiro}, A.}, \bibinfo{author}{{Tomsick}, J.A.},
  \bibinfo{author}{{Nitschelm}, C.H.R.}, \bibinfo{year}{2018}.
\newblock \bibinfo{title}{{Spectroscopic identification of INTEGRAL high-energy
  sources with VLT/ISAAC}}.
\newblock \bibinfo{journal}{\aap} \bibinfo{volume}{618}, \bibinfo{pages}{A150}.
\newblock \DOIprefix\doi{10.1051/0004-6361/201731265},
  \href{http://arxiv.org/abs/1808.09816}{\tt arXiv:1808.09816}.
\bibitem[{{Fossati} et~al.(1998){Fossati}, {Maraschi}, {Celotti}, {Comastri}
  and {Ghisellini}}]{1998MNRAS.299..433F}
\bibinfo{author}{{Fossati}, G.}, \bibinfo{author}{{Maraschi}, L.},
  \bibinfo{author}{{Celotti}, A.}, \bibinfo{author}{{Comastri}, A.},
  \bibinfo{author}{{Ghisellini}, G.}, \bibinfo{year}{1998}.
\newblock \bibinfo{title}{{A unifying view of the spectral energy distributions
  of blazars}}.
\newblock \bibinfo{journal}{\mnras} \bibinfo{volume}{299},
  \bibinfo{pages}{433--448}.
\newblock \href{http://arxiv.org/abs/astro-ph/9804103}{\tt
  arXiv:astro-ph/9804103}.
\bibitem[{Frontera et~al.(1997)Frontera, Costa, Dal~Fiume, Feroci, Nicastro,
  Orlandini, Palazzi and Zavattini}]{Frontera_1997}
\bibinfo{author}{Frontera, F.}, \bibinfo{author}{Costa, E.},
  \bibinfo{author}{Dal~Fiume, D.}, \bibinfo{author}{Feroci, M.},
  \bibinfo{author}{Nicastro, L.}, \bibinfo{author}{Orlandini, M.},
  \bibinfo{author}{Palazzi, E.}, \bibinfo{author}{Zavattini, G.},
  \bibinfo{year}{1997}.
\newblock \bibinfo{title}{The high energy instrument pds on-board the bepposax
  x--ray astronomy satellite}.
\newblock \bibinfo{journal}{Astronomy and Astrophysics Supplement Series}
  \bibinfo{volume}{122}, \bibinfo{pages}{357--369}.
\bibitem[{{Furniss} et~al.(2015){Furniss}, {Noda}, {Boggs}, {Chiang},
  {Christensen}, {Craig}, {Giommi}, {Hailey}, {Harisson}, {Madejski} and
  et~al.}]{2015ApJ...812...65F}
\bibinfo{author}{{Furniss}, A.}, \bibinfo{author}{{Noda}, K.},
  \bibinfo{author}{{Boggs}, S.}, \bibinfo{author}{{Chiang}, J.},
  \bibinfo{author}{{Christensen}, F.}, \bibinfo{author}{{Craig}, W.},
  \bibinfo{author}{{Giommi}, P.}, \bibinfo{author}{{Hailey}, C.},
  \bibinfo{author}{{Harisson}, F.}, \bibinfo{author}{{Madejski}, G.},
  \bibinfo{author}{et~al.}, \bibinfo{year}{2015}.
\newblock \bibinfo{title}{{First NuSTAR Observations of Mrk 501 within a Radio
  to TeV Multi-Instrument Campaign}}.
\newblock \bibinfo{journal}{\apj} \bibinfo{volume}{812}, \bibinfo{pages}{65}.
\newblock \href{http://arxiv.org/abs/1509.04936}{\tt arXiv:1509.04936}.
\bibitem[{{Gao} et~al.(2019){Gao}, {Fedynitch}, {Winter} and
  {Pohl}}]{2019NatAs...3...88G}
\bibinfo{author}{{Gao}, S.}, \bibinfo{author}{{Fedynitch}, A.},
  \bibinfo{author}{{Winter}, W.}, \bibinfo{author}{{Pohl}, M.},
  \bibinfo{year}{2019}.
\newblock \bibinfo{title}{{Modelling the coincident observation of a
  high-energy neutrino and a bright blazar flare}}.
\newblock \bibinfo{journal}{Nature Astronomy} \bibinfo{volume}{3},
  \bibinfo{pages}{88--92}.
\newblock \href{http://arxiv.org/abs/1807.04275}{\tt arXiv:1807.04275}.
\bibitem[{{Ghisellini} et~al.(1998){Ghisellini}, {Celotti}, {Fossati},
  {Maraschi} and {Comastri}}]{1998MNRAS.301..451G}
\bibinfo{author}{{Ghisellini}, G.}, \bibinfo{author}{{Celotti}, A.},
  \bibinfo{author}{{Fossati}, G.}, \bibinfo{author}{{Maraschi}, L.},
  \bibinfo{author}{{Comastri}, A.}, \bibinfo{year}{1998}.
\newblock \bibinfo{title}{{A theoretical unifying scheme for gamma-ray bright
  blazars}}.
\newblock \bibinfo{journal}{\mnras} \bibinfo{volume}{301},
  \bibinfo{pages}{451--468}.
\newblock \href{http://arxiv.org/abs/astro-ph/9807317}{\tt
  arXiv:astro-ph/9807317}.
\bibitem[{{Ghisellini} et~al.(2007){Ghisellini}, {Foschini}, {Tavecchio} and
  {Pian}}]{2007MNRAS.382L..82G}
\bibinfo{author}{{Ghisellini}, G.}, \bibinfo{author}{{Foschini}, L.},
  \bibinfo{author}{{Tavecchio}, F.}, \bibinfo{author}{{Pian}, E.},
  \bibinfo{year}{2007}.
\newblock \bibinfo{title}{{On the 2007 July flare of the blazar 3C 454.3}}.
\newblock \bibinfo{journal}{\mnras} \bibinfo{volume}{382},
  \bibinfo{pages}{L82--L86}.
\newblock \href{http://arxiv.org/abs/0708.0617}{\tt arXiv:0708.0617}.
\bibitem[{{Giann{\'\i}} et~al.(2011){Giann{\'\i}}, {de Rosa}, {Bassani},
  {Bazzano}, {Dean} and {Ubertini}}]{2011MNRAS.411.2137G}
\bibinfo{author}{{Giann{\'\i}}, S.}, \bibinfo{author}{{de Rosa}, A.},
  \bibinfo{author}{{Bassani}, L.}, \bibinfo{author}{{Bazzano}, A.},
  \bibinfo{author}{{Dean}, T.}, \bibinfo{author}{{Ubertini}, P.},
  \bibinfo{year}{2011}.
\newblock \bibinfo{title}{{An X-ray view of the INTEGRAL/IBIS blazars}}.
\newblock \bibinfo{journal}{\mnras} \bibinfo{volume}{411},
  \bibinfo{pages}{2137--2147}.
\newblock \href{http://arxiv.org/abs/1010.5713}{\tt arXiv:1010.5713}.
\bibitem[{{Gilli} et~al.(2007){Gilli}, {Comastri} and
  {Hasinger}}]{2007A&A...463...79G}
\bibinfo{author}{{Gilli}, R.}, \bibinfo{author}{{Comastri}, A.},
  \bibinfo{author}{{Hasinger}, G.}, \bibinfo{year}{2007}.
\newblock \bibinfo{title}{{The synthesis of the cosmic X-ray background in the
  Chandra and XMM-Newton era}}.
\newblock \bibinfo{journal}{\aap} \bibinfo{volume}{463},
  \bibinfo{pages}{79--96}.
\newblock \DOIprefix\doi{10.1051/0004-6361:20066334},
  \href{http://arxiv.org/abs/astro-ph/0610939}{\tt arXiv:astro-ph/0610939}.
\bibitem[{{Giovannini} et~al.(2007){Giovannini}, {Giroletti} and
  {Taylor}}]{2007A&A...474..409G}
\bibinfo{author}{{Giovannini}, G.}, \bibinfo{author}{{Giroletti}, M.},
  \bibinfo{author}{{Taylor}, G.B.}, \bibinfo{year}{2007}.
\newblock \bibinfo{title}{{B2 1144+35B, a giant low power radio galaxy with
  superluminal motion. Orientation and evidence for recurrent activity}}.
\newblock \bibinfo{journal}{\aap} \bibinfo{volume}{474},
  \bibinfo{pages}{409--414}.
\newblock \href{http://arxiv.org/abs/0708.3902}{\tt arXiv:0708.3902}.
\bibitem[{Grebenev et~al.(2012)Grebenev, Lutovinov, Tsygankov and
  Mereminskiy}]{Grebenev_2012}
\bibinfo{author}{Grebenev, S.A.}, \bibinfo{author}{Lutovinov, A.A.},
  \bibinfo{author}{Tsygankov, S.S.}, \bibinfo{author}{Mereminskiy, I.A.},
  \bibinfo{year}{2012}.
\newblock \bibinfo{title}{Deep hard x-ray survey of the large magellanic
  cloud}.
\newblock \bibinfo{journal}{Monthly Notices of the Royal Astronomical Society}
  \bibinfo{volume}{428}, \bibinfo{pages}{50--57}.
\bibitem[{{H.~E.~S.~S. Collaboration} et~al.(2019){H.~E.~S.~S. Collaboration},
  {Abdalla}, {Adam}, {Aharonian}, {Ait Benkhali}, {Ang{\"u}ner}, {Arakawa},
  {Arcaro}, {Armand}, {Ashkar}, {Backes}, {Barbosa Martins}, {Barnard},
  {Becherini}, {Berge}, {Bernl{\"o}hr}, {Blackwell}, {B{\"o}ttcher}, {Boisson},
  {Bolmont}, {Bonnefoy}, {Bregeon}, {Breuhaus}, {Brun}, {Brun}, {Bryan},
  {B{\"u}chele}, {Bulik}, {Bylund}, {Capasso}, {Caroff}, {Carosi}, {Casanova},
  {Cerruti}, {Chand}, {Chandra}, {Chen}, {Colafrancesco}, {Cury{\l}o},
  {Davids}, {Deil}, {Devin}, {deWilt}, {Dirson}, {Djannati-Ata{\"\i}},
  {Dmytriiev}, {Donath}, {Doroshenko}, {Drury}, {Dyks}, {Egberts}, {Emery},
  {Ernenwein}, {Eschbach}, {Feijen}, {Fegan}, {Fiasson}, {Fontaine}, {Funk},
  {F{\"u}{\ss}ling}, {Gabici}, {Gallant}, {Gat{\'e}}, {Giavitto}, {Glawion},
  {Glicenstein}, {Gottschall}, {Grondin}, {Hahn}, {Haupt}, {Heinzelmann},
  {Henri}, {Hermann}, {Hinton}, {Hofmann}, {Hoischen}, {Holch}, {Holler},
  {Horns}, {Huber}, {Iwasaki}, {Jamrozy}, {Jankowsky}, {Jankowsky},
  {Jardin-Blicq}, {Jung-Richardt}, {Kastendieck}, {Katarzy{\'n}ski},
  {Katsuragawa}, {Katz}, {Khangulyan}, {Kh{\'e}lifi}, {King}, {Klepser},
  {Klu{\'z}niak}, {Komin}, {Kosack}, {Kostunin}, {Kraus}, {Lamanna}, {Lau},
  {Lemi{\`e}re}, {Lemoine-Goumard}, {Lenain}, {Leser}, {Levy}, {Lohse},
  {Lypova}, {Mackey}, {Majumdar}, {Malyshev}, {Marandon}, {Marcowith}, {Mares},
  {Mariaud}, {Mart{\'\i}-Devesa}, {Marx}, {Maurin}, {Meintjes}, {Mitchell},
  {Moderski}, {Mohamed}, {Mohrmann}, {Moore}, {Moulin}, {Muller}, {Murach},
  {Nakashima}, {de Naurois}, {Ndiyavala}, {Niederwanger}, {Niemiec}, {Oakes},
  {O'Brien}, {Odaka}, {Ohm}, {de Ona Wilhelmi}, {Ostrowski}, {Oya}, {Panter},
  {Parsons}, {Perennes}, {Petrucci}, {Peyaud}, {Piel}, {Pita}, {Poireau},
  {Priyana Noel}, {Prokhorov}, {Prokoph}, {P{\"u}hlhofer}, {Punch},
  {Quirrenbach}, {Raab}, {Rauth}, {Reimer}, {Reimer}, {Remy}, {Renaud},
  {Rieger}, {Rinchiuso}, {Romoli}, {Rowell}, {Rudak}, {Ruiz-Velasco},
  {Sahakian}, {Saito}, {Sanchez}, {Santangelo}, {Sasaki}, {Schlickeiser},
  {Sch{\"u}ssler}, {Schulz}, {Schutte}, {Schwanke}, {Schwemmer},
  {Seglar-Arroyo}, {Senniappan}, {Seyffert}, {Shafi}, {Shiningayamwe},
  {Simoni}, {Sinha}, {Sol}, {Specovius}, {Spir-Jacob}, {Stawarz}, {Steenkamp},
  {Stegmann}, {Steppa}, {Takahashi}, {Tavernier}, {Taylor}, {Terrier},
  {Tiziani}, {Tluczykont}, {Trichard}, {Tsirou}, {Tsuji}, {Tuffs}, {Uchiyama},
  {van der Walt}, {van Eldik}, {van Rensburg}, {van Soelen}, {Vasileiadis},
  {Veh}, {Venter}, {Vincent}, {Vink}, {Voisin}, {V{\"o}lk}, {Vuillaume},
  {Wadiasingh}, {Wagner}, {White}, {Wierzcholska}, {Yang}, {Yoneda},
  {Zacharias}, {Zanin}, {Zdziarski}, {Zech}, {Ziegler}, {Zorn}, {{\.Z}ywucka}
  and {Meyer}}]{2019A&A...627A.159H}
\bibinfo{author}{{H.~E.~S.~S. Collaboration}}, \bibinfo{author}{{Abdalla}, H.},
  \bibinfo{author}{{Adam}, R.}, \bibinfo{author}{{Aharonian}, F.},
  \bibinfo{author}{{Ait Benkhali}, F.}, \bibinfo{author}{{Ang{\"u}ner}, E.O.},
  \bibinfo{author}{{Arakawa}, M.}, \bibinfo{author}{{Arcaro}, C.},
  \bibinfo{author}{{Armand}, C.}, \bibinfo{author}{{Ashkar}, H.},
  \bibinfo{author}{{Backes}, M.}, \bibinfo{author}{{Barbosa Martins}, V.},
  \bibinfo{author}{{Barnard}, M.}, \bibinfo{author}{{Becherini}, Y.},
  \bibinfo{author}{{Berge}, D.}, \bibinfo{author}{{Bernl{\"o}hr}, K.},
  \bibinfo{author}{{Blackwell}, R.}, \bibinfo{author}{{B{\"o}ttcher}, M.},
  \bibinfo{author}{{Boisson}, C.}, \bibinfo{author}{{Bolmont}, J.},
  \bibinfo{author}{{Bonnefoy}, S.}, \bibinfo{author}{{Bregeon}, J.},
  \bibinfo{author}{{Breuhaus}, M.}, \bibinfo{author}{{Brun}, F.},
  \bibinfo{author}{{Brun}, P.}, \bibinfo{author}{{Bryan}, M.},
  \bibinfo{author}{{B{\"u}chele}, M.}, \bibinfo{author}{{Bulik}, T.},
  \bibinfo{author}{{Bylund}, T.}, \bibinfo{author}{{Capasso}, M.},
  \bibinfo{author}{{Caroff}, S.}, \bibinfo{author}{{Carosi}, A.},
  \bibinfo{author}{{Casanova}, S.}, \bibinfo{author}{{Cerruti}, M.},
  \bibinfo{author}{{Chand}, T.}, \bibinfo{author}{{Chandra}, S.},
  \bibinfo{author}{{Chen}, A.}, \bibinfo{author}{{Colafrancesco}, S.},
  \bibinfo{author}{{Cury{\l}o}, M.}, \bibinfo{author}{{Davids}, I.D.},
  \bibinfo{author}{{Deil}, C.}, \bibinfo{author}{{Devin}, J.},
  \bibinfo{author}{{deWilt}, P.}, \bibinfo{author}{{Dirson}, L.},
  \bibinfo{author}{{Djannati-Ata{\"\i}}, A.}, \bibinfo{author}{{Dmytriiev},
  A.}, \bibinfo{author}{{Donath}, A.}, \bibinfo{author}{{Doroshenko}, V.},
  \bibinfo{author}{{Drury}, L.O.C.}, \bibinfo{author}{{Dyks}, J.},
  \bibinfo{author}{{Egberts}, K.}, \bibinfo{author}{{Emery}, G.},
  \bibinfo{author}{{Ernenwein}, J.P.}, \bibinfo{author}{{Eschbach}, S.},
  \bibinfo{author}{{Feijen}, K.}, \bibinfo{author}{{Fegan}, S.},
  \bibinfo{author}{{Fiasson}, A.}, \bibinfo{author}{{Fontaine}, G.},
  \bibinfo{author}{{Funk}, S.}, \bibinfo{author}{{F{\"u}{\ss}ling}, M.},
  \bibinfo{author}{{Gabici}, S.}, \bibinfo{author}{{Gallant}, Y.A.},
  \bibinfo{author}{{Gat{\'e}}, F.}, \bibinfo{author}{{Giavitto}, G.},
  \bibinfo{author}{{Glawion}, D.}, \bibinfo{author}{{Glicenstein}, J.F.},
  \bibinfo{author}{{Gottschall}, D.}, \bibinfo{author}{{Grondin}, M.H.},
  \bibinfo{author}{{Hahn}, J.}, \bibinfo{author}{{Haupt}, M.},
  \bibinfo{author}{{Heinzelmann}, G.}, \bibinfo{author}{{Henri}, G.},
  \bibinfo{author}{{Hermann}, G.}, \bibinfo{author}{{Hinton}, J.A.},
  \bibinfo{author}{{Hofmann}, W.}, \bibinfo{author}{{Hoischen}, C.},
  \bibinfo{author}{{Holch}, T.L.}, \bibinfo{author}{{Holler}, M.},
  \bibinfo{author}{{Horns}, D.}, \bibinfo{author}{{Huber}, D.},
  \bibinfo{author}{{Iwasaki}, H.}, \bibinfo{author}{{Jamrozy}, M.},
  \bibinfo{author}{{Jankowsky}, D.}, \bibinfo{author}{{Jankowsky}, F.},
  \bibinfo{author}{{Jardin-Blicq}, A.}, \bibinfo{author}{{Jung-Richardt}, I.},
  \bibinfo{author}{{Kastendieck}, M.A.}, \bibinfo{author}{{Katarzy{\'n}ski},
  K.}, \bibinfo{author}{{Katsuragawa}, M.}, \bibinfo{author}{{Katz}, U.},
  \bibinfo{author}{{Khangulyan}, D.}, \bibinfo{author}{{Kh{\'e}lifi}, B.},
  \bibinfo{author}{{King}, J.}, \bibinfo{author}{{Klepser}, S.},
  \bibinfo{author}{{Klu{\'z}niak}, W.}, \bibinfo{author}{{Komin}, N.},
  \bibinfo{author}{{Kosack}, K.}, \bibinfo{author}{{Kostunin}, D.},
  \bibinfo{author}{{Kraus}, M.}, \bibinfo{author}{{Lamanna}, G.},
  \bibinfo{author}{{Lau}, J.}, \bibinfo{author}{{Lemi{\`e}re}, A.},
  \bibinfo{author}{{Lemoine-Goumard}, M.}, \bibinfo{author}{{Lenain}, J.P.},
  \bibinfo{author}{{Leser}, E.}, \bibinfo{author}{{Levy}, C.},
  \bibinfo{author}{{Lohse}, T.}, \bibinfo{author}{{Lypova}, I.},
  \bibinfo{author}{{Mackey}, J.}, \bibinfo{author}{{Majumdar}, J.},
  \bibinfo{author}{{Malyshev}, D.}, \bibinfo{author}{{Marandon}, V.},
  \bibinfo{author}{{Marcowith}, A.}, \bibinfo{author}{{Mares}, A.},
  \bibinfo{author}{{Mariaud}, C.}, \bibinfo{author}{{Mart{\'\i}-Devesa}, G.},
  \bibinfo{author}{{Marx}, R.}, \bibinfo{author}{{Maurin}, G.},
  \bibinfo{author}{{Meintjes}, P.J.}, \bibinfo{author}{{Mitchell}, A.M.W.},
  \bibinfo{author}{{Moderski}, R.}, \bibinfo{author}{{Mohamed}, M.},
  \bibinfo{author}{{Mohrmann}, L.}, \bibinfo{author}{{Moore}, C.},
  \bibinfo{author}{{Moulin}, E.}, \bibinfo{author}{{Muller}, J.},
  \bibinfo{author}{{Murach}, T.}, \bibinfo{author}{{Nakashima}, S.},
  \bibinfo{author}{{de Naurois}, M.}, \bibinfo{author}{{Ndiyavala}, H.},
  \bibinfo{author}{{Niederwanger}, F.}, \bibinfo{author}{{Niemiec}, J.},
  \bibinfo{author}{{Oakes}, L.}, \bibinfo{author}{{O'Brien}, P.},
  \bibinfo{author}{{Odaka}, H.}, \bibinfo{author}{{Ohm}, S.},
  \bibinfo{author}{{de Ona Wilhelmi}, E.}, \bibinfo{author}{{Ostrowski}, M.},
  \bibinfo{author}{{Oya}, I.}, \bibinfo{author}{{Panter}, M.},
  \bibinfo{author}{{Parsons}, R.D.}, \bibinfo{author}{{Perennes}, C.},
  \bibinfo{author}{{Petrucci}, P.O.}, \bibinfo{author}{{Peyaud}, B.},
  \bibinfo{author}{{Piel}, Q.}, \bibinfo{author}{{Pita}, S.},
  \bibinfo{author}{{Poireau}, V.}, \bibinfo{author}{{Priyana Noel}, A.},
  \bibinfo{author}{{Prokhorov}, D.A.}, \bibinfo{author}{{Prokoph}, H.},
  \bibinfo{author}{{P{\"u}hlhofer}, G.}, \bibinfo{author}{{Punch}, M.},
  \bibinfo{author}{{Quirrenbach}, A.}, \bibinfo{author}{{Raab}, S.},
  \bibinfo{author}{{Rauth}, R.}, \bibinfo{author}{{Reimer}, A.},
  \bibinfo{author}{{Reimer}, O.}, \bibinfo{author}{{Remy}, Q.},
  \bibinfo{author}{{Renaud}, M.}, \bibinfo{author}{{Rieger}, F.},
  \bibinfo{author}{{Rinchiuso}, L.}, \bibinfo{author}{{Romoli}, C.},
  \bibinfo{author}{{Rowell}, G.}, \bibinfo{author}{{Rudak}, B.},
  \bibinfo{author}{{Ruiz-Velasco}, E.}, \bibinfo{author}{{Sahakian}, V.},
  \bibinfo{author}{{Saito}, S.}, \bibinfo{author}{{Sanchez}, D.A.},
  \bibinfo{author}{{Santangelo}, A.}, \bibinfo{author}{{Sasaki}, M.},
  \bibinfo{author}{{Schlickeiser}, R.}, \bibinfo{author}{{Sch{\"u}ssler}, F.},
  \bibinfo{author}{{Schulz}, A.}, \bibinfo{author}{{Schutte}, H.},
  \bibinfo{author}{{Schwanke}, U.}, \bibinfo{author}{{Schwemmer}, S.},
  \bibinfo{author}{{Seglar-Arroyo}, M.}, \bibinfo{author}{{Senniappan}, M.},
  \bibinfo{author}{{Seyffert}, A.S.}, \bibinfo{author}{{Shafi}, N.},
  \bibinfo{author}{{Shiningayamwe}, K.}, \bibinfo{author}{{Simoni}, R.},
  \bibinfo{author}{{Sinha}, A.}, \bibinfo{author}{{Sol}, H.},
  \bibinfo{author}{{Specovius}, A.}, \bibinfo{author}{{Spir-Jacob}, M.},
  \bibinfo{author}{{Stawarz}, L.}, \bibinfo{author}{{Steenkamp}, R.},
  \bibinfo{author}{{Stegmann}, C.}, \bibinfo{author}{{Steppa}, C.},
  \bibinfo{author}{{Takahashi}, T.}, \bibinfo{author}{{Tavernier}, T.},
  \bibinfo{author}{{Taylor}, A.M.}, \bibinfo{author}{{Terrier}, R.},
  \bibinfo{author}{{Tiziani}, D.}, \bibinfo{author}{{Tluczykont}, M.},
  \bibinfo{author}{{Trichard}, C.}, \bibinfo{author}{{Tsirou}, M.},
  \bibinfo{author}{{Tsuji}, N.}, \bibinfo{author}{{Tuffs}, R.},
  \bibinfo{author}{{Uchiyama}, Y.}, \bibinfo{author}{{van der Walt}, D.J.},
  \bibinfo{author}{{van Eldik}, C.}, \bibinfo{author}{{van Rensburg}, C.},
  \bibinfo{author}{{van Soelen}, B.}, \bibinfo{author}{{Vasileiadis}, G.},
  \bibinfo{author}{{Veh}, J.}, \bibinfo{author}{{Venter}, C.},
  \bibinfo{author}{{Vincent}, P.}, \bibinfo{author}{{Vink}, J.},
  \bibinfo{author}{{Voisin}, F.}, \bibinfo{author}{{V{\"o}lk}, H.J.},
  \bibinfo{author}{{Vuillaume}, T.}, \bibinfo{author}{{Wadiasingh}, Z.},
  \bibinfo{author}{{Wagner}, S.J.}, \bibinfo{author}{{White}, R.},
  \bibinfo{author}{{Wierzcholska}, A.}, \bibinfo{author}{{Yang}, R.},
  \bibinfo{author}{{Yoneda}, H.}, \bibinfo{author}{{Zacharias}, M.},
  \bibinfo{author}{{Zanin}, R.}, \bibinfo{author}{{Zdziarski}, A.A.},
  \bibinfo{author}{{Zech}, A.}, \bibinfo{author}{{Ziegler}, A.},
  \bibinfo{author}{{Zorn}, J.}, \bibinfo{author}{{{\.Z}ywucka}, N.},
  \bibinfo{author}{{Meyer}, M.}, \bibinfo{year}{2019}.
\newblock \bibinfo{title}{{Constraints on the emission region of <ASTROBJ>3C
  279</ASTROBJ> during strong flares in 2014 and 2015 through VHE
  {\ensuremath{\gamma}}-ray observations with H.E.S.S.}}
\newblock \bibinfo{journal}{\aap} \bibinfo{volume}{627}, \bibinfo{pages}{A159}.
\newblock \DOIprefix\doi{10.1051/0004-6361/201935704},
  \href{http://arxiv.org/abs/1906.04996}{\tt arXiv:1906.04996}.
\bibitem[{{H.~E.~S.~S. Collaboration} et~al.(2011){H.~E.~S.~S. Collaboration},
  {Abramowski}, {Acero}, {Aharonian}, {Akhperjanian}, {Anton}, {Balzer},
  {Barnacka}, {Barres de Almeida}, {Bazer-Bachi}, {Becherini}, {Becker},
  {Behera}, {Bernl{\"o}hr}, {Bochow}, {Boisson}, {Bolmont}, {Bordas}, {Borrel},
  {Brucker}, {Brun}, {Brun}, {Bulik}, {B{\"u}sching}, {Carrigan}, {Casanova},
  {Cerruti}, {Chadwick}, {Charbonnier}, {Chaves}, {Cheesebrough}, {Chounet},
  {Clapson}, {Coignet}, {Colom}, {Conrad}, {Dalton}, {Daniel}, {Davids},
  {Degrange}, {Deil}, {Dickinson}, {Djannati-Ata{\"\i}}, {Domainko}, {Drury},
  {Dubois}, {Dubus}, {Dyks}, {Dyrda}, {Egberts}, {Eger}, {Espigat}, {Fallon},
  {Farnier}, {Fegan}, {Feinstein}, {Fernandes}, {Fiasson}, {Fontaine},
  {F{\"o}rster}, {F{\"u}{\ss}ling}, {Gallant}, {Gast}, {G{\'e}rard}, {Gerbig},
  {Giebels}, {Glicenstein}, {Gl{\"u}ck}, {Goret}, {G{\"o}ring}, {H{\"a}ffner},
  {Hague}, {Hampf}, {Hauser}, {Heinz}, {Heinzelmann}, {Henri}, {Hermann},
  {Hinton}, {Hoffmann}, {Hofmann}, {Hofverberg}, {Holler}, {Horns},
  {Jacholkowska}, {de Jager}, {Jahn}, {Jamrozy}, {Jung}, {Kastendieck},
  {Katarzy{\'n}ski}, {Katz}, {Kaufmann}, {Keogh}, {Khangulyan}, {Kh{\'e}lifi},
  {Klochkov}, {Klu{\'z}niak}, {Kneiske}, {Komin}, {Kosack}, {Kossakowski},
  {Laffon}, {Lamanna}, {Lennarz}, {Lohse}, {Lopatin}, {Lu}, {Marandon},
  {Marcowith}, {Masbou}, {Maurin}, {Maxted}, {McComb}, {Medina}, {M{\'e}hault},
  {Nguyen}, {Moderski}, {Moulin}, {Naumann}, {Naumann-Godo}, {de Naurois},
  {Nedbal}, {Nekrassov}, {Nicholas}, {Niemiec}, {Nolan}, {Ohm}, {Olive}, {de
  O{\~n}a Wilhelmi}, {Opitz}, {Ostrowski}, {Panter}, {Paz Arribas},
  {Pedaletti}, {Pelletier}, {Petrucci}, {Pita}, {P{\"u}hlhofer}, {Punch},
  {Quirrenbach}, {Raue}, {Rayner}, {Reimer}, {Reimer}, {Renaud}, {de los
  Reyes}, {Rieger}, {Ripken}, {Rob}, {Rosier-Lees}, {Rowell}, {Rudak},
  {Rulten}, {Ruppel}, {Ryde}, {Sahakian}, {Santangelo}, {Schlickeiser},
  {Sch{\"o}ck}, {Sch{\"o}nwald}, {Schulz}, {Schwanke}, {Schwarzburg},
  {Schwemmer}, {Shalchi}, {Sikora}, {Skilton}, {Sol}, {Spengler}, {Stawarz},
  {Steenkamp}, {Stegmann}, {Stinzing}, {Stycz}, {Sushch}, {Szostek},
  {Tavernet}, {Terrier}, {Tibolla}, {Tluczykont}, {Valerius}, {van Eldik},
  {Vasileiadis}, {Venter}, {Vialle}, {Viana}, {Vincent}, {Vivier}, {V{\"o}lk},
  {Volpe}, {Vorobiov}, {Vorster}, {Wagner}, {Ward}, {Wierzcholska}, {Zajczyk},
  {Zdziarski}, {Zech}, {Zechlin}, {Burnett} and {Hill}}]{2011A&A...529A..49H}
\bibinfo{author}{{H.~E.~S.~S. Collaboration}}, \bibinfo{author}{{Abramowski},
  A.}, \bibinfo{author}{{Acero}, F.}, \bibinfo{author}{{Aharonian}, F.},
  \bibinfo{author}{{Akhperjanian}, A.G.}, \bibinfo{author}{{Anton}, G.},
  \bibinfo{author}{{Balzer}, A.}, \bibinfo{author}{{Barnacka}, A.},
  \bibinfo{author}{{Barres de Almeida}, U.}, \bibinfo{author}{{Bazer-Bachi},
  A.R.}, \bibinfo{author}{{Becherini}, Y.}, \bibinfo{author}{{Becker}, J.},
  \bibinfo{author}{{Behera}, B.}, \bibinfo{author}{{Bernl{\"o}hr}, K.},
  \bibinfo{author}{{Bochow}, A.}, \bibinfo{author}{{Boisson}, C.},
  \bibinfo{author}{{Bolmont}, J.}, \bibinfo{author}{{Bordas}, P.},
  \bibinfo{author}{{Borrel}, V.}, \bibinfo{author}{{Brucker}, J.},
  \bibinfo{author}{{Brun}, F.}, \bibinfo{author}{{Brun}, P.},
  \bibinfo{author}{{Bulik}, T.}, \bibinfo{author}{{B{\"u}sching}, I.},
  \bibinfo{author}{{Carrigan}, S.}, \bibinfo{author}{{Casanova}, S.},
  \bibinfo{author}{{Cerruti}, M.}, \bibinfo{author}{{Chadwick}, P.M.},
  \bibinfo{author}{{Charbonnier}, A.}, \bibinfo{author}{{Chaves}, R.C.G.},
  \bibinfo{author}{{Cheesebrough}, A.}, \bibinfo{author}{{Chounet}, L.M.},
  \bibinfo{author}{{Clapson}, A.C.}, \bibinfo{author}{{Coignet}, G.},
  \bibinfo{author}{{Colom}, P.}, \bibinfo{author}{{Conrad}, J.},
  \bibinfo{author}{{Dalton}, M.}, \bibinfo{author}{{Daniel}, M.K.},
  \bibinfo{author}{{Davids}, I.D.}, \bibinfo{author}{{Degrange}, B.},
  \bibinfo{author}{{Deil}, C.}, \bibinfo{author}{{Dickinson}, H.J.},
  \bibinfo{author}{{Djannati-Ata{\"\i}}, A.}, \bibinfo{author}{{Domainko}, W.},
  \bibinfo{author}{{Drury}, L.O.C.}, \bibinfo{author}{{Dubois}, F.},
  \bibinfo{author}{{Dubus}, G.}, \bibinfo{author}{{Dyks}, J.},
  \bibinfo{author}{{Dyrda}, M.}, \bibinfo{author}{{Egberts}, K.},
  \bibinfo{author}{{Eger}, P.}, \bibinfo{author}{{Espigat}, P.},
  \bibinfo{author}{{Fallon}, L.}, \bibinfo{author}{{Farnier}, C.},
  \bibinfo{author}{{Fegan}, S.}, \bibinfo{author}{{Feinstein}, F.},
  \bibinfo{author}{{Fernandes}, M.V.}, \bibinfo{author}{{Fiasson}, A.},
  \bibinfo{author}{{Fontaine}, G.}, \bibinfo{author}{{F{\"o}rster}, A.},
  \bibinfo{author}{{F{\"u}{\ss}ling}, M.}, \bibinfo{author}{{Gallant}, Y.A.},
  \bibinfo{author}{{Gast}, H.}, \bibinfo{author}{{G{\'e}rard}, L.},
  \bibinfo{author}{{Gerbig}, D.}, \bibinfo{author}{{Giebels}, B.},
  \bibinfo{author}{{Glicenstein}, J.F.}, \bibinfo{author}{{Gl{\"u}ck}, B.},
  \bibinfo{author}{{Goret}, P.}, \bibinfo{author}{{G{\"o}ring}, D.},
  \bibinfo{author}{{H{\"a}ffner}, S.}, \bibinfo{author}{{Hague}, J.D.},
  \bibinfo{author}{{Hampf}, D.}, \bibinfo{author}{{Hauser}, M.},
  \bibinfo{author}{{Heinz}, S.}, \bibinfo{author}{{Heinzelmann}, G.},
  \bibinfo{author}{{Henri}, G.}, \bibinfo{author}{{Hermann}, G.},
  \bibinfo{author}{{Hinton}, J.A.}, \bibinfo{author}{{Hoffmann}, A.},
  \bibinfo{author}{{Hofmann}, W.}, \bibinfo{author}{{Hofverberg}, P.},
  \bibinfo{author}{{Holler}, M.}, \bibinfo{author}{{Horns}, D.},
  \bibinfo{author}{{Jacholkowska}, A.}, \bibinfo{author}{{de Jager}, O.C.},
  \bibinfo{author}{{Jahn}, C.}, \bibinfo{author}{{Jamrozy}, M.},
  \bibinfo{author}{{Jung}, I.}, \bibinfo{author}{{Kastendieck}, M.A.},
  \bibinfo{author}{{Katarzy{\'n}ski}, K.}, \bibinfo{author}{{Katz}, U.},
  \bibinfo{author}{{Kaufmann}, S.}, \bibinfo{author}{{Keogh}, D.},
  \bibinfo{author}{{Khangulyan}, D.}, \bibinfo{author}{{Kh{\'e}lifi}, B.},
  \bibinfo{author}{{Klochkov}, D.}, \bibinfo{author}{{Klu{\'z}niak}, W.},
  \bibinfo{author}{{Kneiske}, T.}, \bibinfo{author}{{Komin}, N.},
  \bibinfo{author}{{Kosack}, K.}, \bibinfo{author}{{Kossakowski}, R.},
  \bibinfo{author}{{Laffon}, H.}, \bibinfo{author}{{Lamanna}, G.},
  \bibinfo{author}{{Lennarz}, D.}, \bibinfo{author}{{Lohse}, T.},
  \bibinfo{author}{{Lopatin}, A.}, \bibinfo{author}{{Lu}, C.C.},
  \bibinfo{author}{{Marandon}, V.}, \bibinfo{author}{{Marcowith}, A.},
  \bibinfo{author}{{Masbou}, J.}, \bibinfo{author}{{Maurin}, D.},
  \bibinfo{author}{{Maxted}, N.}, \bibinfo{author}{{McComb}, T.J.L.},
  \bibinfo{author}{{Medina}, M.C.}, \bibinfo{author}{{M{\'e}hault}, J.},
  \bibinfo{author}{{Nguyen}, N.}, \bibinfo{author}{{Moderski}, R.},
  \bibinfo{author}{{Moulin}, E.}, \bibinfo{author}{{Naumann}, C.L.},
  \bibinfo{author}{{Naumann-Godo}, M.}, \bibinfo{author}{{de Naurois}, M.},
  \bibinfo{author}{{Nedbal}, D.}, \bibinfo{author}{{Nekrassov}, D.},
  \bibinfo{author}{{Nicholas}, B.}, \bibinfo{author}{{Niemiec}, J.},
  \bibinfo{author}{{Nolan}, S.J.}, \bibinfo{author}{{Ohm}, S.},
  \bibinfo{author}{{Olive}, J.F.}, \bibinfo{author}{{de O{\~n}a Wilhelmi}, E.},
  \bibinfo{author}{{Opitz}, B.}, \bibinfo{author}{{Ostrowski}, M.},
  \bibinfo{author}{{Panter}, M.}, \bibinfo{author}{{Paz Arribas}, M.},
  \bibinfo{author}{{Pedaletti}, G.}, \bibinfo{author}{{Pelletier}, G.},
  \bibinfo{author}{{Petrucci}, P.O.}, \bibinfo{author}{{Pita}, S.},
  \bibinfo{author}{{P{\"u}hlhofer}, G.}, \bibinfo{author}{{Punch}, M.},
  \bibinfo{author}{{Quirrenbach}, A.}, \bibinfo{author}{{Raue}, M.},
  \bibinfo{author}{{Rayner}, S.M.}, \bibinfo{author}{{Reimer}, A.},
  \bibinfo{author}{{Reimer}, O.}, \bibinfo{author}{{Renaud}, M.},
  \bibinfo{author}{{de los Reyes}, R.}, \bibinfo{author}{{Rieger}, F.},
  \bibinfo{author}{{Ripken}, J.}, \bibinfo{author}{{Rob}, L.},
  \bibinfo{author}{{Rosier-Lees}, S.}, \bibinfo{author}{{Rowell}, G.},
  \bibinfo{author}{{Rudak}, B.}, \bibinfo{author}{{Rulten}, C.B.},
  \bibinfo{author}{{Ruppel}, J.}, \bibinfo{author}{{Ryde}, F.},
  \bibinfo{author}{{Sahakian}, V.}, \bibinfo{author}{{Santangelo}, A.},
  \bibinfo{author}{{Schlickeiser}, R.}, \bibinfo{author}{{Sch{\"o}ck}, F.M.},
  \bibinfo{author}{{Sch{\"o}nwald}, A.}, \bibinfo{author}{{Schulz}, A.},
  \bibinfo{author}{{Schwanke}, U.}, \bibinfo{author}{{Schwarzburg}, S.},
  \bibinfo{author}{{Schwemmer}, S.}, \bibinfo{author}{{Shalchi}, A.},
  \bibinfo{author}{{Sikora}, M.}, \bibinfo{author}{{Skilton}, J.L.},
  \bibinfo{author}{{Sol}, H.}, \bibinfo{author}{{Spengler}, G.},
  \bibinfo{author}{{Stawarz}, {\L}.}, \bibinfo{author}{{Steenkamp}, R.},
  \bibinfo{author}{{Stegmann}, C.}, \bibinfo{author}{{Stinzing}, F.},
  \bibinfo{author}{{Stycz}, K.}, \bibinfo{author}{{Sushch}, I.},
  \bibinfo{author}{{Szostek}, A.}, \bibinfo{author}{{Tavernet}, J.P.},
  \bibinfo{author}{{Terrier}, R.}, \bibinfo{author}{{Tibolla}, O.},
  \bibinfo{author}{{Tluczykont}, M.}, \bibinfo{author}{{Valerius}, K.},
  \bibinfo{author}{{van Eldik}, C.}, \bibinfo{author}{{Vasileiadis}, G.},
  \bibinfo{author}{{Venter}, C.}, \bibinfo{author}{{Vialle}, J.P.},
  \bibinfo{author}{{Viana}, A.}, \bibinfo{author}{{Vincent}, P.},
  \bibinfo{author}{{Vivier}, M.}, \bibinfo{author}{{V{\"o}lk}, H.J.},
  \bibinfo{author}{{Volpe}, F.}, \bibinfo{author}{{Vorobiov}, S.},
  \bibinfo{author}{{Vorster}, M.}, \bibinfo{author}{{Wagner}, S.J.},
  \bibinfo{author}{{Ward}, M.}, \bibinfo{author}{{Wierzcholska}, A.},
  \bibinfo{author}{{Zajczyk}, A.}, \bibinfo{author}{{Zdziarski}, A.A.},
  \bibinfo{author}{{Zech}, A.}, \bibinfo{author}{{Zechlin}, H.S.},
  \bibinfo{author}{{Burnett}, T.H.}, \bibinfo{author}{{Hill}, A.B.},
  \bibinfo{year}{2011}.
\newblock \bibinfo{title}{{HESS J1943+213: a candidate extreme BL Lacertae
  object}}.
\newblock \bibinfo{journal}{\aap} \bibinfo{volume}{529}, \bibinfo{pages}{A49}.
\newblock \DOIprefix\doi{10.1051/0004-6361/201116545},
  \href{http://arxiv.org/abs/1103.0763}{\tt arXiv:1103.0763}.
\bibitem[{{Harrison} et~al.(2016){Harrison}, {Aird}, {Civano}, {Lansbury},
  {Mullaney}, {Ballantyne}, {Alexander}, {Stern}, {Ajello}, {Barret}, {Bauer},
  {Balokovi{\'c}}, {Brandt}, {Brightman}, {Boggs}, {Christensen}, {Comastri},
  {Craig}, {Del Moro}, {Forster}, {Gandhi}, {Giommi}, {Grefenstette}, {Hailey},
  {Hickox}, {Hornstrup}, {Kitaguchi}, {Koglin}, {Luo}, {Madsen}, {Mao},
  {Miyasaka}, {Mori}, {Perri}, {Pivovaroff}, {Puccetti}, {Rana}, {Treister},
  {Walton}, {Westergaard}, {Wik}, {Zappacosta}, {Zhang} and
  {Zoglauer}}]{2016ApJ...831..185H}
\bibinfo{author}{{Harrison}, F.A.}, \bibinfo{author}{{Aird}, J.},
  \bibinfo{author}{{Civano}, F.}, \bibinfo{author}{{Lansbury}, G.},
  \bibinfo{author}{{Mullaney}, J.R.}, \bibinfo{author}{{Ballantyne}, D.R.},
  \bibinfo{author}{{Alexander}, D.M.}, \bibinfo{author}{{Stern}, D.},
  \bibinfo{author}{{Ajello}, M.}, \bibinfo{author}{{Barret}, D.},
  \bibinfo{author}{{Bauer}, F.E.}, \bibinfo{author}{{Balokovi{\'c}}, M.},
  \bibinfo{author}{{Brandt}, W.N.}, \bibinfo{author}{{Brightman}, M.},
  \bibinfo{author}{{Boggs}, S.E.}, \bibinfo{author}{{Christensen}, F.E.},
  \bibinfo{author}{{Comastri}, A.}, \bibinfo{author}{{Craig}, W.W.},
  \bibinfo{author}{{Del Moro}, A.}, \bibinfo{author}{{Forster}, K.},
  \bibinfo{author}{{Gandhi}, P.}, \bibinfo{author}{{Giommi}, P.},
  \bibinfo{author}{{Grefenstette}, B.W.}, \bibinfo{author}{{Hailey}, C.J.},
  \bibinfo{author}{{Hickox}, R.C.}, \bibinfo{author}{{Hornstrup}, A.},
  \bibinfo{author}{{Kitaguchi}, T.}, \bibinfo{author}{{Koglin}, J.},
  \bibinfo{author}{{Luo}, B.}, \bibinfo{author}{{Madsen}, K.K.},
  \bibinfo{author}{{Mao}, P.H.}, \bibinfo{author}{{Miyasaka}, H.},
  \bibinfo{author}{{Mori}, K.}, \bibinfo{author}{{Perri}, M.},
  \bibinfo{author}{{Pivovaroff}, M.}, \bibinfo{author}{{Puccetti}, S.},
  \bibinfo{author}{{Rana}, V.}, \bibinfo{author}{{Treister}, E.},
  \bibinfo{author}{{Walton}, D.}, \bibinfo{author}{{Westergaard}, N.J.},
  \bibinfo{author}{{Wik}, D.}, \bibinfo{author}{{Zappacosta}, L.},
  \bibinfo{author}{{Zhang}, W.W.}, \bibinfo{author}{{Zoglauer}, A.},
  \bibinfo{year}{2016}.
\newblock \bibinfo{title}{{The NuSTAR Extragalactic Surveys: The Number Counts
  of Active Galactic Nuclei and the Resolved Fraction of the Cosmic X-Ray
  Background}}.
\newblock \bibinfo{journal}{\apj} \bibinfo{volume}{831}, \bibinfo{pages}{185}.
\newblock \href{http://arxiv.org/abs/1511.04183}{\tt arXiv:1511.04183}.
\bibitem[{{H{\"o}nig}(2019)}]{2019ApJ...884..171H}
\bibinfo{author}{{H{\"o}nig}, S.F.}, \bibinfo{year}{2019}.
\newblock \bibinfo{title}{{Redefining the Torus: A Unifying View of AGNs in the
  Infrared and Submillimeter}}.
\newblock \bibinfo{journal}{\apj} \bibinfo{volume}{884}, \bibinfo{pages}{171}.
\newblock \DOIprefix\doi{10.3847/1538-4357/ab4591},
  \href{http://arxiv.org/abs/1909.08639}{\tt arXiv:1909.08639}.
\bibitem[{{IceCube Collaboration} et~al.(2018){IceCube Collaboration},
  {Aartsen}, {Ackermann}, {Adams}, {Aguilar}, {Ahlers}, {Ahrens}, {Al Samarai},
  {Altmann}, {Andeen} and et~al.}]{2018Sci...361.1378I}
\bibinfo{author}{{IceCube Collaboration}}, \bibinfo{author}{{Aartsen}, M.G.},
  \bibinfo{author}{{Ackermann}, M.}, \bibinfo{author}{{Adams}, J.},
  \bibinfo{author}{{Aguilar}, J.A.}, \bibinfo{author}{{Ahlers}, M.},
  \bibinfo{author}{{Ahrens}, M.}, \bibinfo{author}{{Al Samarai}, I.},
  \bibinfo{author}{{Altmann}, D.}, \bibinfo{author}{{Andeen}, K.},
  \bibinfo{author}{et~al.}, \bibinfo{year}{2018}.
\newblock \bibinfo{title}{{Multimessenger observations of a flaring blazar
  coincident with high-energy neutrino IceCube-170922A}}.
\newblock \bibinfo{journal}{Science} \bibinfo{volume}{361},
  \bibinfo{pages}{eaat1378}.
\newblock \href{http://arxiv.org/abs/1807.08816}{\tt arXiv:1807.08816}.
\bibitem[{Karasev et~al.(2018)Karasev, Lutovinov, Tkachenko, Khorunzhev,
  Krivonos, Medvedev, Pavlinsky, Burenin and Eselevich}]{Karasev_2018}
\bibinfo{author}{Karasev, D.I.}, \bibinfo{author}{Lutovinov, A.A.},
  \bibinfo{author}{Tkachenko, A.Y.}, \bibinfo{author}{Khorunzhev, G.A.},
  \bibinfo{author}{Krivonos, R.A.}, \bibinfo{author}{Medvedev, P.S.},
  \bibinfo{author}{Pavlinsky, M.N.}, \bibinfo{author}{Burenin, R.A.},
  \bibinfo{author}{Eselevich, M.V.}, \bibinfo{year}{2018}.
\newblock \bibinfo{title}{Optical identification of x-ray sources from the
  14-year integral all-sky survey}.
\newblock \bibinfo{journal}{Astronomy Letters} \bibinfo{volume}{44},
  \bibinfo{pages}{522--540}.
\bibitem[{{Keivani} et~al.(2018){Keivani}, {Murase}, {Petropoulou}, {Fox},
  {Cenko}, {Chaty}, {Coleiro}, {DeLaunay}, {Dimitrakoudis}, {Evans}, {Kennea},
  {Marshall}, {Mastichiadis}, {Osborne}, {Santand er}, {Tohuvavohu} and
  {Turley}}]{2018ApJ...864...84K}
\bibinfo{author}{{Keivani}, A.}, \bibinfo{author}{{Murase}, K.},
  \bibinfo{author}{{Petropoulou}, M.}, \bibinfo{author}{{Fox}, D.B.},
  \bibinfo{author}{{Cenko}, S.B.}, \bibinfo{author}{{Chaty}, S.},
  \bibinfo{author}{{Coleiro}, A.}, \bibinfo{author}{{DeLaunay}, J.J.},
  \bibinfo{author}{{Dimitrakoudis}, S.}, \bibinfo{author}{{Evans}, P.A.},
  \bibinfo{author}{{Kennea}, J.A.}, \bibinfo{author}{{Marshall}, F.E.},
  \bibinfo{author}{{Mastichiadis}, A.}, \bibinfo{author}{{Osborne}, J.P.},
  \bibinfo{author}{{Santand er}, M.}, \bibinfo{author}{{Tohuvavohu}, A.},
  \bibinfo{author}{{Turley}, C.F.}, \bibinfo{year}{2018}.
\newblock \bibinfo{title}{{A Multimessenger Picture of the Flaring Blazar TXS
  0506+056: Implications for High-energy Neutrino Emission and Cosmic-Ray
  Acceleration}}.
\newblock \bibinfo{journal}{\apj} \bibinfo{volume}{864}, \bibinfo{pages}{84}.
\newblock \href{http://arxiv.org/abs/1807.04537}{\tt arXiv:1807.04537}.
\bibitem[{{Khorunzhev} et~al.(2012){Khorunzhev}, {Sazonov}, {Burenin} and
  {Tkachenko}}]{2012AstL...38..475K}
\bibinfo{author}{{Khorunzhev}, G.A.}, \bibinfo{author}{{Sazonov}, S.Y.},
  \bibinfo{author}{{Burenin}, R.A.}, \bibinfo{author}{{Tkachenko}, A.Y.},
  \bibinfo{year}{2012}.
\newblock \bibinfo{title}{{Masses and accretion rates of supermassive black
  holes in active galactic nuclei from the INTEGRAL survey}}.
\newblock \bibinfo{journal}{Astronomy Letters} \bibinfo{volume}{38},
  \bibinfo{pages}{475--491}.
\newblock \href{http://arxiv.org/abs/1210.6279}{\tt arXiv:1210.6279}.
\bibitem[{{Krawczynski} et~al.(2004){Krawczynski}, {Hughes}, {Horan},
  {Aharonian}, {Aller}, {Aller}, {Boltwood}, {Buckley}, {Coppi}, {Fossati},
  {G{\"o}tting}, {Holder}, {Horns}, {Kurtanidze}, {Marscher}, {Nikolashvili},
  {Remillard}, {Sadun} and {Schr{\"o}der}}]{2004ApJ...601..151K}
\bibinfo{author}{{Krawczynski}, H.}, \bibinfo{author}{{Hughes}, S.B.},
  \bibinfo{author}{{Horan}, D.}, \bibinfo{author}{{Aharonian}, F.},
  \bibinfo{author}{{Aller}, M.F.}, \bibinfo{author}{{Aller}, H.},
  \bibinfo{author}{{Boltwood}, P.}, \bibinfo{author}{{Buckley}, J.},
  \bibinfo{author}{{Coppi}, P.}, \bibinfo{author}{{Fossati}, G.},
  \bibinfo{author}{{G{\"o}tting}, N.}, \bibinfo{author}{{Holder}, J.},
  \bibinfo{author}{{Horns}, D.}, \bibinfo{author}{{Kurtanidze}, O.M.},
  \bibinfo{author}{{Marscher}, A.P.}, \bibinfo{author}{{Nikolashvili}, M.},
  \bibinfo{author}{{Remillard}, R.A.}, \bibinfo{author}{{Sadun}, A.},
  \bibinfo{author}{{Schr{\"o}der}, M.}, \bibinfo{year}{2004}.
\newblock \bibinfo{title}{{Multiwavelength Observations of Strong Flares from
  the TeV Blazar 1ES 1959+650}}.
\newblock \bibinfo{journal}{\apj} \bibinfo{volume}{601},
  \bibinfo{pages}{151--164}.
\newblock \href{http://arxiv.org/abs/astro-ph/0310158}{\tt
  arXiv:astro-ph/0310158}.
\bibitem[{{Krivonos} et~al.(2007){Krivonos}, {Revnivtsev}, {Lutovinov},
  {Sazonov}, {Churazov} and {Sunyaev}}]{2007A&A...475..775K}
\bibinfo{author}{{Krivonos}, R.}, \bibinfo{author}{{Revnivtsev}, M.},
  \bibinfo{author}{{Lutovinov}, A.}, \bibinfo{author}{{Sazonov}, S.},
  \bibinfo{author}{{Churazov}, E.}, \bibinfo{author}{{Sunyaev}, R.},
  \bibinfo{year}{2007}.
\newblock \bibinfo{title}{{INTEGRAL/IBIS all-sky survey in hard X-rays}}.
\newblock \bibinfo{journal}{\aap} \bibinfo{volume}{475},
  \bibinfo{pages}{775--784}.
\newblock \href{http://arxiv.org/abs/astro-ph/0701836}{\tt
  arXiv:astro-ph/0701836}.
\bibitem[{{Krivonos} et~al.(2015){Krivonos}, {Tsygankov}, {Lutovinov},
  {Revnivtsev}, {Churazov} and {Sunyaev}}]{2015MNRAS.448.3766K}
\bibinfo{author}{{Krivonos}, R.}, \bibinfo{author}{{Tsygankov}, S.},
  \bibinfo{author}{{Lutovinov}, A.}, \bibinfo{author}{{Revnivtsev}, M.},
  \bibinfo{author}{{Churazov}, E.}, \bibinfo{author}{{Sunyaev}, R.},
  \bibinfo{year}{2015}.
\newblock \bibinfo{title}{{INTEGRAL 11-year hard X-ray survey above 100 keV}}.
\newblock \bibinfo{journal}{\mnras} \bibinfo{volume}{448},
  \bibinfo{pages}{3766--3774}.
\newblock \href{http://arxiv.org/abs/1412.1051}{\tt arXiv:1412.1051}.
\bibitem[{{Krivonos} et~al.(2010){Krivonos}, {Tsygankov}, {Revnivtsev},
  {Grebenev}, {Churazov} and {Sunyaev}}]{2010A&A...523A..61K}
\bibinfo{author}{{Krivonos}, R.}, \bibinfo{author}{{Tsygankov}, S.},
  \bibinfo{author}{{Revnivtsev}, M.}, \bibinfo{author}{{Grebenev}, S.},
  \bibinfo{author}{{Churazov}, E.}, \bibinfo{author}{{Sunyaev}, R.},
  \bibinfo{year}{2010}.
\newblock \bibinfo{title}{{INTEGRAL/IBIS 7-year All-Sky Hard X-Ray Survey. II.
  Catalog of sources}}.
\newblock \bibinfo{journal}{\aap} \bibinfo{volume}{523}, \bibinfo{pages}{A61}.
\newblock \DOIprefix\doi{10.1051/0004-6361/201014935},
  \href{http://arxiv.org/abs/1006.4437}{\tt arXiv:1006.4437}.
\bibitem[{{Landi} et~al.(2017){Landi}, {Bassani}, {Bazzano}, {Bird}, {Fiocchi},
  {Malizia}, {Panessa}, {Sguera} and {Ubertini}}]{2017MNRAS.470.1107L}
\bibinfo{author}{{Landi}, R.}, \bibinfo{author}{{Bassani}, L.},
  \bibinfo{author}{{Bazzano}, A.}, \bibinfo{author}{{Bird}, A.J.},
  \bibinfo{author}{{Fiocchi}, M.}, \bibinfo{author}{{Malizia}, A.},
  \bibinfo{author}{{Panessa}, F.}, \bibinfo{author}{{Sguera}, V.},
  \bibinfo{author}{{Ubertini}, P.}, \bibinfo{year}{2017}.
\newblock \bibinfo{title}{{Investigating the X-ray counterparts to unidentified
  sources in the 1000-orbit INTEGRAL/IBIS catalogue}}.
\newblock \bibinfo{journal}{\mnras} \bibinfo{volume}{470},
  \bibinfo{pages}{1107--1120}.
\newblock \DOIprefix\doi{10.1093/mnras/stx908},
  \href{http://arxiv.org/abs/1704.03872}{\tt arXiv:1704.03872}.
\bibitem[{Landi et~al.(2010)Landi, Bassani, Malizia, Stephen, Bazzano, Fiocchi
  and Bird}]{Landi_2010}
\bibinfo{author}{Landi, R.}, \bibinfo{author}{Bassani, L.},
  \bibinfo{author}{Malizia, A.}, \bibinfo{author}{Stephen, J.B.},
  \bibinfo{author}{Bazzano, A.}, \bibinfo{author}{Fiocchi, M.},
  \bibinfo{author}{Bird, A.J.}, \bibinfo{year}{2010}.
\newblock \bibinfo{title}{Swift/xrt observations of unidentified integral/ibis
  sources}.
\newblock \bibinfo{journal}{Monthly Notices of the Royal Astronomical Society}
  \bibinfo{volume}{403}, \bibinfo{pages}{945--959}.
\bibitem[{{Landi} et~al.(2009){Landi}, {Stephen}, {Masetti}, {Grupe},
  {Capitanio}, {Bird}, {Dean}, {Fiocchi} and {Gehrels}}]{2009A&A...493..893L}
\bibinfo{author}{{Landi}, R.}, \bibinfo{author}{{Stephen}, J.B.},
  \bibinfo{author}{{Masetti}, N.}, \bibinfo{author}{{Grupe}, D.},
  \bibinfo{author}{{Capitanio}, F.}, \bibinfo{author}{{Bird}, A.J.},
  \bibinfo{author}{{Dean}, A.J.}, \bibinfo{author}{{Fiocchi}, M.},
  \bibinfo{author}{{Gehrels}, N.}, \bibinfo{year}{2009}.
\newblock \bibinfo{title}{{The AGN nature of three INTEGRAL sources: IGR
  J18249-3243, IGR J19443+2117, and IGR J22292+6647}}.
\newblock \bibinfo{journal}{\aap} \bibinfo{volume}{493},
  \bibinfo{pages}{893--898}.
\newblock \DOIprefix\doi{10.1051/0004-6361:200810503},
  \href{http://arxiv.org/abs/0811.2318}{\tt arXiv:0811.2318}.
\bibitem[{{Lanzuisi} et~al.(2012){Lanzuisi}, {de Rosa}, {Ghisellini},
  {Ubertini}, {Panessa}, {Ajello}, {Bassani}, {Fukazawa} and
  {D'Ammando}}]{2012MNRAS.421..390L}
\bibinfo{author}{{Lanzuisi}, G.}, \bibinfo{author}{{de Rosa}, A.},
  \bibinfo{author}{{Ghisellini}, G.}, \bibinfo{author}{{Ubertini}, P.},
  \bibinfo{author}{{Panessa}, F.}, \bibinfo{author}{{Ajello}, M.},
  \bibinfo{author}{{Bassani}, L.}, \bibinfo{author}{{Fukazawa}, Y.},
  \bibinfo{author}{{D'Ammando}, F.}, \bibinfo{year}{2012}.
\newblock \bibinfo{title}{{Modelling the flaring activity of the high-z, hard
  X-ray-selected blazar IGR J22517+2217}}.
\newblock \bibinfo{journal}{\mnras} \bibinfo{volume}{421},
  \bibinfo{pages}{390--397}.
\newblock \href{http://arxiv.org/abs/1112.0472}{\tt arXiv:1112.0472}.
\bibitem[{Levine et~al.(1984)Levine, Lang, Lewin, Primini, Dobson, Doty,
  Hoffman, Howe, Scheepmaker, Wheaton and et~al.}]{Levine_1984}
\bibinfo{author}{Levine, A.M.}, \bibinfo{author}{Lang, F.L.},
  \bibinfo{author}{Lewin, W.H.G.}, \bibinfo{author}{Primini, F.A.},
  \bibinfo{author}{Dobson, C.A.}, \bibinfo{author}{Doty, J.P.},
  \bibinfo{author}{Hoffman, J.A.}, \bibinfo{author}{Howe, S.K.},
  \bibinfo{author}{Scheepmaker, A.}, \bibinfo{author}{Wheaton, W.A.},
  \bibinfo{author}{et~al.}, \bibinfo{year}{1984}.
\newblock \bibinfo{title}{The heao 1 a-4 catalog of high-energy x-ray sources}.
\newblock \bibinfo{journal}{The Astrophysical Journal Supplement Series}
  \bibinfo{volume}{54}, \bibinfo{pages}{581}.
\bibitem[{{Lichti} et~al.(2008){Lichti}, {Bottacini}, {Ajello}, {Charlot},
  {Collmar}, {Falcone}, {Horan}, {Huber}, {von Kienlin}, {L{\"a}hteenm{\"a}ki},
  {Lindfors}, {Morris}, {Nilsson}, {Petry}, {R{\"u}ger}, {Sillanp{\"a}{\"a}},
  {Spanier} and {Tornikoski}}]{2008A&A...486..721L}
\bibinfo{author}{{Lichti}, G.G.}, \bibinfo{author}{{Bottacini}, E.},
  \bibinfo{author}{{Ajello}, M.}, \bibinfo{author}{{Charlot}, P.},
  \bibinfo{author}{{Collmar}, W.}, \bibinfo{author}{{Falcone}, A.},
  \bibinfo{author}{{Horan}, D.}, \bibinfo{author}{{Huber}, S.},
  \bibinfo{author}{{von Kienlin}, A.}, \bibinfo{author}{{L{\"a}hteenm{\"a}ki},
  A.}, \bibinfo{author}{{Lindfors}, E.}, \bibinfo{author}{{Morris}, D.},
  \bibinfo{author}{{Nilsson}, K.}, \bibinfo{author}{{Petry}, D.},
  \bibinfo{author}{{R{\"u}ger}, M.}, \bibinfo{author}{{Sillanp{\"a}{\"a}}, A.},
  \bibinfo{author}{{Spanier}, F.}, \bibinfo{author}{{Tornikoski}, M.},
  \bibinfo{year}{2008}.
\newblock \bibinfo{title}{{INTEGRAL observations of the blazar Mrk 421 in
  outburst. Results of a multi-wavelength campaign}}.
\newblock \bibinfo{journal}{\aap} \bibinfo{volume}{486},
  \bibinfo{pages}{721--734}.
\newblock \href{http://arxiv.org/abs/0805.2577}{\tt arXiv:0805.2577}.
\bibitem[{{Lubi{\'n}ski} et~al.(2016){Lubi{\'n}ski}, {Beckmann}, {Gibaud},
  {Paltani}, {Papadakis}, {Ricci}, {Soldi}, {T{\"u}rler}, {Walter} and
  {Zdziarski}}]{2016MNRAS.458.2454L}
\bibinfo{author}{{Lubi{\'n}ski}, P.}, \bibinfo{author}{{Beckmann}, V.},
  \bibinfo{author}{{Gibaud}, L.}, \bibinfo{author}{{Paltani}, S.},
  \bibinfo{author}{{Papadakis}, I.E.}, \bibinfo{author}{{Ricci}, C.},
  \bibinfo{author}{{Soldi}, S.}, \bibinfo{author}{{T{\"u}rler}, M.},
  \bibinfo{author}{{Walter}, R.}, \bibinfo{author}{{Zdziarski}, A.A.},
  \bibinfo{year}{2016}.
\newblock \bibinfo{title}{{A comprehensive analysis of the hard X-ray spectra
  of bright Seyfert galaxies}}.
\newblock \bibinfo{journal}{\mnras} \bibinfo{volume}{458},
  \bibinfo{pages}{2454--2475}.
\newblock \href{http://arxiv.org/abs/1602.08402}{\tt arXiv:1602.08402}.
\bibitem[{{Lubi{\'n}ski} et~al.(2010){Lubi{\'n}ski}, {Zdziarski}, {Walter},
  {Paltani}, {Beckmann}, {Soldi}, {Ferrigno} and
  {Courvoisier}}]{2010MNRAS.408.1851L}
\bibinfo{author}{{Lubi{\'n}ski}, P.}, \bibinfo{author}{{Zdziarski}, A.A.},
  \bibinfo{author}{{Walter}, R.}, \bibinfo{author}{{Paltani}, S.},
  \bibinfo{author}{{Beckmann}, V.}, \bibinfo{author}{{Soldi}, S.},
  \bibinfo{author}{{Ferrigno}, C.}, \bibinfo{author}{{Courvoisier}, T.J.L.},
  \bibinfo{year}{2010}.
\newblock \bibinfo{title}{{Extreme flux states of NGC 4151 observed with
  INTEGRAL}}.
\newblock \bibinfo{journal}{\mnras} \bibinfo{volume}{408},
  \bibinfo{pages}{1851--1865}.
\newblock \href{http://arxiv.org/abs/1005.0842}{\tt arXiv:1005.0842}.
\bibitem[{{Luo} et~al.(2017){Luo}, {Brandt}, {Xue}, {Lehmer}, {Alexander},
  {Bauer}, {Vito}, {Yang}, {Basu-Zych}, {Comastri}, {Gilli}, {Gu},
  {Hornschemeier}, {Koekemoer}, {Liu}, {Mainieri}, {Paolillo}, {Ranalli},
  {Rosati}, {Schneider}, {Shemmer}, {Smail}, {Sun}, {Tozzi}, {Vignali} and
  {Wang}}]{2017ApJS..228....2L}
\bibinfo{author}{{Luo}, B.}, \bibinfo{author}{{Brandt}, W.N.},
  \bibinfo{author}{{Xue}, Y.Q.}, \bibinfo{author}{{Lehmer}, B.},
  \bibinfo{author}{{Alexander}, D.M.}, \bibinfo{author}{{Bauer}, F.E.},
  \bibinfo{author}{{Vito}, F.}, \bibinfo{author}{{Yang}, G.},
  \bibinfo{author}{{Basu-Zych}, A.R.}, \bibinfo{author}{{Comastri}, A.},
  \bibinfo{author}{{Gilli}, R.}, \bibinfo{author}{{Gu}, Q.S.},
  \bibinfo{author}{{Hornschemeier}, A.E.}, \bibinfo{author}{{Koekemoer}, A.},
  \bibinfo{author}{{Liu}, T.}, \bibinfo{author}{{Mainieri}, V.},
  \bibinfo{author}{{Paolillo}, M.}, \bibinfo{author}{{Ranalli}, P.},
  \bibinfo{author}{{Rosati}, P.}, \bibinfo{author}{{Schneider}, D.P.},
  \bibinfo{author}{{Shemmer}, O.}, \bibinfo{author}{{Smail}, I.},
  \bibinfo{author}{{Sun}, M.}, \bibinfo{author}{{Tozzi}, P.},
  \bibinfo{author}{{Vignali}, C.}, \bibinfo{author}{{Wang}, J.X.},
  \bibinfo{year}{2017}.
\newblock \bibinfo{title}{{The Chandra Deep Field-South Survey: 7 Ms Source
  Catalogs}}.
\newblock \bibinfo{journal}{\apjs} \bibinfo{volume}{228}, \bibinfo{pages}{2}.
\newblock \href{http://arxiv.org/abs/1611.03501}{\tt arXiv:1611.03501}.
\bibitem[{{MacDonald} and {Mullan}(2017)}]{2017ApJ...850...58M}
\bibinfo{author}{{MacDonald}, J.}, \bibinfo{author}{{Mullan}, D.J.},
  \bibinfo{year}{2017}.
\newblock \bibinfo{title}{{Magnetic Modeling of Inflated Low-mass Stars Using
  Interior Fields No Larger than {\`I}ƒ10 kG}}.
\newblock \bibinfo{journal}{\apj} \bibinfo{volume}{850}, \bibinfo{pages}{58}.
\newblock \href{http://arxiv.org/abs/1708.06994}{\tt arXiv:1708.06994}.
\bibitem[{{Maiolino} and {Rieke}(1995)}]{1995ApJ...454...95M}
\bibinfo{author}{{Maiolino}, R.}, \bibinfo{author}{{Rieke}, G.H.},
  \bibinfo{year}{1995}.
\newblock \bibinfo{title}{{Low-Luminosity and Obscured Seyfert Nuclei in Nearby
  Galaxies}}.
\newblock \bibinfo{journal}{\apj} \bibinfo{volume}{454}, \bibinfo{pages}{95}.
\bibitem[{{Malizia} et~al.(2012){Malizia}, {Bassani}, {Bazzano}, {Bird},
  {Masetti}, {Panessa}, {Stephen} and {Ubertini}}]{2012MNRAS.426.1750M}
\bibinfo{author}{{Malizia}, A.}, \bibinfo{author}{{Bassani}, L.},
  \bibinfo{author}{{Bazzano}, A.}, \bibinfo{author}{{Bird}, A.J.},
  \bibinfo{author}{{Masetti}, N.}, \bibinfo{author}{{Panessa}, F.},
  \bibinfo{author}{{Stephen}, J.B.}, \bibinfo{author}{{Ubertini}, P.},
  \bibinfo{year}{2012}.
\newblock \bibinfo{title}{{The INTEGRAL/IBIS AGN catalogue - I. X-ray
  absorption properties versus optical classification}}.
\newblock \bibinfo{journal}{\mnras} \bibinfo{volume}{426},
  \bibinfo{pages}{1750--1766}.
\newblock \DOIprefix\doi{10.1111/j.1365-2966.2012.21755.x},
  \href{http://arxiv.org/abs/1207.4882}{\tt arXiv:1207.4882}.
\bibitem[{{Malizia} et~al.(2008){Malizia}, {Bassani}, {Bird}, {Landi},
  {Masetti}, {de Rosa}, {Panessa}, {Molina}, {Dean}, {Perri} and
  {Tueller}}]{2008MNRAS.389.1360M}
\bibinfo{author}{{Malizia}, A.}, \bibinfo{author}{{Bassani}, L.},
  \bibinfo{author}{{Bird}, A.J.}, \bibinfo{author}{{Landi}, R.},
  \bibinfo{author}{{Masetti}, N.}, \bibinfo{author}{{de Rosa}, A.},
  \bibinfo{author}{{Panessa}, F.}, \bibinfo{author}{{Molina}, M.},
  \bibinfo{author}{{Dean}, A.J.}, \bibinfo{author}{{Perri}, M.},
  \bibinfo{author}{{Tueller}, J.}, \bibinfo{year}{2008}.
\newblock \bibinfo{title}{{First high-energy observations of narrow-line
  Seyfert 1s with INTEGRAL/IBIS}}.
\newblock \bibinfo{journal}{\mnras} \bibinfo{volume}{389},
  \bibinfo{pages}{1360--1366}.
\newblock \href{http://arxiv.org/abs/0806.4824}{\tt arXiv:0806.4824}.
\bibitem[{{Malizia} et~al.(2003){Malizia}, {Bassani}, {Stephen}, {Di Cocco},
  {Fiore} and {Dean}}]{2003ApJ...589L..17M}
\bibinfo{author}{{Malizia}, A.}, \bibinfo{author}{{Bassani}, L.},
  \bibinfo{author}{{Stephen}, J.B.}, \bibinfo{author}{{Di Cocco}, G.},
  \bibinfo{author}{{Fiore}, F.}, \bibinfo{author}{{Dean}, A.J.},
  \bibinfo{year}{2003}.
\newblock \bibinfo{title}{{BeppoSAX Average Spectra of Seyfert Galaxies}}.
\newblock \bibinfo{journal}{\apj} \bibinfo{volume}{589},
  \bibinfo{pages}{L17--L20}.
\newblock \href{http://arxiv.org/abs/astro-ph/0304133}{\tt
  arXiv:astro-ph/0304133}.
\bibitem[{Malizia et~al.(2007)Malizia, Landi, Bassani, Bird, Molina, De~Rosa,
  Fiocchi, Gehrels, Kennea and Perri}]{Malizia_2007}
\bibinfo{author}{Malizia, A.}, \bibinfo{author}{Landi, R.},
  \bibinfo{author}{Bassani, L.}, \bibinfo{author}{Bird, A.J.},
  \bibinfo{author}{Molina, M.}, \bibinfo{author}{De~Rosa, A.},
  \bibinfo{author}{Fiocchi, M.}, \bibinfo{author}{Gehrels, N.},
  \bibinfo{author}{Kennea, J.}, \bibinfo{author}{Perri, M.},
  \bibinfo{year}{2007}.
\newblock \bibinfo{title}{Swiftxrt observation of 34 newintegralibis agns:
  Discovery of compton‐thick and other peculiar sources}.
\newblock \bibinfo{journal}{The Astrophysical Journal} \bibinfo{volume}{668},
  \bibinfo{pages}{81--86}.
\bibitem[{Malizia et~al.(2016)Malizia, Landi, Molina, Bassani, Bazzano, Bird
  and Ubertini}]{Malizia_2016}
\bibinfo{author}{Malizia, A.}, \bibinfo{author}{Landi, R.},
  \bibinfo{author}{Molina, M.}, \bibinfo{author}{Bassani, L.},
  \bibinfo{author}{Bazzano, A.}, \bibinfo{author}{Bird, A.J.},
  \bibinfo{author}{Ubertini, P.}, \bibinfo{year}{2016}.
\newblock \bibinfo{title}{The integral/ibisagn catalogue: an update}.
\newblock \bibinfo{journal}{Monthly Notices of the Royal Astronomical Society}
  \bibinfo{volume}{460}, \bibinfo{pages}{19--29}.
\bibitem[{{Malizia} et~al.(2014){Malizia}, {Molina}, {Bassani}, {Stephen},
  {Bazzano}, {Ubertini} and {Bird}}]{2014ApJ...782L..25M}
\bibinfo{author}{{Malizia}, A.}, \bibinfo{author}{{Molina}, M.},
  \bibinfo{author}{{Bassani}, L.}, \bibinfo{author}{{Stephen}, J.B.},
  \bibinfo{author}{{Bazzano}, A.}, \bibinfo{author}{{Ubertini}, P.},
  \bibinfo{author}{{Bird}, A.J.}, \bibinfo{year}{2014}.
\newblock \bibinfo{title}{{The INTEGRAL High-energy Cut-off Distribution of
  Type 1 Active Galactic Nuclei}}.
\newblock \bibinfo{journal}{\apj} \bibinfo{volume}{782}, \bibinfo{pages}{L25}.
\newblock \href{http://arxiv.org/abs/1401.3647}{\tt arXiv:1401.3647}.
\bibitem[{{Malizia} et~al.(2009){Malizia}, {Stephen}, {Bassani}, {Bird},
  {Panessa} and {Ubertini}}]{2009MNRAS.399..944M}
\bibinfo{author}{{Malizia}, A.}, \bibinfo{author}{{Stephen}, J.B.},
  \bibinfo{author}{{Bassani}, L.}, \bibinfo{author}{{Bird}, A.J.},
  \bibinfo{author}{{Panessa}, F.}, \bibinfo{author}{{Ubertini}, P.},
  \bibinfo{year}{2009}.
\newblock \bibinfo{title}{{The fraction of Compton-thick sources in an INTEGRAL
  complete AGN sample}}.
\newblock \bibinfo{journal}{\mnras} \bibinfo{volume}{399},
  \bibinfo{pages}{944--951}.
\newblock \href{http://arxiv.org/abs/0906.5544}{\tt arXiv:0906.5544}.
\bibitem[{{Maraschi} and {Haardt}(1997)}]{1997ASPC..121..101M}
\bibinfo{author}{{Maraschi}, L.}, \bibinfo{author}{{Haardt}, F.},
  \bibinfo{year}{1997}.
\newblock \bibinfo{title}{{Disk-Corona Models and X-Ray Emission from Seyfert
  Galaxies}}, in: \bibinfo{editor}{{Wickramasinghe}, D.T.},
  \bibinfo{editor}{{Bicknell}, G.V.}, \bibinfo{editor}{{Ferrario}, L.} (Eds.),
  \bibinfo{booktitle}{IAU Colloq. 163: Accretion Phenomena and Related
  Outflows}, p. \bibinfo{pages}{101}.
\newblock \href{http://arxiv.org/abs/astro-ph/9611048}{\tt
  arXiv:astro-ph/9611048}.
\bibitem[{{Markwardt} et~al.(2005){Markwardt}, {Tueller}, {Skinner}, {Gehrels},
  {Barthelmy} and {Mushotzky}}]{2005ApJ...633L..77M}
\bibinfo{author}{{Markwardt}, C.B.}, \bibinfo{author}{{Tueller}, J.},
  \bibinfo{author}{{Skinner}, G.K.}, \bibinfo{author}{{Gehrels}, N.},
  \bibinfo{author}{{Barthelmy}, S.D.}, \bibinfo{author}{{Mushotzky}, R.F.},
  \bibinfo{year}{2005}.
\newblock \bibinfo{title}{{The Swift/BAT High-Latitude Survey: First Results}}.
\newblock \bibinfo{journal}{\apjl} \bibinfo{volume}{633},
  \bibinfo{pages}{L77--L80}.
\newblock \href{http://arxiv.org/abs/astro-ph/0509860}{\tt
  arXiv:astro-ph/0509860}.
\bibitem[{{Masetti} et~al.(2008a){Masetti}, {Mason}, {Landi}, {Giommi},
  {Bassani}, {Malizia}, {Bird}, {Bazzano}, {Dean}, {Gehrels}, {Palazzi} and
  {Ubertini}}]{2008A&A...480..715M}
\bibinfo{author}{{Masetti}, N.}, \bibinfo{author}{{Mason}, E.},
  \bibinfo{author}{{Landi}, R.}, \bibinfo{author}{{Giommi}, P.},
  \bibinfo{author}{{Bassani}, L.}, \bibinfo{author}{{Malizia}, A.},
  \bibinfo{author}{{Bird}, A.J.}, \bibinfo{author}{{Bazzano}, A.},
  \bibinfo{author}{{Dean}, A.J.}, \bibinfo{author}{{Gehrels}, N.},
  \bibinfo{author}{{Palazzi}, E.}, \bibinfo{author}{{Ubertini}, P.},
  \bibinfo{year}{2008}a.
\newblock \bibinfo{title}{{High-redshift blazar identification for Swift
  J1656.3-3302}}.
\newblock \bibinfo{journal}{\aap} \bibinfo{volume}{480},
  \bibinfo{pages}{715--721}.
\newblock \DOIprefix\doi{10.1051/0004-6361:20078901},
  \href{http://arxiv.org/abs/0801.2976}{\tt arXiv:0801.2976}.
\bibitem[{{Masetti} et~al.(2008b){Masetti}, {Mason}, {Morelli}, {Cellone},
  {McBride}, {Palazzi}, {Bassani}, {Bazzano}, {Bird}, {Charles}, {Dean},
  {Galaz}, {Gehrels}, {Landi}, {Malizia}, {Minniti}, {Panessa}, {Romero},
  {Stephen}, {Ubertini} and {Walter}}]{2008A&A...482..113M}
\bibinfo{author}{{Masetti}, N.}, \bibinfo{author}{{Mason}, E.},
  \bibinfo{author}{{Morelli}, L.}, \bibinfo{author}{{Cellone}, S.A.},
  \bibinfo{author}{{McBride}, V.A.}, \bibinfo{author}{{Palazzi}, E.},
  \bibinfo{author}{{Bassani}, L.}, \bibinfo{author}{{Bazzano}, A.},
  \bibinfo{author}{{Bird}, A.J.}, \bibinfo{author}{{Charles}, P.A.},
  \bibinfo{author}{{Dean}, A.J.}, \bibinfo{author}{{Galaz}, G.},
  \bibinfo{author}{{Gehrels}, N.}, \bibinfo{author}{{Landi}, R.},
  \bibinfo{author}{{Malizia}, A.}, \bibinfo{author}{{Minniti}, D.},
  \bibinfo{author}{{Panessa}, F.}, \bibinfo{author}{{Romero}, G.E.},
  \bibinfo{author}{{Stephen}, J.B.}, \bibinfo{author}{{Ubertini}, P.},
  \bibinfo{author}{{Walter}, R.}, \bibinfo{year}{2008}b.
\newblock \bibinfo{title}{{Unveiling the nature of INTEGRAL objects through
  optical spectroscopy. VI. A multi-observatory identification campaign}}.
\newblock \bibinfo{journal}{\aap} \bibinfo{volume}{482},
  \bibinfo{pages}{113--132}.
\newblock \DOIprefix\doi{10.1051/0004-6361:20079332},
  \href{http://arxiv.org/abs/0802.0988}{\tt arXiv:0802.0988}.
\bibitem[{{Masetti} et~al.(2013){Masetti}, {Parisi}, {Palazzi},
  {Jim{\'e}nez-Bail{\'o}n}, {Chavushyan}, {McBride}, {Rojas}, {Steward},
  {Bassani}, {Bazzano}, {Bird}, {Charles}, {Galaz}, {Landi}, {Malizia},
  {Mason}, {Minniti}, {Morelli}, {Schiavone}, {Stephen} and
  {Ubertini}}]{2013A&A...556A.120M}
\bibinfo{author}{{Masetti}, N.}, \bibinfo{author}{{Parisi}, P.},
  \bibinfo{author}{{Palazzi}, E.}, \bibinfo{author}{{Jim{\'e}nez-Bail{\'o}n},
  E.}, \bibinfo{author}{{Chavushyan}, V.}, \bibinfo{author}{{McBride}, V.},
  \bibinfo{author}{{Rojas}, A.F.}, \bibinfo{author}{{Steward}, L.},
  \bibinfo{author}{{Bassani}, L.}, \bibinfo{author}{{Bazzano}, A.},
  \bibinfo{author}{{Bird}, A.J.}, \bibinfo{author}{{Charles}, P.A.},
  \bibinfo{author}{{Galaz}, G.}, \bibinfo{author}{{Landi}, R.},
  \bibinfo{author}{{Malizia}, A.}, \bibinfo{author}{{Mason}, E.},
  \bibinfo{author}{{Minniti}, D.}, \bibinfo{author}{{Morelli}, L.},
  \bibinfo{author}{{Schiavone}, F.}, \bibinfo{author}{{Stephen}, J.B.},
  \bibinfo{author}{{Ubertini}, P.}, \bibinfo{year}{2013}.
\newblock \bibinfo{title}{{Unveiling the nature of INTEGRAL objects through
  optical spectroscopy. X. A new multi-year, multi-observatory campaign}}.
\newblock \bibinfo{journal}{\aap} \bibinfo{volume}{556}, \bibinfo{pages}{A120}.
\newblock \DOIprefix\doi{10.1051/0004-6361/201322026},
  \href{http://arxiv.org/abs/1307.2898}{\tt arXiv:1307.2898}.
\bibitem[{{Mathur}(2000)}]{2000MNRAS.314L..17M}
\bibinfo{author}{{Mathur}, S.}, \bibinfo{year}{2000}.
\newblock \bibinfo{title}{{Narrow-line Seyfert 1 galaxies and the evolution of
  galaxies and active galaxies}}.
\newblock \bibinfo{journal}{\mnras} \bibinfo{volume}{314},
  \bibinfo{pages}{L17--L20}.
\newblock \href{http://arxiv.org/abs/astro-ph/0003111}{\tt
  arXiv:astro-ph/0003111}.
\bibitem[{Mereminskiy et~al.(2016)Mereminskiy, Krivonos, Lutovinov, Sazonov,
  Revnivtsev and Sunyaev}]{Mereminskiy_2016}
\bibinfo{author}{Mereminskiy, I.A.}, \bibinfo{author}{Krivonos, R.A.},
  \bibinfo{author}{Lutovinov, A.A.}, \bibinfo{author}{Sazonov, S.Y.},
  \bibinfo{author}{Revnivtsev, M.G.}, \bibinfo{author}{Sunyaev, R.A.},
  \bibinfo{year}{2016}.
\newblock \bibinfo{title}{Integral/ibis deep extragalactic survey: M81, lmc and
  3c 273/coma fields}.
\newblock \bibinfo{journal}{Monthly Notices of the Royal Astronomical Society}
  \bibinfo{volume}{459}, \bibinfo{pages}{140--150}.
\bibitem[{{Molina} et~al.(2014){Molina}, {Bassani}, {Malizia}, {Bird},
  {Bazzano}, {Ubertini} and {Venturi}}]{2014A&A...565A...2M}
\bibinfo{author}{{Molina}, M.}, \bibinfo{author}{{Bassani}, L.},
  \bibinfo{author}{{Malizia}, A.}, \bibinfo{author}{{Bird}, A.J.},
  \bibinfo{author}{{Bazzano}, A.}, \bibinfo{author}{{Ubertini}, P.},
  \bibinfo{author}{{Venturi}, T.}, \bibinfo{year}{2014}.
\newblock \bibinfo{title}{{IGR J17488-2338: a newly discovered giant radio
  galaxy}}.
\newblock \bibinfo{journal}{\aap} \bibinfo{volume}{565}, \bibinfo{pages}{A2}.
\newblock \href{http://arxiv.org/abs/1403.1400}{\tt arXiv:1403.1400}.
\bibitem[{{Molina} et~al.(2013){Molina}, {Bassani}, {Malizia}, {Stephen},
  {Bird}, {Bazzano} and {Ubertini}}]{2013MNRAS.433.1687M}
\bibinfo{author}{{Molina}, M.}, \bibinfo{author}{{Bassani}, L.},
  \bibinfo{author}{{Malizia}, A.}, \bibinfo{author}{{Stephen}, J.B.},
  \bibinfo{author}{{Bird}, A.J.}, \bibinfo{author}{{Bazzano}, A.},
  \bibinfo{author}{{Ubertini}, P.}, \bibinfo{year}{2013}.
\newblock \bibinfo{title}{{Hard-X-ray spectra of active galactic nuclei in the
  INTEGRAL complete sample}}.
\newblock \bibinfo{journal}{\mnras} \bibinfo{volume}{433},
  \bibinfo{pages}{1687--1700}.
\newblock \href{http://arxiv.org/abs/1305.2722}{\tt arXiv:1305.2722}.
\bibitem[{{Molina} et~al.(2006){Molina}, {Malizia}, {Bassani}, {Bird}, {Dean},
  {Landi}, {de Rosa}, {Walter}, {Barlow} and {Clark}}]{2006MNRAS.371..821M}
\bibinfo{author}{{Molina}, M.}, \bibinfo{author}{{Malizia}, A.},
  \bibinfo{author}{{Bassani}, L.}, \bibinfo{author}{{Bird}, A.J.},
  \bibinfo{author}{{Dean}, A.J.}, \bibinfo{author}{{Landi}, R.},
  \bibinfo{author}{{de Rosa}, A.}, \bibinfo{author}{{Walter}, R.},
  \bibinfo{author}{{Barlow}, E.J.}, \bibinfo{author}{{Clark}, D.J.},
  \bibinfo{year}{2006}.
\newblock \bibinfo{title}{{INTEGRAL observations of active galactic nuclei
  obscured by the Galactic plane}}.
\newblock \bibinfo{journal}{\mnras} \bibinfo{volume}{371},
  \bibinfo{pages}{821--828}.
\newblock \href{http://arxiv.org/abs/astro-ph/0606488}{\tt
  arXiv:astro-ph/0606488}.
\bibitem[{{Molina} et~al.(2019){Molina}, {Malizia}, {Bassani}, {Ursini},
  {Bazzano} and {Ubertini}}]{2019MNRAS.484.2735M}
\bibinfo{author}{{Molina}, M.}, \bibinfo{author}{{Malizia}, A.},
  \bibinfo{author}{{Bassani}, L.}, \bibinfo{author}{{Ursini}, F.},
  \bibinfo{author}{{Bazzano}, A.}, \bibinfo{author}{{Ubertini}, P.},
  \bibinfo{year}{2019}.
\newblock \bibinfo{title}{{Swift/XRT-NuSTAR spectra of type 1 AGN: confirming
  INTEGRAL results on the high-energy cut-off}}.
\newblock \bibinfo{journal}{\mnras} \bibinfo{volume}{484},
  \bibinfo{pages}{2735--2746}.
\newblock \href{http://arxiv.org/abs/1901.04188}{\tt arXiv:1901.04188}.
\bibitem[{{Molina} et~al.(2015){Molina}, {Venturi}, {Malizia}, {Bassani},
  {Dallacasa}, {Lal}, {Bird} and {Ubertini}}]{2015MNRAS.451.2370M}
\bibinfo{author}{{Molina}, M.}, \bibinfo{author}{{Venturi}, T.},
  \bibinfo{author}{{Malizia}, A.}, \bibinfo{author}{{Bassani}, L.},
  \bibinfo{author}{{Dallacasa}, D.}, \bibinfo{author}{{Lal}, D.V.},
  \bibinfo{author}{{Bird}, A.J.}, \bibinfo{author}{{Ubertini}, P.},
  \bibinfo{year}{2015}.
\newblock \bibinfo{title}{{IGR J14488-4008: an X-ray peculiar giant radio
  galaxy discovered by INTEGRAL}}.
\newblock \bibinfo{journal}{\mnras} \bibinfo{volume}{451},
  \bibinfo{pages}{2370--2375}.
\newblock \href{http://arxiv.org/abs/1505.03684}{\tt arXiv:1505.03684}.
\bibitem[{{Murase}(2017)}]{2017nacs.book...15M}
\bibinfo{author}{{Murase}, K.}, \bibinfo{year}{2017}.
\newblock \bibinfo{title}{{Active Galactic Nuclei as High-Energy Neutrino
  Sources}}.
\newblock pp. \bibinfo{pages}{15--31}.
\bibitem[{Oh et~al.(2018)Oh, Koss, Markwardt, Schawinski, Baumgartner,
  Barthelmy, Cenko, Gehrels, Mushotzky, Petulante and et~al.}]{Oh_2018}
\bibinfo{author}{Oh, K.}, \bibinfo{author}{Koss, M.},
  \bibinfo{author}{Markwardt, C.B.}, \bibinfo{author}{Schawinski, K.},
  \bibinfo{author}{Baumgartner, W.H.}, \bibinfo{author}{Barthelmy, S.D.},
  \bibinfo{author}{Cenko, S.B.}, \bibinfo{author}{Gehrels, N.},
  \bibinfo{author}{Mushotzky, R.}, \bibinfo{author}{Petulante, A.},
  \bibinfo{author}{et~al.}, \bibinfo{year}{2018}.
\newblock \bibinfo{title}{The 105-month swift-bat all-sky hard x-ray survey}.
\newblock \bibinfo{journal}{The Astrophysical Journal Supplement Series}
  \bibinfo{volume}{235}, \bibinfo{pages}{4}.
\bibitem[{{Osterbrock} and {Pogge}(1985)}]{1985ApJ...297..166O}
\bibinfo{author}{{Osterbrock}, D.E.}, \bibinfo{author}{{Pogge}, R.W.},
  \bibinfo{year}{1985}.
\newblock \bibinfo{title}{{The spectra of narrow-line Seyfert 1 galaxies.}}
\newblock \bibinfo{journal}{\apj} \bibinfo{volume}{297},
  \bibinfo{pages}{166--176}.
\bibitem[{{Padovani} et~al.(2012){Padovani}, {Giommi} and
  {Rau}}]{2012MNRAS.422L..48P}
\bibinfo{author}{{Padovani}, P.}, \bibinfo{author}{{Giommi}, P.},
  \bibinfo{author}{{Rau}, A.}, \bibinfo{year}{2012}.
\newblock \bibinfo{title}{{The discovery of high-power high synchrotron peak
  blazars}}.
\newblock \bibinfo{journal}{\mnras} \bibinfo{volume}{422},
  \bibinfo{pages}{L48--L52}.
\newblock \href{http://arxiv.org/abs/1202.2236}{\tt arXiv:1202.2236}.
\bibitem[{{Paltani} et~al.(2008){Paltani}, {Walter}, {McHardy}, {Dwelly},
  {Steiner} and {Courvoisier}}]{2008A&A...485..707P}
\bibinfo{author}{{Paltani}, S.}, \bibinfo{author}{{Walter}, R.},
  \bibinfo{author}{{McHardy}, I.M.}, \bibinfo{author}{{Dwelly}, T.},
  \bibinfo{author}{{Steiner}, C.}, \bibinfo{author}{{Courvoisier}, T.J.L.},
  \bibinfo{year}{2008}.
\newblock \bibinfo{title}{{A deep INTEGRAL hard X-ray survey of the 3C 273/Coma
  region}}.
\newblock \bibinfo{journal}{\aap} \bibinfo{volume}{485},
  \bibinfo{pages}{707--718}.
\newblock \DOIprefix\doi{10.1051/0004-6361:200809450},
  \href{http://arxiv.org/abs/0805.0537}{\tt arXiv:0805.0537}.
\bibitem[{{Panessa} et~al.(2008){Panessa}, {Bassani}, {de Rosa}, {Bird},
  {Dean}, {Fiocchi}, {Malizia}, {Molina}, {Ubertini} and
  {Walter}}]{2008A&A...483..151P}
\bibinfo{author}{{Panessa}, F.}, \bibinfo{author}{{Bassani}, L.},
  \bibinfo{author}{{de Rosa}, A.}, \bibinfo{author}{{Bird}, A.J.},
  \bibinfo{author}{{Dean}, A.J.}, \bibinfo{author}{{Fiocchi}, M.},
  \bibinfo{author}{{Malizia}, A.}, \bibinfo{author}{{Molina}, M.},
  \bibinfo{author}{{Ubertini}, P.}, \bibinfo{author}{{Walter}, R.},
  \bibinfo{year}{2008}.
\newblock \bibinfo{title}{{The broad-band XMM-Newton and INTEGRAL spectra of
  bright type 1 Seyfert galaxies}}.
\newblock \bibinfo{journal}{\aap} \bibinfo{volume}{483},
  \bibinfo{pages}{151--160}.
\newblock \href{http://arxiv.org/abs/0803.0896}{\tt arXiv:0803.0896}.
\bibitem[{{Panessa} et~al.(2016){Panessa}, {Bassani}, {Landi}, {Bazzano},
  {Dallacasa}, {La Franca}, {Malizia}, {Venturi} and
  {Ubertini}}]{2016MNRAS.461.3153P}
\bibinfo{author}{{Panessa}, F.}, \bibinfo{author}{{Bassani}, L.},
  \bibinfo{author}{{Landi}, R.}, \bibinfo{author}{{Bazzano}, A.},
  \bibinfo{author}{{Dallacasa}, D.}, \bibinfo{author}{{La Franca}, F.},
  \bibinfo{author}{{Malizia}, A.}, \bibinfo{author}{{Venturi}, T.},
  \bibinfo{author}{{Ubertini}, P.}, \bibinfo{year}{2016}.
\newblock \bibinfo{title}{{The column density distribution of hard X-ray radio
  galaxies}}.
\newblock \bibinfo{journal}{\mnras} \bibinfo{volume}{461},
  \bibinfo{pages}{3153--3164}.
\newblock \href{http://arxiv.org/abs/1606.05504}{\tt arXiv:1606.05504}.
\bibitem[{{Panessa} et~al.(2011){Panessa}, {de Rosa}, {Bassani}, {Bazzano},
  {Bird}, {Landi}, {Malizia}, {Miniutti}, {Molina} and
  {Ubertini}}]{2011MNRAS.417.2426P}
\bibinfo{author}{{Panessa}, F.}, \bibinfo{author}{{de Rosa}, A.},
  \bibinfo{author}{{Bassani}, L.}, \bibinfo{author}{{Bazzano}, A.},
  \bibinfo{author}{{Bird}, A.}, \bibinfo{author}{{Landi}, R.},
  \bibinfo{author}{{Malizia}, A.}, \bibinfo{author}{{Miniutti}, G.},
  \bibinfo{author}{{Molina}, M.}, \bibinfo{author}{{Ubertini}, P.},
  \bibinfo{year}{2011}.
\newblock \bibinfo{title}{{Narrow-line Seyfert 1 galaxies at hard X-rays}}.
\newblock \bibinfo{journal}{\mnras} \bibinfo{volume}{417},
  \bibinfo{pages}{2426--2439}.
\newblock \DOIprefix\doi{10.1111/j.1365-2966.2011.19268.x},
  \href{http://arxiv.org/abs/1106.3023}{\tt arXiv:1106.3023}.
\bibitem[{{Panessa} et~al.(2015){Panessa}, {Tarchi}, {Castangia}, {Maiorano},
  {Bassani}, {Bicknell}, {Bazzano}, {Bird}, {Malizia} and
  {Ubertini}}]{2015MNRAS.447.1289P}
\bibinfo{author}{{Panessa}, F.}, \bibinfo{author}{{Tarchi}, A.},
  \bibinfo{author}{{Castangia}, P.}, \bibinfo{author}{{Maiorano}, E.},
  \bibinfo{author}{{Bassani}, L.}, \bibinfo{author}{{Bicknell}, G.},
  \bibinfo{author}{{Bazzano}, A.}, \bibinfo{author}{{Bird}, A.J.},
  \bibinfo{author}{{Malizia}, A.}, \bibinfo{author}{{Ubertini}, P.},
  \bibinfo{year}{2015}.
\newblock \bibinfo{title}{{The 1.4-GHz radio properties of hard X-ray-selected
  AGN}}.
\newblock \bibinfo{journal}{\mnras} \bibinfo{volume}{447},
  \bibinfo{pages}{1289--1298}.
\newblock \DOIprefix\doi{10.1093/mnras/stu2455},
  \href{http://arxiv.org/abs/1411.7829}{\tt arXiv:1411.7829}.
\bibitem[{{Perola} et~al.(2002){Perola}, {Matt}, {Cappi}, {Fiore}, {Guainazzi},
  {Maraschi}, {Petrucci} and {Piro}}]{2002A&A...389..802P}
\bibinfo{author}{{Perola}, G.C.}, \bibinfo{author}{{Matt}, G.},
  \bibinfo{author}{{Cappi}, M.}, \bibinfo{author}{{Fiore}, F.},
  \bibinfo{author}{{Guainazzi}, M.}, \bibinfo{author}{{Maraschi}, L.},
  \bibinfo{author}{{Petrucci}, P.O.}, \bibinfo{author}{{Piro}, L.},
  \bibinfo{year}{2002}.
\newblock \bibinfo{title}{{Compton reflection and iron fluorescence in BeppoSAX
  observations of Seyfert type 1 galaxies}}.
\newblock \bibinfo{journal}{\aap} \bibinfo{volume}{389},
  \bibinfo{pages}{802--811}.
\newblock \href{http://arxiv.org/abs/astro-ph/0205045}{\tt
  arXiv:astro-ph/0205045}.
\bibitem[{{Petrucci} et~al.(2001){Petrucci}, {Haardt}, {Maraschi}, {Grandi},
  {Malzac}, {Matt}, {Nicastro}, {Piro}, {Perola} and {De
  Rosa}}]{2001ApJ...556..716P}
\bibinfo{author}{{Petrucci}, P.O.}, \bibinfo{author}{{Haardt}, F.},
  \bibinfo{author}{{Maraschi}, L.}, \bibinfo{author}{{Grandi}, P.},
  \bibinfo{author}{{Malzac}, J.}, \bibinfo{author}{{Matt}, G.},
  \bibinfo{author}{{Nicastro}, F.}, \bibinfo{author}{{Piro}, L.},
  \bibinfo{author}{{Perola}, G.C.}, \bibinfo{author}{{De Rosa}, A.},
  \bibinfo{year}{2001}.
\newblock \bibinfo{title}{{Testing Comptonization Models Using BeppoSAX
  Observations of Seyfert 1 Galaxies}}.
\newblock \bibinfo{journal}{\apj} \bibinfo{volume}{556},
  \bibinfo{pages}{716--726}.
\newblock \href{http://arxiv.org/abs/astro-ph/0101219}{\tt
  arXiv:astro-ph/0101219}.
\bibitem[{{Pian} et~al.(2005){Pian}, {Foschini}, {Beckmann},
  {Sillanp{\"a}{\"a}}, {Soldi}, {Tagliaferri}, {Takalo}, {Barr}, {Ghisellini},
  {Malaguti}, {Maraschi}, {Palumbo}, {Treves}, {Courvoisier}, {Di Cocco},
  {Gehrels}, {Giommi}, {Hudec}, {Lindfors}, {Marcowith}, {Nilsson}, {Pasanen},
  {Pursimo}, {Raiteri}, {Savolainen}, {Sikora}, {Tornikoski}, {Tosti},
  {T{\"u}rler}, {Valtaoja}, {Villata} and {Walter}}]{2005A&A...429..427P}
\bibinfo{author}{{Pian}, E.}, \bibinfo{author}{{Foschini}, L.},
  \bibinfo{author}{{Beckmann}, V.}, \bibinfo{author}{{Sillanp{\"a}{\"a}}, A.},
  \bibinfo{author}{{Soldi}, S.}, \bibinfo{author}{{Tagliaferri}, G.},
  \bibinfo{author}{{Takalo}, L.}, \bibinfo{author}{{Barr}, P.},
  \bibinfo{author}{{Ghisellini}, G.}, \bibinfo{author}{{Malaguti}, G.},
  \bibinfo{author}{{Maraschi}, L.}, \bibinfo{author}{{Palumbo}, G.G.C.},
  \bibinfo{author}{{Treves}, A.}, \bibinfo{author}{{Courvoisier}, T.J.L.},
  \bibinfo{author}{{Di Cocco}, G.}, \bibinfo{author}{{Gehrels}, N.},
  \bibinfo{author}{{Giommi}, P.}, \bibinfo{author}{{Hudec}, R.},
  \bibinfo{author}{{Lindfors}, E.}, \bibinfo{author}{{Marcowith}, A.},
  \bibinfo{author}{{Nilsson}, K.}, \bibinfo{author}{{Pasanen}, M.},
  \bibinfo{author}{{Pursimo}, T.}, \bibinfo{author}{{Raiteri}, C.M.},
  \bibinfo{author}{{Savolainen}, T.}, \bibinfo{author}{{Sikora}, M.},
  \bibinfo{author}{{Tornikoski}, M.}, \bibinfo{author}{{Tosti}, G.},
  \bibinfo{author}{{T{\"u}rler}, M.}, \bibinfo{author}{{Valtaoja}, E.},
  \bibinfo{author}{{Villata}, M.}, \bibinfo{author}{{Walter}, R.},
  \bibinfo{year}{2005}.
\newblock \bibinfo{title}{{INTEGRAL observations of the field of the BL
  Lacertae object S5 0716+714}}.
\newblock \bibinfo{journal}{\aap} \bibinfo{volume}{429},
  \bibinfo{pages}{427--431}.
\newblock \href{http://arxiv.org/abs/astro-ph/0408580}{\tt
  arXiv:astro-ph/0408580}.
\bibitem[{{Pian} et~al.(2006){Pian}, {Foschini}, {Beckmann}, {Soldi},
  {T{\"u}rler}, {Gehrels}, {Ghisellini}, {Giommi}, {Maraschi}, {Pursimo},
  {Raiteri}, {Tagliaferri}, {Tornikoski}, {Tosti}, {Treves}, {Villata}, {Barr},
  {Courvoisier}, {Di Cocco}, {Hudec}, {Fuhrmann}, {Malaguti}, {Persic},
  {Tavecchio} and {Walter}}]{2006A&A...449L..21P}
\bibinfo{author}{{Pian}, E.}, \bibinfo{author}{{Foschini}, L.},
  \bibinfo{author}{{Beckmann}, V.}, \bibinfo{author}{{Soldi}, S.},
  \bibinfo{author}{{T{\"u}rler}, M.}, \bibinfo{author}{{Gehrels}, N.},
  \bibinfo{author}{{Ghisellini}, G.}, \bibinfo{author}{{Giommi}, P.},
  \bibinfo{author}{{Maraschi}, L.}, \bibinfo{author}{{Pursimo}, T.},
  \bibinfo{author}{{Raiteri}, C.M.}, \bibinfo{author}{{Tagliaferri}, G.},
  \bibinfo{author}{{Tornikoski}, M.}, \bibinfo{author}{{Tosti}, G.},
  \bibinfo{author}{{Treves}, A.}, \bibinfo{author}{{Villata}, M.},
  \bibinfo{author}{{Barr}, P.}, \bibinfo{author}{{Courvoisier}, T.J.L.},
  \bibinfo{author}{{Di Cocco}, G.}, \bibinfo{author}{{Hudec}, R.},
  \bibinfo{author}{{Fuhrmann}, L.}, \bibinfo{author}{{Malaguti}, G.},
  \bibinfo{author}{{Persic}, M.}, \bibinfo{author}{{Tavecchio}, F.},
  \bibinfo{author}{{Walter}, R.}, \bibinfo{year}{2006}.
\newblock \bibinfo{title}{{INTEGRAL observations of the blazar 3C 454.3 in
  outburst}}.
\newblock \bibinfo{journal}{\aap} \bibinfo{volume}{449},
  \bibinfo{pages}{L21--L25}.
\newblock \href{http://arxiv.org/abs/astro-ph/0602268}{\tt
  arXiv:astro-ph/0602268}.
\bibitem[{{Pian} et~al.(2014){Pian}, {T{\"u}rler}, {Fiocchi}, {Boissay},
  {Bazzano}, {Foschini}, {Tavecchio}, {Bianchin}, {Castignani}, {Ferrigno},
  {Raiteri}, {Villata}, {Beckmann}, {D'Ammand o}, {Hudec}, {Malaguti},
  {Maraschi}, {Pursimo}, {Romano}, {Soldi}, {Stamerra}, {Treves}, {Ubertini},
  {Vercellone} and {Walter}}]{2014A&A...570A..77P}
\bibinfo{author}{{Pian}, E.}, \bibinfo{author}{{T{\"u}rler}, M.},
  \bibinfo{author}{{Fiocchi}, M.}, \bibinfo{author}{{Boissay}, R.},
  \bibinfo{author}{{Bazzano}, A.}, \bibinfo{author}{{Foschini}, L.},
  \bibinfo{author}{{Tavecchio}, F.}, \bibinfo{author}{{Bianchin}, V.},
  \bibinfo{author}{{Castignani}, G.}, \bibinfo{author}{{Ferrigno}, C.},
  \bibinfo{author}{{Raiteri}, C.M.}, \bibinfo{author}{{Villata}, M.},
  \bibinfo{author}{{Beckmann}, V.}, \bibinfo{author}{{D'Ammand o}, F.},
  \bibinfo{author}{{Hudec}, R.}, \bibinfo{author}{{Malaguti}, G.},
  \bibinfo{author}{{Maraschi}, L.}, \bibinfo{author}{{Pursimo}, T.},
  \bibinfo{author}{{Romano}, P.}, \bibinfo{author}{{Soldi}, S.},
  \bibinfo{author}{{Stamerra}, A.}, \bibinfo{author}{{Treves}, A.},
  \bibinfo{author}{{Ubertini}, P.}, \bibinfo{author}{{Vercellone}, S.},
  \bibinfo{author}{{Walter}, R.}, \bibinfo{year}{2014}.
\newblock \bibinfo{title}{{An active state of the BL Lacertae object Markarian
  421 detected by INTEGRAL in April 2013}}.
\newblock \bibinfo{journal}{\aap} \bibinfo{volume}{570}, \bibinfo{pages}{A77}.
\newblock \href{http://arxiv.org/abs/1307.0558}{\tt arXiv:1307.0558}.
\bibitem[{{Pian} et~al.(2011){Pian}, {Ubertini}, {Bazzano}, {Beckmann},
  {Eckert}, {Ghisellini}, {Pursimo}, {Tagliaferri}, {Tavecchio}, {T{\"u}rler},
  {Bianchi}, {Bianchin}, {Hudec}, {Maraschi}, {Raiteri}, {Soldi}, {Treves} and
  {Villata}}]{2011A&A...526A.125P}
\bibinfo{author}{{Pian}, E.}, \bibinfo{author}{{Ubertini}, P.},
  \bibinfo{author}{{Bazzano}, A.}, \bibinfo{author}{{Beckmann}, V.},
  \bibinfo{author}{{Eckert}, D.}, \bibinfo{author}{{Ghisellini}, G.},
  \bibinfo{author}{{Pursimo}, T.}, \bibinfo{author}{{Tagliaferri}, G.},
  \bibinfo{author}{{Tavecchio}, F.}, \bibinfo{author}{{T{\"u}rler}, M.},
  \bibinfo{author}{{Bianchi}, S.}, \bibinfo{author}{{Bianchin}, V.},
  \bibinfo{author}{{Hudec}, R.}, \bibinfo{author}{{Maraschi}, L.},
  \bibinfo{author}{{Raiteri}, C.M.}, \bibinfo{author}{{Soldi}, S.},
  \bibinfo{author}{{Treves}, A.}, \bibinfo{author}{{Villata}, M.},
  \bibinfo{year}{2011}.
\newblock \bibinfo{title}{{INTEGRAL observations of the GeV blazar PKS 1502+106
  and the hard X-ray bright Seyfert galaxy Mkn 841}}.
\newblock \bibinfo{journal}{\aap} \bibinfo{volume}{526}, \bibinfo{pages}{A125}.
\newblock \href{http://arxiv.org/abs/1011.3224}{\tt arXiv:1011.3224}.
\bibitem[{{Pian} et~al.(1998){Pian}, {Vacanti}, {Tagliaferri}, {Ghisellini},
  {Maraschi}, {Treves}, {Urry}, {Fiore}, {Giommi}, {Palazzi}, {Chiappetti} and
  {Sambruna}}]{1998ApJ...492L..17P}
\bibinfo{author}{{Pian}, E.}, \bibinfo{author}{{Vacanti}, G.},
  \bibinfo{author}{{Tagliaferri}, G.}, \bibinfo{author}{{Ghisellini}, G.},
  \bibinfo{author}{{Maraschi}, L.}, \bibinfo{author}{{Treves}, A.},
  \bibinfo{author}{{Urry}, C.M.}, \bibinfo{author}{{Fiore}, F.},
  \bibinfo{author}{{Giommi}, P.}, \bibinfo{author}{{Palazzi}, E.},
  \bibinfo{author}{{Chiappetti}, L.}, \bibinfo{author}{{Sambruna}, R.M.},
  \bibinfo{year}{1998}.
\newblock \bibinfo{title}{{BeppoSAX Observations of Unprecedented Synchrotron
  Activity in the BL Lacertae Object Markarian 501}}.
\newblock \bibinfo{journal}{\apjl} \bibinfo{volume}{492},
  \bibinfo{pages}{L17--L20}.
\newblock \href{http://arxiv.org/abs/astro-ph/9710331}{\tt
  arXiv:astro-ph/9710331}.
\bibitem[{{Pounds} et~al.(1995){Pounds}, {Done} and
  {Osborne}}]{1995MNRAS.277L...5P}
\bibinfo{author}{{Pounds}, K.A.}, \bibinfo{author}{{Done}, C.},
  \bibinfo{author}{{Osborne}, J.P.}, \bibinfo{year}{1995}.
\newblock \bibinfo{title}{{RE 1034+39: a high-state Seyfert galaxy?}}
\newblock \bibinfo{journal}{\mnras} \bibinfo{volume}{277},
  \bibinfo{pages}{L5--L10}.
\bibitem[{{Ranalli} et~al.(2013){Ranalli}, {Comastri}, {Vignali}, {Carrera},
  {Cappelluti}, {Gilli}, {Puccetti}, {Brand t}, {Brunner}, {Brusa},
  {Georgantopoulos}, {Iwasawa} and {Mainieri}}]{2013A&A...555A..42R}
\bibinfo{author}{{Ranalli}, P.}, \bibinfo{author}{{Comastri}, A.},
  \bibinfo{author}{{Vignali}, C.}, \bibinfo{author}{{Carrera}, F.J.},
  \bibinfo{author}{{Cappelluti}, N.}, \bibinfo{author}{{Gilli}, R.},
  \bibinfo{author}{{Puccetti}, S.}, \bibinfo{author}{{Brand t}, W.N.},
  \bibinfo{author}{{Brunner}, H.}, \bibinfo{author}{{Brusa}, M.},
  \bibinfo{author}{{Georgantopoulos}, I.}, \bibinfo{author}{{Iwasawa}, K.},
  \bibinfo{author}{{Mainieri}, V.}, \bibinfo{year}{2013}.
\newblock \bibinfo{title}{{The XMM deep survey in the CDF-S. III. Point source
  catalogue and number counts in the hard X-rays}}.
\newblock \bibinfo{journal}{\aap} \bibinfo{volume}{555}, \bibinfo{pages}{A42}.
\newblock \href{http://arxiv.org/abs/1304.5717}{\tt arXiv:1304.5717}.
\bibitem[{Ricci et~al.(2017)Ricci, Trakhtenbrot, Koss, Ueda, Schawinski, Oh,
  Lamperti, Mushotzky, Treister, Ho and et~al.}]{Ricci_2017}
\bibinfo{author}{Ricci, C.}, \bibinfo{author}{Trakhtenbrot, B.},
  \bibinfo{author}{Koss, M.J.}, \bibinfo{author}{Ueda, Y.},
  \bibinfo{author}{Schawinski, K.}, \bibinfo{author}{Oh, K.},
  \bibinfo{author}{Lamperti, I.}, \bibinfo{author}{Mushotzky, R.},
  \bibinfo{author}{Treister, E.}, \bibinfo{author}{Ho, L.C.},
  \bibinfo{author}{et~al.}, \bibinfo{year}{2017}.
\newblock \bibinfo{title}{The close environments of accreting massive black
  holes are shaped by radiative feedback}.
\newblock \bibinfo{journal}{Nature} \bibinfo{volume}{549},
  \bibinfo{pages}{488--491}.
\newblock \URLprefix \url{http://dx.doi.org/10.1038/nature23906},
  \DOIprefix\doi{10.1038/nature23906}.
\bibitem[{{Ricci} et~al.(2015){Ricci}, {Ueda}, {Koss}, {Trakhtenbrot}, {Bauer}
  and {Gandhi}}]{2015ApJ...815L..13R}
\bibinfo{author}{{Ricci}, C.}, \bibinfo{author}{{Ueda}, Y.},
  \bibinfo{author}{{Koss}, M.J.}, \bibinfo{author}{{Trakhtenbrot}, B.},
  \bibinfo{author}{{Bauer}, F.E.}, \bibinfo{author}{{Gandhi}, P.},
  \bibinfo{year}{2015}.
\newblock \bibinfo{title}{{Compton-thick Accretion in the Local Universe}}.
\newblock \bibinfo{journal}{\apjl} \bibinfo{volume}{815}, \bibinfo{pages}{L13}.
\newblock \href{http://arxiv.org/abs/1603.04852}{\tt arXiv:1603.04852}.
\bibitem[{{Righi} et~al.(2019){Righi}, {Tavecchio} and
  {Inoue}}]{2019MNRAS.483L.127R}
\bibinfo{author}{{Righi}, C.}, \bibinfo{author}{{Tavecchio}, F.},
  \bibinfo{author}{{Inoue}, S.}, \bibinfo{year}{2019}.
\newblock \bibinfo{title}{{Neutrino emission from BL Lac objects: the role of
  radiatively inefficient accretion flows}}.
\newblock \bibinfo{journal}{\mnras} \bibinfo{volume}{483},
  \bibinfo{pages}{L127--L131}.
\newblock \href{http://arxiv.org/abs/1807.10506}{\tt arXiv:1807.10506}.
\bibitem[{{Risaliti} et~al.(1999){Risaliti}, {Maiolino} and
  {Salvati}}]{1999ApJ...522..157R}
\bibinfo{author}{{Risaliti}, G.}, \bibinfo{author}{{Maiolino}, R.},
  \bibinfo{author}{{Salvati}, M.}, \bibinfo{year}{1999}.
\newblock \bibinfo{title}{{The Distribution of Absorbing Column Densities among
  Seyfert 2 Galaxies}}.
\newblock \bibinfo{journal}{\apj} \bibinfo{volume}{522},
  \bibinfo{pages}{157--164}.
\newblock \href{http://arxiv.org/abs/astro-ph/9902377}{\tt
  arXiv:astro-ph/9902377}.
\bibitem[{{Rojas} et~al.(2017){Rojas}, {Masetti}, {Minniti},
  {Jim{\'e}nez-Bail{\'o}n}, {Chavushyan}, {Hau}, {McBride}, {Bassani},
  {Bazzano}, {Bird}, {Galaz}, {Gavignaud}, {Landi}, {Malizia}, {Morelli},
  {Palazzi}, {Pati{\~n}o-{\'A}lvarez}, {Stephen} and
  {Ubertini}}]{2017A&A...602A.124R}
\bibinfo{author}{{Rojas}, A.F.}, \bibinfo{author}{{Masetti}, N.},
  \bibinfo{author}{{Minniti}, D.}, \bibinfo{author}{{Jim{\'e}nez-Bail{\'o}n},
  E.}, \bibinfo{author}{{Chavushyan}, V.}, \bibinfo{author}{{Hau}, G.},
  \bibinfo{author}{{McBride}, V.A.}, \bibinfo{author}{{Bassani}, L.},
  \bibinfo{author}{{Bazzano}, A.}, \bibinfo{author}{{Bird}, A.J.},
  \bibinfo{author}{{Galaz}, G.}, \bibinfo{author}{{Gavignaud}, I.},
  \bibinfo{author}{{Landi}, R.}, \bibinfo{author}{{Malizia}, A.},
  \bibinfo{author}{{Morelli}, L.}, \bibinfo{author}{{Palazzi}, E.},
  \bibinfo{author}{{Pati{\~n}o-{\'A}lvarez}, V.}, \bibinfo{author}{{Stephen},
  J.B.}, \bibinfo{author}{{Ubertini}, P.}, \bibinfo{year}{2017}.
\newblock \bibinfo{title}{{The nature of 50 Palermo Swift-BAT hard X-ray
  objects through optical spectroscopy}}.
\newblock \bibinfo{journal}{\aap} \bibinfo{volume}{602}, \bibinfo{pages}{A124}.
\newblock \DOIprefix\doi{10.1051/0004-6361/201629463},
  \href{http://arxiv.org/abs/1702.01629}{\tt arXiv:1702.01629}.
\bibitem[{{Sambruna} et~al.(1996){Sambruna}, {Maraschi} and
  {Urry}}]{1996ApJ...463..444S}
\bibinfo{author}{{Sambruna}, R.M.}, \bibinfo{author}{{Maraschi}, L.},
  \bibinfo{author}{{Urry}, C.M.}, \bibinfo{year}{1996}.
\newblock \bibinfo{title}{{On the Spectral Energy Distributions of Blazars}}.
\newblock \bibinfo{journal}{\apj} \bibinfo{volume}{463}, \bibinfo{pages}{444}.
\newblock \DOIprefix\doi{10.1086/177260}.
\bibitem[{{Sazonov} et~al.(2015){Sazonov}, {Churazov} and
  {Krivonos}}]{2015MNRAS.454.1202S}
\bibinfo{author}{{Sazonov}, S.}, \bibinfo{author}{{Churazov}, E.},
  \bibinfo{author}{{Krivonos}, R.}, \bibinfo{year}{2015}.
\newblock \bibinfo{title}{{Does the obscured AGN fraction really depend on
  luminosity?}}
\newblock \bibinfo{journal}{\mnras} \bibinfo{volume}{454},
  \bibinfo{pages}{1202--1220}.
\newblock \href{http://arxiv.org/abs/1509.01259}{\tt arXiv:1509.01259}.
\bibitem[{{Sazonov} et~al.(2005){Sazonov}, {Churazov}, {Revnivtsev},
  {Vikhlinin} and {Sunyaev}}]{2005A&A...444L..37S}
\bibinfo{author}{{Sazonov}, S.}, \bibinfo{author}{{Churazov}, E.},
  \bibinfo{author}{{Revnivtsev}, M.}, \bibinfo{author}{{Vikhlinin}, A.},
  \bibinfo{author}{{Sunyaev}, R.}, \bibinfo{year}{2005}.
\newblock \bibinfo{title}{{Identification of 8 INTEGRAL hard X-ray sources with
  Chandra}}.
\newblock \bibinfo{journal}{\aap} \bibinfo{volume}{444},
  \bibinfo{pages}{L37--L40}.
\newblock \DOIprefix\doi{10.1051/0004-6361:200500205},
  \href{http://arxiv.org/abs/astro-ph/0508593}{\tt arXiv:astro-ph/0508593}.
\bibitem[{{Sazonov} et~al.(2010){Sazonov}, {Churazov}, {Krivonos}, {Revnivtsev}
  and {Sunyaev}}]{2010int..workE...6S}
\bibinfo{author}{{Sazonov}, S.}, \bibinfo{author}{{Churazov}, E.M.},
  \bibinfo{author}{{Krivonos}, R.}, \bibinfo{author}{{Revnivtsev}, M.},
  \bibinfo{author}{{Sunyaev}, R.}, \bibinfo{year}{2010}.
\newblock \bibinfo{title}{{Statistical properties of local AGN based on the
  INTEGRAL/IBIS 7-year all-sky hard X-ray survey}}, in:
  \bibinfo{booktitle}{Eighth Integral Workshop. The Restless Gamma-ray Universe
  (INTEGRAL 2010)}, p.~\bibinfo{pages}{6}.
\bibitem[{{Sazonov} et~al.(2008a){Sazonov}, {Krivonos}, {Revnivtsev},
  {Churazov} and {Sunyaev}}]{2008A&A...482..517S}
\bibinfo{author}{{Sazonov}, S.}, \bibinfo{author}{{Krivonos}, R.},
  \bibinfo{author}{{Revnivtsev}, M.}, \bibinfo{author}{{Churazov}, E.},
  \bibinfo{author}{{Sunyaev}, R.}, \bibinfo{year}{2008}a.
\newblock \bibinfo{title}{{Cumulative hard X-ray spectrum of local AGN: a link
  to the cosmic X-ray background}}.
\newblock \bibinfo{journal}{\aap} \bibinfo{volume}{482},
  \bibinfo{pages}{517--527}.
\newblock \href{http://arxiv.org/abs/0708.3215}{\tt arXiv:0708.3215}.
\bibitem[{{Sazonov} et~al.(2008b){Sazonov}, {Revnivtsev}, {Burenin},
  {Churazov}, {Sunyaev}, {Forman} and {Murray}}]{2008A&A...487..509S}
\bibinfo{author}{{Sazonov}, S.}, \bibinfo{author}{{Revnivtsev}, M.},
  \bibinfo{author}{{Burenin}, R.}, \bibinfo{author}{{Churazov}, E.},
  \bibinfo{author}{{Sunyaev}, R.}, \bibinfo{author}{{Forman}, W.R.},
  \bibinfo{author}{{Murray}, S.S.}, \bibinfo{year}{2008}b.
\newblock \bibinfo{title}{{Discovery of heavily-obscured AGN among seven
  INTEGRAL hard X-ray sources observed by Chandra}}.
\newblock \bibinfo{journal}{\aap} \bibinfo{volume}{487},
  \bibinfo{pages}{509--517}.
\newblock \href{http://arxiv.org/abs/0802.0928}{\tt arXiv:0802.0928}.
\bibitem[{{Sazonov} et~al.(2007){Sazonov}, {Revnivtsev}, {Krivonos}, {Churazov}
  and {Sunyaev}}]{2007A&A...462...57S}
\bibinfo{author}{{Sazonov}, S.}, \bibinfo{author}{{Revnivtsev}, M.},
  \bibinfo{author}{{Krivonos}, R.}, \bibinfo{author}{{Churazov}, E.},
  \bibinfo{author}{{Sunyaev}, R.}, \bibinfo{year}{2007}.
\newblock \bibinfo{title}{{Hard X-ray luminosity function and absorption
  distribution of nearby AGN: INTEGRAL all-sky survey}}.
\newblock \bibinfo{journal}{\aap} \bibinfo{volume}{462},
  \bibinfo{pages}{57--66}.
\newblock \href{http://arxiv.org/abs/astro-ph/0608418}{\tt
  arXiv:astro-ph/0608418}.
\bibitem[{{Sazonov} et~al.(2012){Sazonov}, {Willner}, {Goulding}, {Hickox},
  {Gorjian}, {Werner}, {Churazov}, {Krivonos}, {Revnivtsev}, {Sunyaev},
  {Jones}, {Murray}, {Vikhlinin}, {Fabian} and {Forman}}]{2012ApJ...757..181S}
\bibinfo{author}{{Sazonov}, S.}, \bibinfo{author}{{Willner}, S.P.},
  \bibinfo{author}{{Goulding}, A.D.}, \bibinfo{author}{{Hickox}, R.C.},
  \bibinfo{author}{{Gorjian}, V.}, \bibinfo{author}{{Werner}, M.W.},
  \bibinfo{author}{{Churazov}, E.}, \bibinfo{author}{{Krivonos}, R.},
  \bibinfo{author}{{Revnivtsev}, M.}, \bibinfo{author}{{Sunyaev}, R.},
  \bibinfo{author}{{Jones}, C.}, \bibinfo{author}{{Murray}, S.S.},
  \bibinfo{author}{{Vikhlinin}, A.}, \bibinfo{author}{{Fabian}, A.C.},
  \bibinfo{author}{{Forman}, W.R.}, \bibinfo{year}{2012}.
\newblock \bibinfo{title}{{Contribution of the Accretion Disk, Hot Corona, and
  Obscuring Torus to the Luminosity of Seyfert Galaxies: INTEGRAL and Spitzer
  Observations}}.
\newblock \bibinfo{journal}{\apj} \bibinfo{volume}{757}, \bibinfo{pages}{181}.
\newblock \href{http://arxiv.org/abs/1208.1612}{\tt arXiv:1208.1612}.
\bibitem[{{Sazonov} et~al.(2004a){Sazonov}, {Ostriker} and
  {Sunyaev}}]{2004MNRAS.347..144S}
\bibinfo{author}{{Sazonov}, S.Y.}, \bibinfo{author}{{Ostriker}, J.P.},
  \bibinfo{author}{{Sunyaev}, R.A.}, \bibinfo{year}{2004}a.
\newblock \bibinfo{title}{{Quasars: the characteristic spectrum and the induced
  radiative heating}}.
\newblock \bibinfo{journal}{\mnras} \bibinfo{volume}{347},
  \bibinfo{pages}{144--156}.
\newblock \href{http://arxiv.org/abs/astro-ph/0305233}{\tt
  arXiv:astro-ph/0305233}.
\bibitem[{{Sazonov} and {Revnivtsev}(2004)}]{2004A&A...423..469S}
\bibinfo{author}{{Sazonov}, S.Y.}, \bibinfo{author}{{Revnivtsev}, M.G.},
  \bibinfo{year}{2004}.
\newblock \bibinfo{title}{{Statistical properties of local active galactic
  nuclei inferred from the RXTE 3-20 keV all-sky survey}}.
\newblock \bibinfo{journal}{\aap} \bibinfo{volume}{423},
  \bibinfo{pages}{469--480}.
\newblock \href{http://arxiv.org/abs/astro-ph/0402415}{\tt
  arXiv:astro-ph/0402415}.
\bibitem[{{Sazonov} et~al.(2004b){Sazonov}, {Revnivtsev}, {Lutovinov},
  {Sunyaev} and {Grebenev}}]{2004A&A...421L..21S}
\bibinfo{author}{{Sazonov}, S.Y.}, \bibinfo{author}{{Revnivtsev}, M.G.},
  \bibinfo{author}{{Lutovinov}, A.A.}, \bibinfo{author}{{Sunyaev}, R.A.},
  \bibinfo{author}{{Grebenev}, S.A.}, \bibinfo{year}{2004}b.
\newblock \bibinfo{title}{{Broadband X-ray spectrum of GRS 1734-292, a luminous
  Seyfert 1 galaxy behind the Galactic Center}}.
\newblock \bibinfo{journal}{\aap} \bibinfo{volume}{421},
  \bibinfo{pages}{L21--L24}.
\newblock \href{http://arxiv.org/abs/astro-ph/0405259}{\tt
  arXiv:astro-ph/0405259}.
\bibitem[{{Semena} et~al.(2019){Semena}, {Sazonov} and
  {Krivonos}}]{2019AstL...45..490S}
\bibinfo{author}{{Semena}, A.N.}, \bibinfo{author}{{Sazonov}, S.Y.},
  \bibinfo{author}{{Krivonos}, R.A.}, \bibinfo{year}{2019}.
\newblock \bibinfo{title}{{Spectral Properties of Heavily Obscured Seyfert
  Galaxies from the INTEGRAL All-Sky Survey}}.
\newblock \bibinfo{journal}{Astronomy Letters} \bibinfo{volume}{45},
  \bibinfo{pages}{490--520}.
\newblock \DOIprefix\doi{10.1134/S1063773719080085}.
\bibitem[{{Sinha} et~al.(2015){Sinha}, {Shukla}, {Misra}, {Chitnis}, {Rao} and
  {Acharya}}]{2015A&A...580A.100S}
\bibinfo{author}{{Sinha}, A.}, \bibinfo{author}{{Shukla}, A.},
  \bibinfo{author}{{Misra}, R.}, \bibinfo{author}{{Chitnis}, V.R.},
  \bibinfo{author}{{Rao}, A.R.}, \bibinfo{author}{{Acharya}, B.S.},
  \bibinfo{year}{2015}.
\newblock \bibinfo{title}{{Underlying particle spectrum of Mkn 421 during the
  huge X-ray flare in April 2013}}.
\newblock \bibinfo{journal}{\aap} \bibinfo{volume}{580}, \bibinfo{pages}{A100}.
\newblock \href{http://arxiv.org/abs/1506.03629}{\tt arXiv:1506.03629}.
\bibitem[{{Smith} et~al.(1988){Smith}, {Elston}, {Berriman}, {Allen} and
  {Balonek}}]{1988ApJ...326L..39S}
\bibinfo{author}{{Smith}, P.S.}, \bibinfo{author}{{Elston}, R.},
  \bibinfo{author}{{Berriman}, G.}, \bibinfo{author}{{Allen}, R.G.},
  \bibinfo{author}{{Balonek}, T.J.}, \bibinfo{year}{1988}.
\newblock \bibinfo{title}{{Evidence for Accretion Disks in Highly Polarized
  Quasars}}.
\newblock \bibinfo{journal}{\apjl} \bibinfo{volume}{326}, \bibinfo{pages}{L39}.
\newblock \DOIprefix\doi{10.1086/185119}.
\bibitem[{{Soldi} et~al.(2005){Soldi}, {Beckmann}, {Bassani}, {Courvoisier},
  {Landi}, {Malizia}, {Dean}, {de Rosa}, {Fabian} and
  {Walter}}]{2005A&A...444..431S}
\bibinfo{author}{{Soldi}, S.}, \bibinfo{author}{{Beckmann}, V.},
  \bibinfo{author}{{Bassani}, L.}, \bibinfo{author}{{Courvoisier}, T.J.L.},
  \bibinfo{author}{{Landi}, R.}, \bibinfo{author}{{Malizia}, A.},
  \bibinfo{author}{{Dean}, A.J.}, \bibinfo{author}{{de Rosa}, A.},
  \bibinfo{author}{{Fabian}, A.C.}, \bibinfo{author}{{Walter}, R.},
  \bibinfo{year}{2005}.
\newblock \bibinfo{title}{{INTEGRAL observations of six AGN in the Galactic
  Plane}}.
\newblock \bibinfo{journal}{\aap} \bibinfo{volume}{444},
  \bibinfo{pages}{431--441}.
\newblock \DOIprefix\doi{10.1051/0004-6361:20053875},
  \href{http://arxiv.org/abs/astro-ph/0509123}{\tt arXiv:astro-ph/0509123}.
\bibitem[{{Soldi} et~al.(2006){Soldi}, {Beckmann}, {Bassani}, {Courvoisier},
  {Landi}, {Malizia}, {Dean}, {de Rosa}, {Fabian} and
  {Walter}}]{2006ESASP.604..667S}
\bibinfo{author}{{Soldi}, S.}, \bibinfo{author}{{Beckmann}, V.},
  \bibinfo{author}{{Bassani}, L.}, \bibinfo{author}{{Courvoisier}, T.J.L.},
  \bibinfo{author}{{Landi}, R.}, \bibinfo{author}{{Malizia}, A.},
  \bibinfo{author}{{Dean}, A.J.}, \bibinfo{author}{{de Rosa}, A.},
  \bibinfo{author}{{Fabian}, A.C.}, \bibinfo{author}{{Walter}, R.},
  \bibinfo{year}{2006}.
\newblock \bibinfo{title}{{INTEGRAL Observations of Six AGN in the Galactic
  Plane}}, in: \bibinfo{editor}{{Wilson}, A.} (Ed.), \bibinfo{booktitle}{The
  X-ray Universe 2005}, p. \bibinfo{pages}{667}.
\bibitem[{{Soldi} et~al.(2008){Soldi}, {T{\"u}rler}, {Paltani}, {Aller},
  {Aller}, {Burki}, {Chernyakova}, {L{\"a}hteenm{\"a}ki}, {McHardy} and
  {Robson}}]{2008A&A...486..411S}
\bibinfo{author}{{Soldi}, S.}, \bibinfo{author}{{T{\"u}rler}, M.},
  \bibinfo{author}{{Paltani}, S.}, \bibinfo{author}{{Aller}, H.D.},
  \bibinfo{author}{{Aller}, M.F.}, \bibinfo{author}{{Burki}, G.},
  \bibinfo{author}{{Chernyakova}, M.}, \bibinfo{author}{{L{\"a}hteenm{\"a}ki},
  A.}, \bibinfo{author}{{McHardy}, I.M.}, \bibinfo{author}{{Robson}, E.I.},
  \bibinfo{year}{2008}.
\newblock \bibinfo{title}{{The multiwavelength variability of 3C 273}}.
\newblock \bibinfo{journal}{\aap} \bibinfo{volume}{486},
  \bibinfo{pages}{411--425}.
\newblock \href{http://arxiv.org/abs/0805.3411}{\tt arXiv:0805.3411}.
\bibitem[{{Steffen} et~al.(2003){Steffen}, {Barger}, {Cowie}, {Mushotzky} and
  {Yang}}]{2003ApJ...596L..23S}
\bibinfo{author}{{Steffen}, A.T.}, \bibinfo{author}{{Barger}, A.J.},
  \bibinfo{author}{{Cowie}, L.L.}, \bibinfo{author}{{Mushotzky}, R.F.},
  \bibinfo{author}{{Yang}, Y.}, \bibinfo{year}{2003}.
\newblock \bibinfo{title}{{The Changing Active Galactic Nucleus Population}}.
\newblock \bibinfo{journal}{\apjl} \bibinfo{volume}{596},
  \bibinfo{pages}{L23--L26}.
\newblock \href{http://arxiv.org/abs/astro-ph/0308238}{\tt
  arXiv:astro-ph/0308238}.
\bibitem[{{Stephen} et~al.(2005){Stephen}, {Bassani}, {Molina}, {Malizia},
  {Bazzano}, {Ubertini}, {Dean}, {Bird}, {Lebrun}, {Much} and
  {Walter}}]{2005A&A...432L..49S}
\bibinfo{author}{{Stephen}, J.B.}, \bibinfo{author}{{Bassani}, L.},
  \bibinfo{author}{{Molina}, M.}, \bibinfo{author}{{Malizia}, A.},
  \bibinfo{author}{{Bazzano}, A.}, \bibinfo{author}{{Ubertini}, P.},
  \bibinfo{author}{{Dean}, A.J.}, \bibinfo{author}{{Bird}, A.J.},
  \bibinfo{author}{{Lebrun}, F.}, \bibinfo{author}{{Much}, R.},
  \bibinfo{author}{{Walter}, R.}, \bibinfo{year}{2005}.
\newblock \bibinfo{title}{{Using the ROSAT Bright Source Catalogue to find
  counterparts for IBIS/ISGRI survey sources}}.
\newblock \bibinfo{journal}{\aap} \bibinfo{volume}{432},
  \bibinfo{pages}{L49--L52}.
\newblock \href{http://arxiv.org/abs/astro-ph/0502158}{\tt
  arXiv:astro-ph/0502158}.
\bibitem[{{Subrahmanyan} et~al.(1996){Subrahmanyan}, {Saripalli} and
  {Hunstead}}]{1996MNRAS.279..257S}
\bibinfo{author}{{Subrahmanyan}, R.}, \bibinfo{author}{{Saripalli}, L.},
  \bibinfo{author}{{Hunstead}, R.W.}, \bibinfo{year}{1996}.
\newblock \bibinfo{title}{{Morphologies in megaparsec-size powerful radio
  galaxies}}.
\newblock \bibinfo{journal}{\mnras} \bibinfo{volume}{279},
  \bibinfo{pages}{257--274}.
\bibitem[{{The Fermi-LAT collaboration}(2019)}]{2019arXiv190510771T}
\bibinfo{author}{{The Fermi-LAT collaboration}}, \bibinfo{year}{2019}.
\newblock \bibinfo{title}{{The Fourth Catalog of Active Galactic Nuclei
  Detected by the Fermi Large Area Telescope}}.
\newblock \bibinfo{journal}{arXiv e-prints} ,
  \bibinfo{pages}{arXiv:1905.10771}\href{http://arxiv.org/abs/1905.10771}{\tt
  arXiv:1905.10771}.
\bibitem[{{Ubertini} et~al.(2003){Ubertini}, {Lebrun}, {Di Cocco}, {Bazzano},
  {Bird}, {Broenstad}, {Goldwurm}, {La Rosa}, {Labanti}, {Laurent}, {Mirabel},
  {Quadrini}, {Ramsey}, {Reglero}, {Sabau}, {Sacco}, {Staubert}, {Vigroux},
  {Weisskopf} and {Zdziarski}}]{2003A&A...411L.131U}
\bibinfo{author}{{Ubertini}, P.}, \bibinfo{author}{{Lebrun}, F.},
  \bibinfo{author}{{Di Cocco}, G.}, \bibinfo{author}{{Bazzano}, A.},
  \bibinfo{author}{{Bird}, A.J.}, \bibinfo{author}{{Broenstad}, K.},
  \bibinfo{author}{{Goldwurm}, A.}, \bibinfo{author}{{La Rosa}, G.},
  \bibinfo{author}{{Labanti}, C.}, \bibinfo{author}{{Laurent}, P.},
  \bibinfo{author}{{Mirabel}, I.F.}, \bibinfo{author}{{Quadrini}, E.M.},
  \bibinfo{author}{{Ramsey}, B.}, \bibinfo{author}{{Reglero}, V.},
  \bibinfo{author}{{Sabau}, L.}, \bibinfo{author}{{Sacco}, B.},
  \bibinfo{author}{{Staubert}, R.}, \bibinfo{author}{{Vigroux}, L.},
  \bibinfo{author}{{Weisskopf}, M.C.}, \bibinfo{author}{{Zdziarski}, A.A.},
  \bibinfo{year}{2003}.
\newblock \bibinfo{title}{{IBIS: The Imager on-board INTEGRAL}}.
\newblock \bibinfo{journal}{\aap} \bibinfo{volume}{411},
  \bibinfo{pages}{L131--L139}.
\newblock \DOIprefix\doi{10.1051/0004-6361:20031224}.
\bibitem[{{Ubertini} et~al.(2009){Ubertini}, {Sguera}, {Stephen}, {Bassani},
  {Bazzano} and {Bird}}]{2009ApJ...706L...7U}
\bibinfo{author}{{Ubertini}, P.}, \bibinfo{author}{{Sguera}, V.},
  \bibinfo{author}{{Stephen}, J.B.}, \bibinfo{author}{{Bassani}, L.},
  \bibinfo{author}{{Bazzano}, A.}, \bibinfo{author}{{Bird}, A.J.},
  \bibinfo{year}{2009}.
\newblock \bibinfo{title}{{The Fermi/LAT Sky as Seen by INTEGRAL/IBIS}}.
\newblock \bibinfo{journal}{\apjl} \bibinfo{volume}{706},
  \bibinfo{pages}{L7--L11}.
\newblock \DOIprefix\doi{10.1088/0004-637X/706/1/L7},
  \href{http://arxiv.org/abs/0910.1738}{\tt arXiv:0910.1738}.
\bibitem[{{Ueda} et~al.(2014){Ueda}, {Akiyama}, {Hasinger}, {Miyaji} and
  {Watson}}]{2014ApJ...786..104U}
\bibinfo{author}{{Ueda}, Y.}, \bibinfo{author}{{Akiyama}, M.},
  \bibinfo{author}{{Hasinger}, G.}, \bibinfo{author}{{Miyaji}, T.},
  \bibinfo{author}{{Watson}, M.G.}, \bibinfo{year}{2014}.
\newblock \bibinfo{title}{{Toward the Standard Population Synthesis Model of
  the X-Ray Background: Evolution of X-Ray Luminosity and Absorption Functions
  of Active Galactic Nuclei Including Compton-thick Populations}}.
\newblock \bibinfo{journal}{\apj} \bibinfo{volume}{786}, \bibinfo{pages}{104}.
\newblock \href{http://arxiv.org/abs/1402.1836}{\tt arXiv:1402.1836}.
\bibitem[{{Ueda} et~al.(2003){Ueda}, {Akiyama}, {Ohta} and
  {Miyaji}}]{2003ApJ...598..886U}
\bibinfo{author}{{Ueda}, Y.}, \bibinfo{author}{{Akiyama}, M.},
  \bibinfo{author}{{Ohta}, K.}, \bibinfo{author}{{Miyaji}, T.},
  \bibinfo{year}{2003}.
\newblock \bibinfo{title}{{Cosmological Evolution of the Hard X-Ray Active
  Galactic Nucleus Luminosity Function and the Origin of the Hard X-Ray
  Background}}.
\newblock \bibinfo{journal}{\apj} \bibinfo{volume}{598},
  \bibinfo{pages}{886--908}.
\newblock \href{http://arxiv.org/abs/astro-ph/0308140}{\tt
  arXiv:astro-ph/0308140}.
\bibitem[{Urry and Padovani(1995)}]{Urry_1995}
\bibinfo{author}{Urry, C.M.}, \bibinfo{author}{Padovani, P.},
  \bibinfo{year}{1995}.
\newblock \bibinfo{title}{Unified schemes for radio-loud active galactic
  nuclei}.
\newblock \bibinfo{journal}{Publications of the Astronomical Society of the
  Pacific} \bibinfo{volume}{107}, \bibinfo{pages}{803}.
\bibitem[{{Ursini} et~al.(2018a){Ursini}, {Bassani}, {Panessa}, {Bazzano},
  {Bird}, {Malizia} and {Ubertini}}]{2018MNRAS.474.5684U}
\bibinfo{author}{{Ursini}, F.}, \bibinfo{author}{{Bassani}, L.},
  \bibinfo{author}{{Panessa}, F.}, \bibinfo{author}{{Bazzano}, A.},
  \bibinfo{author}{{Bird}, A.J.}, \bibinfo{author}{{Malizia}, A.},
  \bibinfo{author}{{Ubertini}, P.}, \bibinfo{year}{2018}a.
\newblock \bibinfo{title}{{Where are Compton-thick radio galaxies? A hard X-ray
  view of three candidates}}.
\newblock \bibinfo{journal}{\mnras} \bibinfo{volume}{474},
  \bibinfo{pages}{5684--5693}.
\newblock \href{http://arxiv.org/abs/1712.01300}{\tt arXiv:1712.01300}.
\bibitem[{{Ursini} et~al.(2018b){Ursini}, {Bassani}, {Panessa}, {Bird},
  {Bruni}, {Fiocchi}, {Malizia}, {Saripalli} and
  {Ubertini}}]{2018MNRAS.481.4250U}
\bibinfo{author}{{Ursini}, F.}, \bibinfo{author}{{Bassani}, L.},
  \bibinfo{author}{{Panessa}, F.}, \bibinfo{author}{{Bird}, A.J.},
  \bibinfo{author}{{Bruni}, G.}, \bibinfo{author}{{Fiocchi}, M.},
  \bibinfo{author}{{Malizia}, A.}, \bibinfo{author}{{Saripalli}, L.},
  \bibinfo{author}{{Ubertini}, P.}, \bibinfo{year}{2018}b.
\newblock \bibinfo{title}{{Hard X-ray-selected giant radio galaxies - I. The
  X-ray properties and radio connection}}.
\newblock \bibinfo{journal}{\mnras} \bibinfo{volume}{481},
  \bibinfo{pages}{4250--4260}.
\newblock \href{http://arxiv.org/abs/1809.05297}{\tt arXiv:1809.05297}.
\bibitem[{{Vercellone} et~al.(2011){Vercellone}, {Striani}, {Vittorini},
  {Donnarumma}, {Pacciani}, {Pucella}, {Tavani}, {Raiteri}, {Villata},
  {Romano}, {Fiocchi}, {Bazzano}, {Bianchin}, {Ferrigno}, {Maraschi}, {Pian},
  {T{\"u}rler}, {Ubertini}, {Bulgarelli}, {Chen}, {Giuliani}, {Longo},
  {Barbiellini}, {Cardillo}, {Cattaneo}, {Del Monte}, {Evangelista}, {Feroci},
  {Ferrari}, {Fuschino}, {Gianotti}, {Giusti}, {Lazzarotto}, {Pellizzoni},
  {Piano}, {Pilia}, {Rapisarda}, {Rappoldi}, {Sabatini}, {Soffitta},
  {Trifoglio}, {Trois}, {Giommi}, {Lucarelli}, {Pittori}, {Santolamazza},
  {Verrecchia}, {Agudo}, {Aller}, {Aller}, {Arkharov}, {Bach}, {Berdyugin},
  {Borman}, {Chigladze}, {Efimov}, {Efimova}, {G{\'o}mez}, {Gurwell},
  {McHardy}, {Joshi}, {Kimeridze}, {Krajci}, {Kurtanidze}, {Kurtanidze},
  {Larionov}, {Lindfors}, {Molina}, {Morozova}, {Nazarov}, {Nikolashvili},
  {Nilsson}, {Pasanen}, {Reinthal}, {Ros}, {Sadun}, {Sakamoto}, {Sallum},
  {Sergeev}, {Schwartz}, {Sigua}, {Sillanp{\"a}{\"a}}, {Sokolovsky},
  {Strelnitski}, {Takalo}, {Taylor} and {Walker}}]{2011ApJ...736L..38V}
\bibinfo{author}{{Vercellone}, S.}, \bibinfo{author}{{Striani}, E.},
  \bibinfo{author}{{Vittorini}, V.}, \bibinfo{author}{{Donnarumma}, I.},
  \bibinfo{author}{{Pacciani}, L.}, \bibinfo{author}{{Pucella}, G.},
  \bibinfo{author}{{Tavani}, M.}, \bibinfo{author}{{Raiteri}, C.M.},
  \bibinfo{author}{{Villata}, M.}, \bibinfo{author}{{Romano}, P.},
  \bibinfo{author}{{Fiocchi}, M.}, \bibinfo{author}{{Bazzano}, A.},
  \bibinfo{author}{{Bianchin}, V.}, \bibinfo{author}{{Ferrigno}, C.},
  \bibinfo{author}{{Maraschi}, L.}, \bibinfo{author}{{Pian}, E.},
  \bibinfo{author}{{T{\"u}rler}, M.}, \bibinfo{author}{{Ubertini}, P.},
  \bibinfo{author}{{Bulgarelli}, A.}, \bibinfo{author}{{Chen}, A.W.},
  \bibinfo{author}{{Giuliani}, A.}, \bibinfo{author}{{Longo}, F.},
  \bibinfo{author}{{Barbiellini}, G.}, \bibinfo{author}{{Cardillo}, M.},
  \bibinfo{author}{{Cattaneo}, P.W.}, \bibinfo{author}{{Del Monte}, E.},
  \bibinfo{author}{{Evangelista}, Y.}, \bibinfo{author}{{Feroci}, M.},
  \bibinfo{author}{{Ferrari}, A.}, \bibinfo{author}{{Fuschino}, F.},
  \bibinfo{author}{{Gianotti}, F.}, \bibinfo{author}{{Giusti}, M.},
  \bibinfo{author}{{Lazzarotto}, F.}, \bibinfo{author}{{Pellizzoni}, A.},
  \bibinfo{author}{{Piano}, G.}, \bibinfo{author}{{Pilia}, M.},
  \bibinfo{author}{{Rapisarda}, M.}, \bibinfo{author}{{Rappoldi}, A.},
  \bibinfo{author}{{Sabatini}, S.}, \bibinfo{author}{{Soffitta}, P.},
  \bibinfo{author}{{Trifoglio}, M.}, \bibinfo{author}{{Trois}, A.},
  \bibinfo{author}{{Giommi}, P.}, \bibinfo{author}{{Lucarelli}, F.},
  \bibinfo{author}{{Pittori}, C.}, \bibinfo{author}{{Santolamazza}, P.},
  \bibinfo{author}{{Verrecchia}, F.}, \bibinfo{author}{{Agudo}, I.},
  \bibinfo{author}{{Aller}, H.D.}, \bibinfo{author}{{Aller}, M.F.},
  \bibinfo{author}{{Arkharov}, A.A.}, \bibinfo{author}{{Bach}, U.},
  \bibinfo{author}{{Berdyugin}, A.}, \bibinfo{author}{{Borman}, G.A.},
  \bibinfo{author}{{Chigladze}, R.}, \bibinfo{author}{{Efimov}, Y.S.},
  \bibinfo{author}{{Efimova}, N.V.}, \bibinfo{author}{{G{\'o}mez}, J.L.},
  \bibinfo{author}{{Gurwell}, M.A.}, \bibinfo{author}{{McHardy}, I.M.},
  \bibinfo{author}{{Joshi}, M.}, \bibinfo{author}{{Kimeridze}, G.N.},
  \bibinfo{author}{{Krajci}, T.}, \bibinfo{author}{{Kurtanidze}, O.M.},
  \bibinfo{author}{{Kurtanidze}, S.O.}, \bibinfo{author}{{Larionov}, V.M.},
  \bibinfo{author}{{Lindfors}, E.}, \bibinfo{author}{{Molina}, S.N.},
  \bibinfo{author}{{Morozova}, D.A.}, \bibinfo{author}{{Nazarov}, S.V.},
  \bibinfo{author}{{Nikolashvili}, M.G.}, \bibinfo{author}{{Nilsson}, K.},
  \bibinfo{author}{{Pasanen}, M.}, \bibinfo{author}{{Reinthal}, R.},
  \bibinfo{author}{{Ros}, J.A.}, \bibinfo{author}{{Sadun}, A.C.},
  \bibinfo{author}{{Sakamoto}, T.}, \bibinfo{author}{{Sallum}, S.},
  \bibinfo{author}{{Sergeev}, S.G.}, \bibinfo{author}{{Schwartz}, R.D.},
  \bibinfo{author}{{Sigua}, L.A.}, \bibinfo{author}{{Sillanp{\"a}{\"a}}, A.},
  \bibinfo{author}{{Sokolovsky}, K.V.}, \bibinfo{author}{{Strelnitski}, V.},
  \bibinfo{author}{{Takalo}, L.}, \bibinfo{author}{{Taylor}, B.},
  \bibinfo{author}{{Walker}, G.}, \bibinfo{year}{2011}.
\newblock \bibinfo{title}{{The Brightest Gamma-Ray Flaring Blazar in the Sky:
  AGILE and Multi-wavelength Observations of 3C 454.3 During 2010 November}}.
\newblock \bibinfo{journal}{\apjl} \bibinfo{volume}{736}, \bibinfo{pages}{L38}.
\newblock \href{http://arxiv.org/abs/1106.5162}{\tt arXiv:1106.5162}.
\bibitem[{{We{\.z}gowiec} et~al.(2016){We{\.z}gowiec}, {Jamrozy} and
  {Mack}}]{2016AcA....66...85W}
\bibinfo{author}{{We{\.z}gowiec}, M.}, \bibinfo{author}{{Jamrozy}, M.},
  \bibinfo{author}{{Mack}, K.H.}, \bibinfo{year}{2016}.
\newblock \bibinfo{title}{{1.4-GHz Observations of Extended Giant Radio
  Galaxies}}.
\newblock \bibinfo{journal}{\actaa} \bibinfo{volume}{66},
  \bibinfo{pages}{85--119}.
\newblock \href{http://arxiv.org/abs/1604.05142}{\tt arXiv:1604.05142}.
\bibitem[{Woo et~al.(2019)Woo, Cho, Gallo, Hodges-Kluck, Le, Shin, Son and
  Horst}]{Woo_2019}
\bibinfo{author}{Woo, J.H.}, \bibinfo{author}{Cho, H.}, \bibinfo{author}{Gallo,
  E.}, \bibinfo{author}{Hodges-Kluck, E.}, \bibinfo{author}{Le, H.A.N.},
  \bibinfo{author}{Shin, J.}, \bibinfo{author}{Son, D.},
  \bibinfo{author}{Horst, J.C.}, \bibinfo{year}{2019}.
\newblock \bibinfo{title}{A 10,000-solar-mass black hole in the nucleus of a
  bulgeless dwarf galaxy}.
\newblock \bibinfo{journal}{Nature Astronomy} .
\bibitem[{Zdziarski(1998)}]{Zdziarski_1998}
\bibinfo{author}{Zdziarski, A.A.}, \bibinfo{year}{1998}.
\newblock \bibinfo{title}{Hot accretion discs with thermal comptonization and
  advection in luminous black hole sources}.
\newblock \bibinfo{journal}{Monthly Notices of the Royal Astronomical Society}
  \bibinfo{volume}{296}, \bibinfo{pages}{L51--L55}.
\bibitem[{{Zurita Heras} et~al.(2009){Zurita Heras}, {Chaty} and
  {Tomsick}}]{2009A&A...502..787Z}
\bibinfo{author}{{Zurita Heras}, J.A.}, \bibinfo{author}{{Chaty}, S.},
  \bibinfo{author}{{Tomsick}, J.A.}, \bibinfo{year}{2009}.
\newblock \bibinfo{title}{{Infrared identification of IGR
  J09026-4812{\enskip}as a Seyfert 1 galaxy}}.
\newblock \bibinfo{journal}{\aap} \bibinfo{volume}{502},
  \bibinfo{pages}{787--790}.
\newblock \DOIprefix\doi{10.1051/0004-6361/200912359},
  \href{http://arxiv.org/abs/0905.3309}{\tt arXiv:0905.3309}.

\end{thebibliography}

\end{document}